\documentclass[journal]{IEEEtran}

\usepackage{multirow}
\usepackage{graphicx}
\usepackage{array}
\usepackage{listings}
\usepackage{enumitem}
\usepackage{xcolor,colortbl}

\usepackage[hidelinks]{hyperref}

\usepackage{makecell}

\usepackage{amssymb}
\usepackage{pifont}

\usepackage{fancyhdr} 
\usepackage{xcolor}

\usepackage{cite}

\newcounter{mysubsubsection}[subsection]
\newcommand{\mysubsubsection}[1]{\refstepcounter{mysubsubsection}\par\smallskip\textbf{\themysubsubsection) #1:}}
\newcommand{\myparagraph}[1]{\smallskip\textbf{#1}} 

\definecolor{listingmauve}{RGB}{224,176,255}

\newcommand{\polyc}[1]{
                      {\footnotesize
                       \newline
                       \centerline{#1}
                      }
                     }
\newcolumntype{M}[1]{>{\centering\arraybackslash}m{#1}}

\setlist{leftmargin=5.5mm}

\definecolor{listinggreen}{rgb}{0,0.6,0}
\definecolor{listinggray}{rgb}{0.5,0.5,0.5}
\definecolor{listingmauve}{rgb}{0.58,0,0.82}
\definecolor{listingkeywordcolor}{rgb}{1.0,0.4,0.0}
\definecolor{listinglightgray}{rgb}{0.8863,0.8863,0.8863}

\newcommand{\insight}[1]{\textit{#1}}

\begin{document}

\title{Hardware Acceleration of Sparse and Irregular Tensor Computations of ML Models:\\ A Survey and Insights}

\author{Shail~Dave,~\IEEEmembership{Student Member,~IEEE,}
        Riyadh~Baghdadi,~\IEEEmembership{Member,~IEEE,}
        Tony~Nowatzki,~\IEEEmembership{Member,~IEEE,}
        Sasikanth~Avancha,~\IEEEmembership{Member,~IEEE,}
        Aviral~Shrivastava,~\IEEEmembership{Senior Member,~IEEE,}
        and~Baoxin~Li,~\IEEEmembership{Senior Member,~IEEE}
\thanks{Shail Dave, Aviral Shrivastava, and Baoxin Li are with the School of Computing Informatics and Decision Systems Engineering at Arizona State University, Tempe, AZ, 85281 USA (e-mail: shail.dave@asu.edu; aviral.shrivastava@asu.edu; baoxin.li@asu.edu).}
\thanks{Riyadh Baghdadi is with the Computer Science and Artificial Intelligence Laboratory at Massachusetts Institute of Technology, Cambridge, MA 02139 USA (e-mail: baghdadi@mit.edu).}%
\thanks{Tony Nowatzki is with the School of Computer Science at University of California, Los Angeles, CA 90095 USA (e-mail: tjn@cs.ucla.edu).}%
\thanks{Sasikanth Avancha is with the Parallel Computing Lab at Intel Labs, Bangalore, India (email:sasikanth.avancha@intel.com).}
\thanks{\textcopyright  2021 IEEE.  Personal use of this material is permitted.  Permission from IEEE must be obtained for all other uses, in any current or future media, including reprinting/republishing this material for advertising or promotional purposes, creating new collective works, for resale or redistribution to servers or lists, or reuse of any copyrighted component of this work in other works.}        
}


\maketitle

\begin{abstract}
Machine learning (ML) models are widely used in many important domains. For efficiently processing these computational- and memory-intensive applications, tensors of these over-parameterized models are compressed by leveraging sparsity, size reduction, and quantization of tensors. Unstructured sparsity and tensors with varying dimensions yield irregular computation, communication, and memory access patterns; processing them on hardware accelerators in a conventional manner does not inherently leverage acceleration opportunities. This paper provides a comprehensive survey on the efficient execution of sparse and irregular tensor computations of ML models on hardware accelerators. In particular, it discusses enhancement modules in the architecture design and the software support; categorizes different hardware designs and acceleration techniques and analyzes them in terms of hardware and execution costs; analyzes achievable accelerations for recent DNNs; highlights further opportunities in terms of hardware/software/model co-design optimizations (inter/intra-module). The takeaways from this paper include: understanding the key challenges in accelerating sparse, irregular-shaped, and quantized tensors; understanding enhancements in accelerator systems for supporting their efficient computations; analyzing trade-offs in opting for a specific design choice for encoding, storing, extracting, communicating, computing, and load-balancing the non-zeros; understanding how structured sparsity can improve storage efficiency and balance computations; understanding how to compile and map models with sparse tensors on the accelerators; understanding recent design trends for efficient accelerations and further opportunities. 
\end{abstract}

\vspace{-0.45cm}

\IEEEpeerreviewmaketitle

\vspace{-0.25cm}
\section{Introduction}

Machine learning (ML) models implement intelligence in computing systems. Different ML models are widely used in several important domains including computer vision (object classification \cite{krizhevsky2012imagenet, he2016deep, tan2019efficientnet} and detection \cite{redmon2018yolov3, chen2017rethinking, tran2015c3d}), natural language processing  \cite{vaswani2017attention, devlin2019bert, brown2020language}, media generation \cite{goodfellow2014generative}, recommendation systems \cite{park2018deep, dlrm}, medical diagnosis \cite{litjens2017survey}, large-scale scientific computing \cite{kurth2018exascale}, embedded systems \cite{rezk2020recurrent}, mobile and edge processing \cite{zhang2019deep, banbury2020benchmarking}, and even for designing or optimizing hardware and software systems \cite{dean20201, olukotun2018designing}. Domain-customized accelerators can significantly speed up their execution in an energy-efficient manner \cite{chen2016eyeriss, fleischer2018scalable, jouppi2017datacenter, yang2018dnn}. However, the computational and memory requirements for processing these models have surged drastically \cite{openaiblog}. Moreover, ML models can be deeper and larger, which improves learning accuracy, but significant redundancy may exist in these often over-parameterized models \cite{han2015learning, srivastava2014dropout}. Therefore, recent techniques for efficient learning and inference have proposed compressing tensors of ML models. Tensors are compressed by inducing and leveraging: \textit{(a) sparsity (zero values in tensors)} \cite{han2015deep, wen2016learning, frankle2018lottery, mishra2017fine, han2020survey}, \textit{(b) size reduction (tensor decomposition, dimension reduction, and shape reduction)} \cite{tan2019efficientnet, sandler2018mobilenetv2, iandola2016squeezenet, cichocki2015tensor, ragged_tensors}, and \textit{(c) quantization (precision lowering and leveraging value similarity)} \cite{krishnamoorthi2018quantizing, han2015deep}. With significantly lowered computational, storage, and communication requirements, efficient processing of \textit{compressed tensors (sparse, size-reduced, and quantized)} offers notable acceleration and energy efficiency opportunities \cite{zhou2018cambricon, ren2019admm, narang2017exploring, yang2017designing}.

\begin{figure*}[t!]
\centering
\centerline{\includegraphics[width=0.95\linewidth]{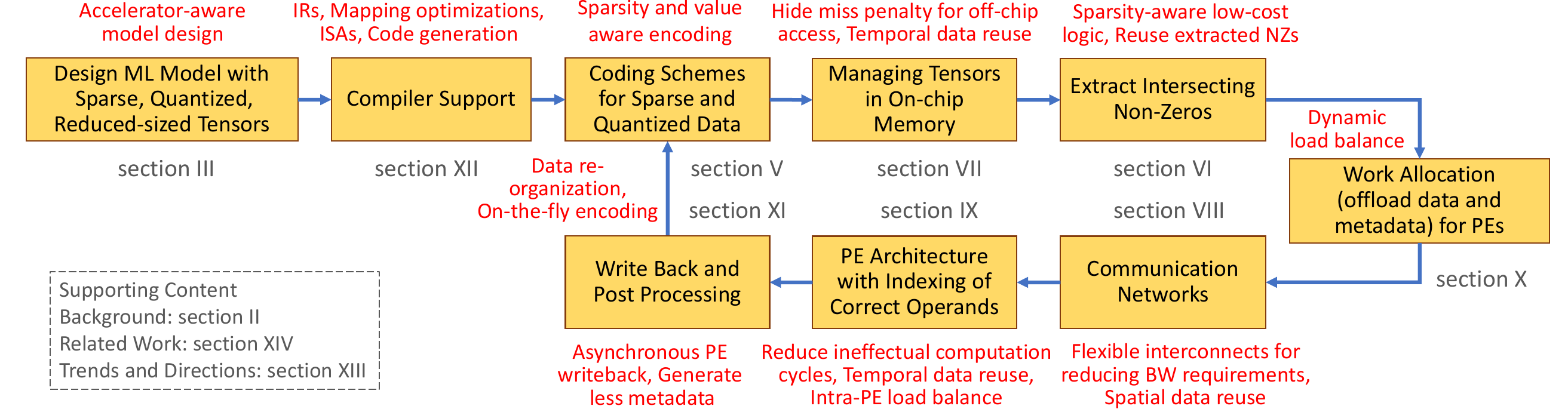}}
\caption{Overview of the accelerator system for processing sparse and irregular tensor computations. (Section \ref{sec::overview-accel-compressed-tensors} provides further discussion.)}
\label{fig::sparse-accel-block-diagram}
\end{figure*}

Hardware accelerators can efficiently process tensor computations of ML models. In particular, coarse-grain spatial architectures are a common choice for hardware accelerator designs. They contain an array of processing elements (PEs) with local registers/memory and shared memory. These accelerators feature interconnects like mesh or multicast for communicating data to PEs and reusing the data spatially, which reduces the accesses to the memory hierarchy. With simple PE designs and effective spatial and temporal management of the data and computations, such architectures achieve high speedups and energy-efficiency \cite{fleischer2018scalable, jouppi2017datacenter, chen2016eyeriss}.

Special mechanisms are needed to exploit the acceleration benefits due to tensor sparsity, size reduction, and quantization. This is because, while hardware accelerators for ML can process low-precision tensors, they inherently cannot benefit from sparsity \cite{zhang2016cambricon, han2016eie}. They are designed for performing structured computations with regular memory accesses and communication patterns. Without special support for sparse tensors, they fetch all the data, including zero values from memory and feed into PEs, thereby wasting the execution time. Sparsity, especially unstructured, induces irregularity in processing since non-zeros (NZs) or blocks of NZs are scattered across tensors. So, leveraging sparsity necessitates additional mechanisms to store, extract, communicate, compute, and load-balance the NZs and the corresponding hardware or software support. The goal of exploiting sparsity is to exploit all forms of sparsity possible to considerably reduce computation, communication, and storage of zeros while avoiding adding performance, power, and area overheads. Exploiting sparsity effectively depends on tailoring the data encoding and extraction, dataflow, memory banking structure, interconnect design, and write-back mechanisms. Further, it requires new representations and enables new opportunities for hardware/software/model co-designs. In this survey, we mainly discuss different accelerator designs that have leveraged the sparsity of different tensors and different opportunities for performance gains and energy efficiency. 
Tensor decomposition and dimension reduction yield tensors of various sizes and asymmetric shapes \cite{tan2019efficientnet, chen2019eyeriss}. Dataflow mechanisms for executing layers of the models are typically optimized well for some commonly used layers (symmetric dimensions). They often become ill-suited for processing tensors with reduced dimensions \cite{chen2019eyeriss} and different functionality. So, we describe how configurable designs and flexible dataflows can help to achieve efficient execution. Sparse tensors quantized with value sharing require additional support to index a dictionary for obtaining shared values. The survey also discusses how accelerators leverage value similarity across inputs, weights, or outputs and support variable bit-widths of sparse tensors.

\textbf{Contributions:} This paper provides a comprehensive survey of different techniques for efficiently executing sparse and irregular tensor computations of the compact ML models on hardware accelerators. It describes corresponding enhancements in the hardware architecture and the required software support. In specific, 
\begin{itemize}
    \item For inference and training of different ML models, we summarize various sources of the sparsity of tensors.
    \item We highlight challenges in accelerating computations of sparse (especially unstructured) and irregular-shaped tensors (e.g., dot product, convolution, and matrix multiplication) on spatial-architecture-based hardware accelerators that execute with dataflow mechanisms.
    \item We present an overview of the accelerator system along with the different hardware/software modules for sparse and irregular computations, their interfacing, and the execution flow. We provide an in-depth discussion of the need of each module, different design choices, and qualitative analysis of the different choices.
    \item We survey different accelerator systems and execution techniques for sparse tensors of ML models and provide taxonomies to categorize them based on the various hardware/software aspects of the designs.
     \item We analyze how variations in sparsity and tensor shapes of different models impact the storage efficiency of different sparsity-encodings and the reuse of tensors. 
    \item For designing these accelerator modules and overall accelerator system, we discuss recent trends and outline further opportunities for hardware/software/model co-designs.
\end{itemize} 

\textbf{Paper organization:}  
\begin{itemize}[leftmargin=*]
    \item Section \ref{sec::background} provides a brief background on different ML models, hardware accelerators for their tensor computations, and the need for further efficiency by reducing computation, storage, and communication requirements.
    \item Section \ref{sec::tensor-compression} discusses tensor compression and opportunities due to sparse, size-reduced, and quantized tensors and why their efficient processing  requires special support.
    \item Section \ref{sec::overview-accel-compressed-tensors} provides an overview of the accelerator system with enhanced architectural modules and software support for sparse and irregular tensor computations (Fig. \ref{fig::sparse-accel-block-diagram}). It also presents a case study of accelerations of recent, sparse DNNs and analyzes execution bottlenecks. In-depth discussions of individual modules follow through sections \ref{sec::sparse-data-coding}--\ref{sec::software-support}. Opportunities for optimizing each module further are discussed at the end of corresponding sections or subsections.
    \item Section \ref{sec::sparse-data-coding} illustrates common sparse data encodings, analyzes their implications in terms of storage and coding overheads, and describes the group-wise encoding of tensors.
    \item Section \ref{sec::NZ-data-extraction} discusses techniques for extracting matching NZs from tensors for computations. It analyzes the advantages and limitations of the centralized and in-PE extractions.
    \item Section \ref{sec::memory-management} discusses managing non-coherent, multi-banked, global scratchpad and hiding the memory access latency behind computations. It also discusses data reuse of the sparse tensors and cross-layer reuse opportunities. 
    \item Section \ref{sec::comm-networks} discusses interconnect designs for distributing data from memory and reducing partial outputs, their bandwidth requirements, spatial data reuse, and their configurability to support multiple dataflows for execution. 
    \item Section \ref{sec::PEArch} describes sparsity-aware dataflows and pipelined PE architecture including tailoring functional units for sparsity, bit-adaptive computing, and leveraging value similarity. 
    \item Section \ref{sec::load-balance} discusses sources of the inter-PE and intra-PE imbalance due to sparsity and their impact, software-directed balancing, and hardware structures for dynamic balancing. 
    \item Section \ref{sec::post-processing} describes different write-back mechanisms for collecting data from PEs and assembling the data locally in PEs or on a central module. It also discusses data layout transformations and on-the-fly encoding of sparse outputs.
    \item Section \ref{sec::software-support} discusses compiler support for targeting hardware accelerators, including intermediate representations for deep learning models, compiler optimizations and their automation, and ISAs and code generation for accelerators.
    \item Section \ref{sec::future-directions} describes recent trends and future directions in terms of developing tools and techniques for systematic exploration of hardware/software/model co-designs.
    \item Section \ref{sec::related-works} discusses relevant surveys that describe additional details (domain-specific models, tensor compression techniques, etc.) and can be useful to readers. 
\end{itemize}

\section{Background: Need for Efficient Execution of ML Models on Hardware Accelerators}
\label{sec::background}

\subsection{Domain-Specific Machine Learning Models}

Learning through ML models can be supervised (where labeled data is available), unsupervised (training samples are unlabeled), or semi-supervised. We refer non-expert readers to surveys \cite{sze2017efficient, pouyanfar2018survey, alom2019state} for a detailed discussion on different learning approaches and inference and training of various models. Discussions through this survey mainly focus on accelerating different \textit{deep neural networks (DNNs)} that are commonly used for supervised learning. 

\textbf{Convolutional neural networks (CNNs)} are used for object classification and detection in image processing, video analysis, and autonomous vehicle systems. CNNs majorly consist of many \textit{convolution layers (CONV)} and a \textit{few fully-connected (FC)} layers. Early CONV layers capture low-level features from the images (e.g., edges and corners), which are used for constructing high-level features (e.g., shapes) by subsequent layers. Finally, the classifier aka FC layer determines the type of the objects \cite{sze2017efficient}.

\textbf{Sequence-to-sequence models} include recurrent neural networks (RNNs), gated recurrent units (GRU), long-short term memory (LSTM) \cite{rezk2020recurrent}, and attention mechanisms \cite{vaswani2017attention, devlin2019bert}. These models are used for \textit{natural language processing (NLP)} and media processing tasks. They essentially use unidirectional or bidirectional recurrent cells at their core and process \textit{multi-layer perceptrons (MLP)} aka FC structures. 

\textbf{Models for semantic segmentation and language translation} use encoder-decoder structures with convolutions \cite{chen2017rethinking, kurth2018exascale}, recurrent cells, or attention layers \cite{devlin2019bert}, respectively.

\textbf{Generative adversarial networks (GANs)} \cite{goodfellow2014generative} are used by media generation applications. GANs use generators and discriminative networks that consist of convolution layers. 

\textbf{Graph neural networks (GNNs)} and other graph learning models \cite{hamilton2017representation} are used for applications such as text classification and translation, node classification and link predictions in large social graphs, etc. They learn graph properties and infer about unforeseen information. To achieve this objective, each node contains an embedding feature vector with the information mixture about own and neighborhood features. The nodes then recurrently aggregate features of local neighbors, perform neural network computations on aggregated data (e.g., MLP for down-scaling embeddings), and update their embeddings. 

\textbf{Recommendation system models} consist of embedding layers (look-ups and matrix operations) \cite{dlrm}, CNNs for object detection and video understanding, and RNNs for processing language models \cite{park2018deep}. 

Primitives like MLP or GEMM (general matrix multiply) and CONV are at the core of many models and dominate the execution. So, ML frameworks like PyTorch \cite{paszke2019pytorch}, TensorFlow \cite{abadi2016tensorflow}, and Intel MKL \cite{wang2014intel} provide efficient implementations of these primitives for execution on commodity hardware (CPUs, GPUs, FPGAs) or even specialized accelerators. So, our discussions mainly focus on efficiently accelerating tensor computations of MLP, CONV, and RNN operators.

\subsection{Hardware Accelerators for Machine Learning}

\begin{figure}[t]
\centering
\centerline{\includegraphics[width=\linewidth]{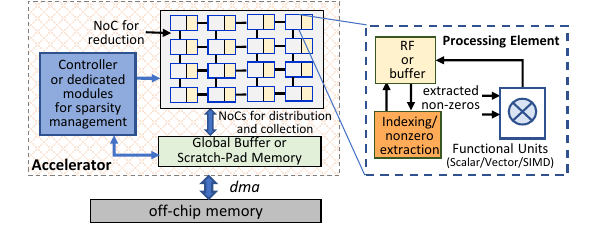}}
\caption{Abstract accelerator design for processing sparse tensors of machine learning applications. Execution of applications require explicit management of computational, communication, and memory resources.}
\label{fig::ML-accelerators}
\end{figure}

In the “new golden age of computer architecture”, recent research efforts and commercial solutions have extensively demonstrated that domain-customized hardware accelerators significantly speed up the execution of ML models in an energy-efficient way \cite{chen2016eyeriss, fleischer2018scalable, jouppi2017datacenter, chen2014dadiannao, xDNN, fowers2018configurable, suda2016throughput}. Typically, these specialized solutions feature \textit{spatial architectures}, which are those that expose low-level aspects of the hardware's interconnect and storage to the hardware-software interface. Spatial architectures can be coarse-grained or fine-grained. Coarse-grained architectures feature arrays of interconnected PEs, and fine-grained designs are realized by programming FPGAs. \textit{Coarse-grained spatial architectures} are a common implementation choice for designing hardware accelerators for ML \cite{chen2016eyeriss, fleischer2018scalable, fleming2019apparatus, jouppi2017datacenter, yang2018dnn}. As Fig. \ref{fig::ML-accelerators} illustrates, the accelerator comprises an array of PEs that may contain private register files (RFs) and shared buffers or a scratchpad memory. PEs are simple in design (functional units with little local control), and the shared scratchpad is non-coherent with software-directed execution. Therefore, these accelerators are a few orders of magnitude more power-efficient than out-of-order CPU or GPU cores \cite{chen2016eyeriss, jouppi2017datacenter, fleischer2018scalable}. They lead to highly energy-efficient execution of ML models that are compute-intensive and memory-intensive. Performance-critical tensor computations of ML models are relatively simple operations like element-wise or tensor additions and multiplications. So, they can be processed efficiently with structured computations on the PE-array. Moreover, private and shared memories of PEs enable high temporal reuse of the data \cite{dave2019dmazerunner, kwon2019understanding}; with efficient data management, PEs can be continuously engaged in tensor computations while the data is communicated via memories \cite{chen2016eyeriss}. Additionally, interconnects like mesh or multicast enable data communication among PEs and spatial reuse of the data, lowering the accesses to off-chip memory. Thus, with minimized execution time, spatial-architecture-based hardware accelerators yield very high throughput and low latency for processing ML models.

\subsection{Need for Further Efficient Execution} 
\label{sec::need-for-compression}

\begin{figure}[t]
  \centering
  \includegraphics[width=\linewidth]{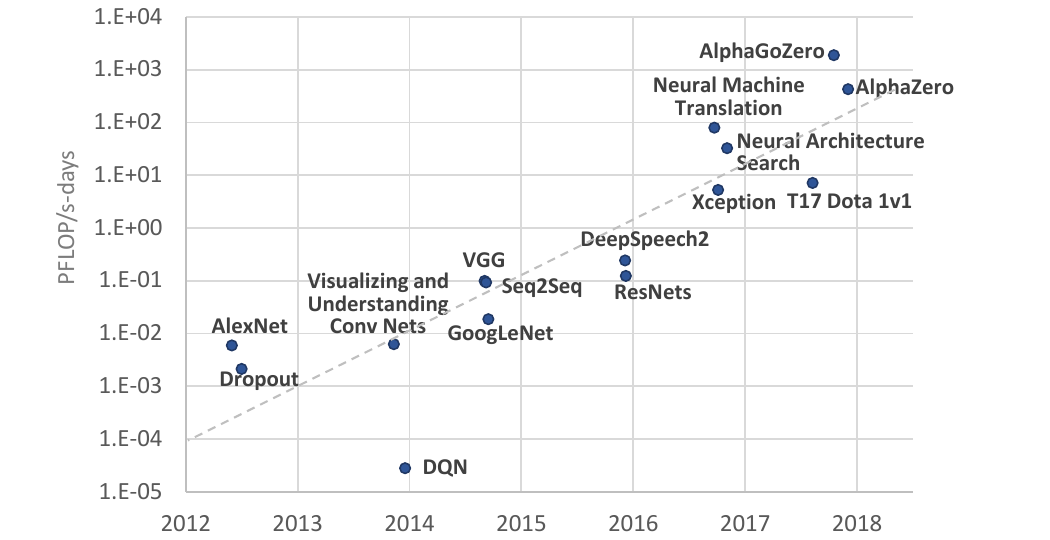}
  \caption{Computation requirements for the training of AI algorithms almost double every few months (Figure adopted from \cite{openaiblog}).}
  \label{fig::ML-model-compute-req}
\end{figure}

With recent advances in the development of ML models, their computational and memory requirements have increased drastically \cite{openaiblog, dean20201}. Fig. \ref{fig::ML-model-compute-req} provides an overview of this dramatic surge. One major reason is the rise of deeper models. For example, for processing ImageNet images, AlexNet \cite{krizhevsky2012imagenet} contained five CONV and three FC layers (eight parameter layers) with the model size of 61M parameters (weights and bias) and computation of 724 MFLOPs. DNNs like ResNet-101 \cite{he2016deep} achieved higher classification accuracy but contained 100+ parameter layers and required processing about 7.6 GFLOPs per image. \insight{Memory requirements for NLP models have increased massively}, e.g., from 50M--100M parameters (Transformer \cite{vaswani2017attention}, 2017) to 175 billion (GPT-3 \cite{brown2020language}, 2020).

While deeper and larger models achieve high efficiency for various tasks \cite{sze2017efficient}, they consume high execution time, energy, and memory. Previous studies showed that \insight{significant data redundancy exists in these often over-parameterized models} \cite{srivastava2014dropout, han2015learning}. So, researchers have developed techniques that compress tensors and obtain compact models, reducing computational, communication, and storage requirements significantly.

\section{Acceleration Opportunities due to Compact Models and the Need for Special Support}
\label{sec::tensor-compression}

Efficiency of executing ML models can be improved further by drastically reducing computation, communication, and memory requirements. This can be achieved by compressing tensors of ML models. Tensors are compressed by inducing and leveraging: (a) sparsity (zero values) \cite{han2015deep, wen2016learning, frankle2018lottery, mishra2017fine}, (b) size reduction (tensor decomposition, dimension reduction, and shape reduction) \cite{tan2019efficientnet, sandler2018mobilenetv2, iandola2016squeezenet, cichocki2015tensor, mishra2017fine, ragged_tensors}, and (c) quantization (precision lowering and value similarity) \cite{han2015deep, krishnamoorthi2018quantizing}. Previous techniques have achieved highly compact models without incurring accuracy loss. For example, after applying pruning, quantization, and Huffman encoding, Deep Compression \cite{han2015deep} reduced the model size of AlexNet and VGG-16 by 35$\times$ and 49$\times$ (e.g., from 552 MB to 11.3 MB), respectively. Accelerator-aware designs can compress the model further. For AlexNet and GoogLeNet models, \cite{yang2017designing} pruned 91\% and 66\% of weights and reduced computational requirements by 6.63$\times$ and 3.43$\times$, respectively. ADMM-NN \cite{ren2019admm} applied weight pruning and quantization, thereby reducing the model size of AlexNet, VGG-16, and ResNet-50 (with up to 0.2\% accuracy loss) by 99$\times$, 66.5$\times$ and 25.3$\times$, respectively.

This section describes various sources of tensor sparsity which is either inherent or induced by model architecture or regularization. It describes how sparsity reduces computations, storage, and communication requirements. It also discusses techniques for reducing the size and quantization of the tensors, and how they offer advantages in terms of storage/performance/energy-efficiency. Then, it describes how compression techniques may induce irregularity in the processing and why special support is needed for efficiently processing the compressed tensors on hardware accelerators.

\begin{figure}[t]
  \centering
  \includegraphics[width=\linewidth]{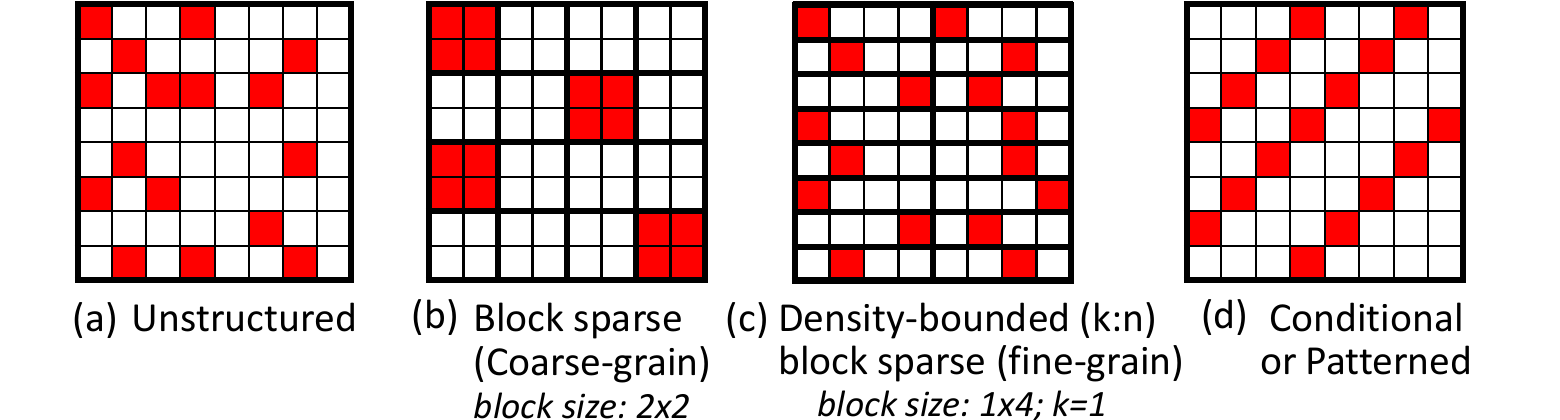}
  \caption{Common sparsity structures (e.g., for a 75\% sparse 8$\times$8 matrix).}
  \label{fig::sparsity-structures}
\end{figure}

\subsection{Opportunities Due to Sparse Tensors}

\mysubsubsection{Sparsity Structures} Inherent sparsity is usually unstructured (e.g., of activations, gradients, or tensors of scientific computing applications), where NZ elements are randomly scattered (shaded elements in Fig. \ref{fig::sparsity-structures}a). Applying ReLU, dropout, quantization, or fine-grain pruning also induces unstructured sparsity in input activations ($IA$) or weights ($W$). For improving execution efficiency, pruning techniques or model operators induce structured sparsity. For example, weights can be pruned in coarse-grain blocks where block shape can vary from 1-D (vector) to n-D for an n-dimension tensor \cite{wen2016learning, srivastava2019joint, zhou2018cambricon, yu2017scalpel}. Fig. \ref{fig::sparsity-structures}b shows 4$\times$4 blocks for a block-sparse tensor, where each block contains all zeros or all NZs. With larger blocks, techniques often prune entire dimensions (e.g., channels or filters in CNN models) \cite{wen2016learning}. The selection of block size and shape depends on task accuracy requirements. Alternatively, tensors are  sparsified with density bounded blocks (Fig. \ref{fig::sparsity-structures}c), where each n-D block contains a fixed ($k$) number of NZs \cite{kang2019accelerator, ampere, liu2020systolic}. It equally scatters NZs throughout the tensor. NZs are located arbitrarily in the whole block, or a fixed number of NZs can be induced across each dimension of the block. Values of $k$ can be selected based on the sensitivity of the pruning to accuracy. For example, analysis of \cite{kang2019accelerator} showed that for VGG-16 and ResNet-50, about 12 out of 16 elements can be pruned without any accuracy loss, and about 10 out of 16 elements for compact models like MobileNetV1 and SqueezeNetV1. To preserve accuracy while achieving high sparsity, a mixture of blocks (with different block sizes or sparsity) can also be introduced \cite{vooturi2018hierarchical}. Lastly, tensors can be pruned in patterns or conditionally with sophisticated rules (e.g., diagonally, as Fig. \ref{fig::sparsity-structures}d shows).

\mysubsubsection{Sources of Sparsity}
Tensors of different ML models can be sparse due to multiple reasons:

$\bullet$ CNNs use the ReLU activation function \cite{krizhevsky2012imagenet} that clamps negative values to zero. So, sparsity of input activations ($IA$-sparsity) can be 40\% in CNNs, on average \cite{cao2019seernet} and higher in later layers (about up to 70\% \cite{cao2019seernet, li2019squeezeflow}). Cao et al. \cite{cao2019seernet} reported that max-pooling can amplify it, e.g., up to 80\% for VGG-16 layers. Lee et al. \cite{lee20197} showed that IA-sparsity eliminated about 40\% and 55\% of the multiply-and-accumulate (MAC) operations during CNN training and inference, respectively. For recent compact models like MobileNetV2 \cite{sandler2018mobilenetv2}, IA-sparsity eliminates about 20\% of the MACs.

$\bullet$ Neural networks use drop-out layers to avoid overfitting. After applying the drop-out, only partial activations are retained \cite{srivastava2014dropout}. Dropping the activations induces sparsity \cite{srivastava2014dropout}. 

$\bullet$ Pruning techniques remove unimportant weights and alleviate the overfitting of the model while maintaining the classification accuracy. Typically, weights with the least significant values can be safely pruned \cite{han2015learning, han2020survey} (in training or post-training). Pruning can bring regularity in the learning of the model and can even increase accuracy slightly \cite{kang2019accelerator, zhou2018cambricon}. Pruning algorithms introduce significant sparsity, e.g., more than 60\% weights of CONV and more than 90\% of the weights of FC layers can be removed \cite{han2015learning} ($W$-sparsity). For recent compact models like MobileNetV2 and EfficientNetB0, W-sparsity can be 50\%--93\% (80\%--85\% in point-wise convolutions) \cite{elsen2020fast}, which reduces MACs by 2.5$\times$--4.2$\times$. Similarly, more than 80\% weights of RNN, GRU, or LSTMs can be pruned \cite{narang2017exploring, han2017ese, zhu2017prune}, especially for medium or large models, without significantly increasing error rate. For NLP models Transformers \cite{vaswani2017attention} and BERT \cite{devlin2019bert}, recent techniques induce 80\% \cite{gale2019state} and 93\% \cite{huggingface} W-sparsity, which reduces total MACs by about 4.8$\times$ and 12.3$\times$, respectively. Besides, regularization of the models (e.g., L1 or group-lasso based) can induce unstructured or structured W-sparsity \cite{wen2016learning}. 

Pruning of activations is also shown as effective \cite{albericio2016cnvlutin, yang2019dasnet, reagen2016minerva, georgiadis2019accelerating, shi2017speeding}. DasNet \cite{yang2019dasnet} reported eliminating about 27\% and 12\% MACs by activation sparsification for AlexNet and MobileNet. It achieved 79\% IA-sparsity for AlexNet FC layers along with pruning 81\% weights, without dropping top-1 accuracy. Similarly, MASR \cite{gupta2019masr} refactored batch normalization, achieving about 60\% IA-sparsity for RNNs. For attention-based NLP models, SpAtten \cite{wang2020spatten} pruned unimportant tokens and heads. It reported reducing computations and DRAM accesses by up to 3.8$\times$ and 1.1$\times$, respectively, without accuracy loss.

$\bullet$ CNNs use Atrous (dilated) convolutions where filters are upsampled by inserting zeros between weights \cite{chen2017rethinking}. 

$\bullet$ GANs use transposed convolution in a degenerator network, where input data is upscaled first by inserting zeros between values, and then convolution is applied. For transposed convolutions in different GANs, about 60\% MACs can be zero \cite{yazdanbakhsh2018ganax}. Additional sparsity is introduced when GANs are forced to forget generating specific objects \cite{radford2015unsupervised}.

$\bullet$ Input data for object detection tasks can be inherently sparse, as only specific regions of frames are valid \cite{acorns}. For example, object detection models of autonomous driving systems process 3D LiDAR data by constructing point clouds and projecting them from the bird's eye view (top view) \cite{ren2018sbnet, engelcke2017vote3deep}. The resultant images are then fed to object detection algorithms for locating the regions of interest. Recent techniques have reported that the sparsity of the input data for object detection can be 80\% or more \cite{acorns, ren2018sbnet}.

$\bullet$ For efficient communication in distributed training, \emph{gradients} ($Grad$) are sparsified and compressed. E.g., \textit{Grad-sparsity} can be 99\%+ for \textit{computer vision (CV)} or language processing tasks \cite{lin2018deep} and 95\%--99\% for recommendation models \cite{gupta2020fast}.

$\bullet$ Input data for the tasks of recommendation systems (e.g., user-item matrix) can be inherently highly sparse, e.g., from 95\% \cite{he2017neural} to 99\% \cite{park2018deep}. Recommendation models compute dot products on dense-sparse or sparse-sparse data \cite{acun2020understanding, gupta2020fast}.

$\bullet$ GNNs process large graphs, e.g., with thousands of vertices. Depending on the real-world interactions of objects (vertices), data contain high (e.g., 75\%--99\%) or hyper (99\%+) unstructured sparsity \cite{hygcn, geng2020awb}. For example, in processing large graphs with GCNs, many features of vertices are local and lead to zeros in adjacency matrices for remote nodes \cite{geng2020awb}. GNN computations involve aggregation on sparse data and multiplications of dense matrices with dense or sparse matrices \cite{geng2020awb, zeng2020graphact}, which are often processed on separate modules of the accelerator (e.g., in HyGCN \cite{hygcn} and EnGN \cite{liang2020engn}).

$\bullet$ Text corpus in text analytics applications leads to high sparsity since each document contains only a fraction of the words from the vocabulary. Such analytics applications include PCA for dimensionality reduction of the sparse data, support vector machines and regression for classification, collaborative filtering for the recommendation, and k-means for clustering the data \cite{mishra2017fine}. These operations involve multiplications of sparse matrices with dense or sparse vectors, where the matrix sparsity can vary from 67\% to 99\% \cite{mishra2017fine}.

While we describe leveraging sparsity for ML models, applications of many domains, including linear algebra, graph processing, and scientific computing \cite{duff1977survey, hegde2019extensor}, can be accelerated by exploiting sparsity.

\mysubsubsection{Advantages}
Sparsity allows (i) \textit{eliminating ineffectual computations}, i.e., reduces execution time and energy by processing only NZs, (ii) \textit{reducing storage} by encoding only NZ values, so more data fits in on-chip memory and off-chip memory accesses (extremely energy-consuming \cite{han2016eie, horowitz20141}) are reduced, and (iii) improving speedup due to \textit{reduced communication requirements} for data-intensive ML models.  

\subsection{Opportunities Due to Size-Reduced Tensors}

Symmetric or high-dimensional tensors have large sizes and their processing requires more computation and memory. So, ML models are designed to reduce such requirements by using group or parallel operators \cite{krizhevsky2012imagenet, xie2017aggregated}, 1$\times$1 or point-wise convolutions (PW-CONV) \cite{szegedy2015going, iandola2016squeezenet}, or dimensionality reduction with PCA \cite{cichocki2015tensor, mishra2017fine}. Moreover, tensors can be decomposed with spatial factorization \cite{szegedy2016rethinking, cichocki2015tensor}, depth-wise separation for convolutions \cite{sandler2018mobilenetv2, tan2019efficientnet}, or low-rank approximations \cite{cichocki2015tensor}. Further, tensors can be ragged \cite{ragged_tensors} to eliminate the need for structured or rectangular shapes. While these transformations significantly reduce storage and computations, they make tensors irregular-shaped (asymmetric). 

\subsection{Opportunities Due to Quantized Tensors}

Quantization includes precision lowering \cite{krishnamoorthi2018quantizing} and leveraging value similarity \cite{han2015deep, hegde2018ucnn, riera2018computation}. Precision lowering allows representing tensors (weights, activations, gradients, weight updates) at much lower bit-width (e.g., 8b or lower for inference and 8b/16b for learning). Moreover, elements with similar values can be clustered and approximated by sharing common values (centroids of clusters). Further, similar values of outputs are reused with memoization (partially or the entire layer). In general, significant redundancy exists in tensor elements (particularly in the parameters of large models), and a successfully trained model is generalized and immune to noisy data. So, the error induced by quantization or approximation may often be tolerated by a well-trained model \cite{quantization8bitPeter}. It can also obviate over-fitting caused otherwise by excessive precision, thereby bringing generality in learning \cite{kalamkar2019study}. For compensating accuracy drop due to quantization, learning algorithms fine-tune the model or use quantization-aware training \cite{krishnamoorthi2018quantizing}. Thus, quantization or approximation techniques typically do not degrade inference accuracy \cite{lee20197} or trade it off for notable execution efficiency \cite{cao2019seernet, hubara2017quantized, lee2017lognet}. 

Quantization significantly reduces storage requirements and accesses to off-chip memory. It also \textit{reduces area and power}, since for quantized tensors, functional units can be simpler and energy-efficient  (e.g., int8 multiplier consumes 20$\times$ less energy than FP32 multiplier \cite{horowitz20141} for a 45 nm process). Bus sizes can be smaller as bandwidth requirements are reduced. 

Thus, with sparse, size-reduced, and quantized tensors, compact models can achieve higher accuracy as models with uncompressed tensors, while becoming amenable for deployment at the edge, mobile, or online-learning platforms \cite{han2015deep, banbury2020benchmarking} due to scope for low latency, energy, and storage. So, leveraging such opportunities is crucial for further accelerations.

\subsection{Need for Special Support to Accelerate Sparse and Irregular Tensor Computations}
\label{sec::need-special-support}

Hardware accelerators efficiently process different models \cite{jouppi2017datacenter, chen2014diannao, fleischer2018scalable}. But, they inherently cannot benefit from the sparsity because all the data, including the zero values of activations, weights, and gradients, have to be fetched from memory and communicated to PEs; PEs are also unable to skip ineffectual computations, wasting the execution time. Sparsity, especially unstructured, induces irregularity in processing since NZs or blocks of NZs are scattered across the tensor. Therefore, leveraging sparsity necessitates additional mechanisms to store, extract, communicate, compute, and load-balance the NZs, and corresponding hardware and software support \cite{zhang2016cambricon, qin2020sigma}. Different sparsity levels and patterns from various sources lead to unique challenges and solutions in hardware/software co-design. Therefore, our discussions throughout this survey mainly focus on exploiting tensor sparsity for accelerating compact models.

Tensor dimension reduction and tensor decomposition make tensors irregular-shaped (asymmetric), and they may also modify the functionality of the computational primitives, e.g., depthwise convolution (DW-CONV). Since execution on hardware accelerators is typically well-optimized for processing symmetric tensors with a specific dataflow mechanism, these shape transformations and supporting different functionality (e.g., DW-CONV, randomized or approximated matrix multiply \cite{adelman2018faster}) may introduce irregularity in processing requirements. To sustain high utilization of computational resources, it requires additional support including configurable hardware architectures and flexible mappings of the functionality onto architectural resources \cite{chen2019eyeriss, qin2020sigma, zhang2019snap}. 

Hardware accelerators have supported low-precision tensors of fixed bit-widths, and even more recently, tensors with mixed precision \cite{lee20197}. However, when sparse tensors are quantized with value sharing, it requires indexing the codebook through indices for approximated elements \cite{han2015deep}. Such irregular accesses are handled by implementing separate indirection tables in the pipelined hardware datapath \cite{han2016eie, zhou2018cambricon}. Moreover, value similarity is leveraged further by reusing computations with memoized outputs, which requires additional processing. Further, supporting different bit-widths of various sparse tensors of different models requires configurable architectures for \emph{bit-adaptive} computing \cite{albericio2017bit, moons201714, dadu2019towards}.  

To sum up, compressed tensors lead to sparse and irregular computations. Their efficient accelerations require special support, which is described in the next section. The appendix describes that exploiting sparsity (especially unstructured) is relatively hard for execution on CPUs and GPUs; with special support, hardware accelerators can achieve notable gains.

\section{Accelerator Design for Efficient Sparse and Irregular Tensor Computations}
\label{sec::overview-accel-compressed-tensors}

\subsection{Overview}

To efficiently process sparse and irregular tensor computations, designers of the accelerator systems can integrate special hardware or software modules. It enables orchestration of the structured computations while processing the tensors in compressed formats. Consequently, it can lead to efficient utilization of the accelerator resources and allows exploiting acceleration opportunities. Fig. \ref{fig::sparse-accel-block-diagram} provides an overview of the accelerator system equipped with such modules. This section briefly describes these system modules.

\begin{table}
\centering

\caption{Accelerators for Processing Sparse Tensors.}
\label{tab:overview-sparse-accel}
\begin{tabular}{|c|m{4.85cm}|}
\hline
\thead{Objective} & \thead{Techniques} \\ \hline
\makecell{Compressed data \\ in off-chip \\ memory (storage)} 
& \cite{chen2016eyeriss, han2016eie, zhang2016cambricon, zhang2019compact, judd2017cnvlutin2, zheng2018kernelxform, struharik2018conna, parashar2017scnn, zhang2019snap, kang2019accelerator, mishra2017fine, moons201714, lee20197, chen2019eyeriss, qin2020sigma, gondimalla2019sparten, zhou2018cambricon, li2019squeezeflow, hegde2019extensor, yuan2018sticker, choles2018parsecore, yavits2017accelerator, nvdla, aimar2018nullhop, han2017ese, page2017sparcnet, lu2018spwa, lu2019efficient} \\ \hline  

\makecell{Compressed data \\ in on-chip \\ memory (storage)} & \cite{han2016eie, zhang2016cambricon, zhang2019compact, zheng2018kernelxform, struharik2018conna, albericio2016cnvlutin, judd2017cnvlutin2, parashar2017scnn, zhang2019snap, kang2019accelerator, mishra2017fine, lee20197, chen2019eyeriss, qin2020sigma, gondimalla2019sparten, zhou2018cambricon, li2019squeezeflow, hegde2019extensor, yuan2018sticker, choles2018parsecore, yavits2017accelerator, aimar2018nullhop, han2017ese, lu2019efficient, lu2018spwa, gao2018deltarnn}  \\ \hline 

\makecell{Skip processing \\ zeros \\ (energy efficiency)} & \cite{chen2016eyeriss, han2016eie, zhang2016cambricon, asgari2019eridanus, struharik2018conna, kung2018adaptive, albericio2016cnvlutin, judd2017cnvlutin2,  parashar2017scnn, zhang2019snap, kang2019accelerator, yin2017thinker, lee2018stitch, mishra2017fine, moons201714, lee20197, chen2019eyeriss, kim2017zena, qin2020sigma, gondimalla2019sparten, zhou2018cambricon, li2019squeezeflow, hegde2019extensor, yuan2018sticker, jang2019mnnfast, mcdanel2019full, choles2018parsecore, yavits2017accelerator, aimar2018nullhop, han2017ese, lu2019efficient, page2017sparcnet, whatmough2018dnn, lu2018spwa, gao2018deltarnn, venkatesh2017accelerating} \\ \hline , 

\makecell{Reduce ineffectual \\ computation cycles \\ (performance \& energy)} & \cite{han2016eie, zhang2016cambricon, zheng2018kernelxform, struharik2018conna, albericio2016cnvlutin, judd2017cnvlutin2,  parashar2017scnn, zhang2019snap, kang2019accelerator, lee2018stitch, mishra2017fine, lee20197, chen2019eyeriss, kim2017zena, qin2020sigma, gondimalla2019sparten, zhou2018cambricon, li2019squeezeflow, hegde2019extensor, yuan2018sticker, choles2018parsecore, yavits2017accelerator, aimar2018nullhop, han2017ese, lu2019efficient, lu2018spwa, gao2018deltarnn, venkatesh2017accelerating} \\ \hline 

\makecell{Load balancing \\ (performance)} & \cite{han2016eie, asgari2019eridanus, kung2018adaptive, kang2019accelerator, lee20197, chen2019eyeriss, kim2017zena, gondimalla2019sparten, zhou2018cambricon, li2019squeezeflow, choles2018parsecore, han2017ese, lu2018spwa} \\ \hline  
\end{tabular}%
\end{table}

\begin{table*}[!t]
\centering

\caption{Accelerator Systems Leveraging Sparsity of Different Tensors for Different ML Models.}
\label{tab:target-sparsity}
\begin{tabular}{|c|c|m{1.5cm}|m{10cm}|}
\hline

\multirow{3}{*}{\makecell{Dynamicity \\ of Sparsity}} 
& Static & \multicolumn{2}{l|}{\cite{zheng2018kernelxform,  zhang2016cambricon, asgari2019eridanus, kung2018adaptive, kang2019accelerator, yavits2017accelerator, han2017ese, lu2019efficient, page2017sparcnet, lu2018spwa, liu2020systolic}} \\ \cline{2-4} 
& \multirow{2}{*}{Dynamic} 
& \multicolumn{2}{l|}{\cite{zhang2019compact, struharik2018conna, chen2016eyeriss, han2016eie, albericio2016cnvlutin, judd2017cnvlutin2, parashar2017scnn, zhang2019snap, chen2019eyeriss, mishra2017fine, lee20197, moons201714, yuan2018sticker, li2019squeezeflow, qin2020sigma, hegde2019extensor, gondimalla2019sparten, zhou2018cambricon, choles2018parsecore, wang2020spatten}} \\ 
& & \multicolumn{2}{l|}{\cite{aimar2018nullhop, gao2018deltarnn, yin2017thinker, lee2018stitch, kim2017zena, jang2019mnnfast, mcdanel2019full,  whatmough2018dnn, jang2019mnnfast, ham20203}}  \\ \hline

\multirow{4}{*}{\makecell{Tensors Treated \\ as Sparse}} & \multirow{2}{*}{Weight} & Unstructured & \cite{zheng2018kernelxform,  zhang2016cambricon, lu2019efficient, page2017sparcnet, lu2018spwa, kung2018adaptive}, \cite{he2020sparse} \\ \cline{3-4} 
&  & Structured &   \cite{han2017ese, kang2019accelerator, asgari2019eridanus, ampere, choles2018parsecore, zhou2018cambricon, srivastava2019joint, liu2020systolic, shi2020csb}  \\ \cline{2-4}
& Activation & \multicolumn{2}{l|}{\cite{zhang2019compact, chen2016eyeriss, albericio2016cnvlutin, lee20197, yin2017thinker, aimar2018nullhop, whatmough2018dnn, gao2018deltarnn, wang2020spatten, jang2019mnnfast, ham20203}} \\ \cline{2-4}
& Both & \multicolumn{2}{l|}{\cite{struharik2018conna, han2016eie, judd2017cnvlutin2, parashar2017scnn, zhang2019snap, lee2018stitch, chen2019eyeriss, mishra2017fine, moons201714, yuan2018sticker, li2019squeezeflow, qin2020sigma, hegde2019extensor, gondimalla2019sparten, zhou2018cambricon, kim2017zena, yavits2017accelerator, mcdanel2019full, choles2018parsecore, venkatesh2017accelerating}} \\ \hline

\multirow{5}{*}{\makecell{Primitive \\ Operation}} 
& \makecell{Matrix-Vector Multiply} & \multicolumn{2}{l|}{\cite{han2016eie, struharik2018conna, zhang2016cambricon, kang2019accelerator, zhou2018cambricon, yavits2017accelerator, hegde2019extensor, gondimalla2019sparten, zhang2019snap, lee2018stitch, chen2019eyeriss, mishra2017fine, lee20197, whatmough2018dnn}} \\ \cline{2-4} 
& \makecell{Matrix-Matrix Multiply} &  \multicolumn{2}{l|}{\cite{zhang2016cambricon, asgari2019eridanus, kung2018adaptive, mcdanel2019full, hegde2019extensor, gondimalla2019sparten, qin2020sigma, venkatesh2017accelerating}} \\ \cline{2-4} 
& \makecell{Convolution} &  \multicolumn{2}{l|}{\cite{zhang2016cambricon, parashar2017scnn, zheng2018kernelxform, zhang2019compact, struharik2018conna, chen2016eyeriss, albericio2016cnvlutin, judd2017cnvlutin2, zhang2019snap, kang2019accelerator, zhou2018cambricon, choles2018parsecore, gondimalla2019sparten, li2019squeezeflow, lee2018stitch, chen2019eyeriss, lee20197, moons201714, yuan2018sticker, aimar2018nullhop, lu2019efficient, page2017sparcnet, lu2018spwa}} \\ \cline{2-4}
& \makecell{Recurrent / Attention Layer} &  \multicolumn{2}{l|}{\cite{gupta2019masr, yin2017thinker, han2017ese, gao2018deltarnn, wang2020spatten, jang2019mnnfast, ham20203, shi2020csb}} \\ \hline
\multicolumn{2}{|l|}{Accelerators for Learning} & \multicolumn{2}{l|}{\cite{qin2020sigma, lee20197}} \\ \hline
\end{tabular}
\end{table*}

Sparse, size-reduced, and quantized tensors of ML models offer various opportunities for storage, performance, and energy efficiency. Hence, several accelerators have provided marginal or comprehensive support and leveraged some or all the opportunities. Table \ref{tab:overview-sparse-accel} lists such common objectives and corresponding accelerator solutions that meet these objectives. 

Different accelerators for inference and learning exploit $W$-sparsity, $IA$-sparsity, or both, which impacts acceleration gains \cite{kim2017zena}. Several accelerators, including Cambricon-X \cite{zhang2016cambricon}, exploit only static sparsity (Table \ref{tab:target-sparsity}), e.g., when locations of zeros in weights are known beforehand for inference. Static sparsity allows offline encoding and data transformations for arranging structured computations (e.g., for systolic arrays \cite{asgari2019eridanus, he2020sparse, liu2020systolic}). Recent accelerators, including ZENA \cite{kim2017zena}, SNAP \cite{zhang2019snap}, and EyerissV2 \cite{chen2019eyeriss}, leverage dynamic sparsity also. It requires determining locations of intersecting NZs in both tensors at run-time to feed functional units, on-the-fly decoding (encoding) NZs, and often balancing computations on PEs. Table \ref{tab:target-sparsity} lists different accelerators that support static and dynamic sparsity of tensors. Now, we describe different hardware and software aspects of the accelerator system that help in leveraging sparsity effectively.

\textbf{Sparsity encodings:} Sparse tensors are compressed using encodings, where only NZ values are stored in a "data" tensor and one or more "metadata" tensors encode locations of NZs. Section \ref{sec::sparse-data-coding} discusses different formats and associated costs for encoding and decoding. For different sparsity levels, it analyzes their effectiveness in terms of storage efficiency. E.g., tensors can be compressed by 1.8$\times$ and 2.8$\times$ for 50\% and 70\% sparsity (bitmap or RLC-2) and 7.6$\times$ and 55$\times$--60$\times$ for 90\% (RLC-4) and 99\% sparsity (CSC or RLC-7). Structured sparsity (coarse-grain block-sparse) can alleviate the overheads of metadata and fine-grained data extraction by encoding indices for only large dense blocks. For accelerating ML models, sparse tensors are also quantized i.e., their precisions are lowered (typically int8 or int16 for inference \cite{chen2019eyeriss, parashar2017scnn, kim2017zena} and FP16 for learning \cite{lee20197, qin2020sigma}) and often approximated by clustering data of similar values \cite{han2016eie, zhou2018cambricon, zhang2019compact}. Therefore, encoded sparse data contains quantized values of NZs. 

\textbf{NZ detection and data extraction:} In processing sparse tensors of different primitives, corresponding elements of the weight and activation tensors are multiplied and accumulated. Depending on the sparsity, accelerators need to use data extraction logic that decodes compressed tensors, search within a window of NZs or index the buffer, and obtain matching pairs of NZs to feed the functional units for computation. Section \ref{sec::NZ-data-extraction} provides a taxonomy of different data extraction mechanisms and analyzes their implications for various sparsity levels. Up to moderate $IA$-sparsity and high $W$-sparsity, these indexing or intersection-based mechanisms efficiently extract sufficient NZs at every cycle for keeping functional units engaged. For efficient compute-bounded executions at such sparsity, accelerators reported achieving near-ideal speedups (e.g., about 80\%--97\% of the speedup corresponding to reduced operations, i.e., \textit{sparsity-speedup ratio}) \cite{zhang2016cambricon, kim2017zena, han2016eie}. However, extraction becomes challenging at high (e.g., 90\%+) or hyper sparsity as NZs are scattered at distant locations \cite{geng2020awb}, and execution is usually memory-bounded with low arithmetic intensity. Section \ref{sec::NZ-data-extraction} also discusses sharing of the data extraction mechanism among PEs or employing in PEs. Then, it discusses opportunities for further optimizations.

\textbf{Memory management:} Compressed tensors are often stored in the shared on-chip memory that is non-coherent, multi-banked, and often non-unified. For a pre-determined sequence of execution, a controller or PEs initiates the accesses between off-chip and on-chip memory; their latency needs to be hidden behind computations on PEs. Section \ref{sec::memory-management} discusses corresponding memory architectures and techniques for hiding miss penalty for sparse tensors via double-buffering or asynchronous computation and memory accesses. It describes the data reuse opportunities for various sparsity and dimensions of tensors of common DNNs and how sparsity lowers the reuse. It also discusses techniques that leverage cross-layer reuse of intermediate output layers and reduce latency. 

\textbf{Communication networks:} Once tensor blocks are fetched from memory, they are distributed to appropriate PEs via interconnect networks (often one per operand). Efficient designs ensure that sufficient data can be fed to PEs while they perform computations. Reuse is leveraged spatially by multicast or mesh networks that communicate common data blocks to multiple PEs. It lowers accesses to memory hierarchy and communication latency. However, spatial reuse opportunities vary depending on the sparsity, NZ extraction mechanism, and mapping of the functionality on the accelerator. Section \ref{sec::comm-networks} discusses different designs for distributing sparse and quantized tensors and reducing partial outputs. It also describes challenges in executing inter-PE communications that may become unstructured due to sparsity and the temporal and spatial mechanisms for reduction/collection of the outputs. It describes how configurable designs support various communication patterns for different sparsity, reuse, and functionality. 

\textbf{PE architecture:} Several accelerators consist of scalar PEs with fused MAC units (e.g., EIE \cite{han2016eie}, LNPU \cite{lee20197}, and Envision \cite{moons201714}). Others contain SIMD PEs (multiple functional units) (e.g., EyerissV2 \cite{chen2019eyeriss}) or vector PEs consisting of multiplier-arrays and adder-trees (e.g., Cambricon-X \cite{zhang2016cambricon} and SNAP \cite{zhang2019snap}). PE architectures either directly process pairs of matching NZs extracted from tensors or use hardware logic for data extraction or coordinate computation (Fig. \ref{fig::ML-accelerators}). Effectively utilizing functional units can be challenging for variations in sparsity, precisions, and functionality, and it may require configurable designs. Section \ref{sec::PEArch} provides corresponding discussions and describes sparsity-aware dataflow mechanisms (mapping of tensor computations on accelerator resources) used by different accelerators. It also describes how accelerators have leveraged value similarity of tensors and the corresponding modifications in the PE architecture.

\textbf{Load balancing:} Depending on the distribution of zeros, the execution may end up with processing a different amount of NZs on different PEs or their functional units, which creates inter-PE or intra-PE load imbalance. Section \ref{sec::load-balance} analyzes such sources of the imbalance and introduces a taxonomy of different load balancing techniques. Accelerators achieve load balance through either software techniques (e.g., structured pruning or data reorganization) or by providing a hardware module for dynamic work balance (through asynchronous execution or work sharing), which provides further accelerations. For example, ZENA \cite{kim2017zena} leveraged the sparsity of both activation and weight tensors for AlexNet and VGG-16 models and reported about 32\% additional performance gains through load balancing. Dynamic load balancing can provide notable speedups for high, unstructured sparsity \cite{geng2020awb}.

\textbf{Write-back and post-processing:} Tensor elements produced by PEs need to be collected, post-processed for further operations, and written back to the memory. PEs in different accelerators either write back sequentially or asynchronously through a shared bus or via point-to-point links. In addition, accelerators usually contain a post-processing unit that re-organizes the data (as per the dataflow mechanism of the current and next layer of the model) and encodes sparse output on the fly. Section \ref{sec::post-processing} discusses such mechanisms. 

\textbf{Compilation support:} It is important to support the execution of various ML models on accelerators and easier programming of models from ML libraries. Section \ref{sec::software-support} discusses compiler support for sparse models and hardware accelerators. It discusses polyhedral and non-polyhedral intermediate representations and their implications on the compiler’s ability to represent the code and apply code transformations. It describes challenges in supporting sparse tensors and DNN compilers that facilitate sparse tensor computations. Then, it discusses compiler optimizations including common loop optimizations and those specific to hardware intrinsics. It also describes semi-automatic optimizations for transforming the loops and data layout and automatic optimizations using cost models. Finally, it discusses ISAs used by accelerators and their code generation by using libraries of high-level primitives.

\subsection{Case Study: Acceleration of DNNs and Bottleneck Analysis}
\label{sec::sparse-dnn-acceleration-case-study}

\begin{table}[!t]
\centering
\caption{Sparsity of Some Popular DNNs.}
\label{tab:sparsity-dnn-models}
\resizebox{0.48\textwidth}{!}{
\addtolength{\tabcolsep}{-3pt}
\begin{tabular}{|c|c|c|c|c|c|c|c|}
\hline
\multirow{2}{*}{Model} & \multirow{2}{*}{Domain} & \multirow{2}{*}{Dataset} & \multirow{2}{*}{\makecell{GOps \\ (dense)}} & \multicolumn{3}{c|}{Sparsity \%} & \multirow{2}{*}{\makecell{Sparse\\ Model}} \\ \cline{5-7}
 &  &  &  & IA & W & Ops &  \\ \hline
MobileNetV2 \cite{sandler2018mobilenetv2} & CV & ImageNet & 0.3 & 34 & 52 & 81 & \cite{elsen2020fast} \\ \hline
EfficientNetB0 \cite{lu2019efficient} & CV & ImageNet & 0.5 & 0 & 68 & 60 & \cite{elsen2020fast} \\ \hline
Transformer \cite{vaswani2017attention} & NLP & WMT En-De & 4.6 & 0 & 79 & 79 & \cite{gale2019state} \\ \hline
\makecell{BERT-base-uncased \cite{devlin2019bert}} & NLP & \makecell{SQuAD} & 9.3 & 0 & 92 & 92 & \cite{huggingface} \\ \hline
\end{tabular}
\addtolength{\tabcolsep}{3pt}
}
\end{table}

\begin{table*}[!t]
\centering
\caption{Architectural features of analyzed accelerators for sparse DNNs.}
\label{tab:microarch-features-sparse-DNN-accelerators}
\addtolength{\tabcolsep}{-3pt}
\begin{tabular}{|c|c|c|c|c|c|c|c|c|c|c|c|c|c|c|c|}
\hline
\multirow{3}{*}{ID} & \multirow{3}{*}{\begin{tabular}[c]{@{}c@{}}Reference \\ Architecture\end{tabular}} & \multirow{3}{*}{\begin{tabular}[c]{@{}c@{}}Supported \\ Operators\end{tabular}} & \multicolumn{2}{c|}{\multirow{2}{*}{\begin{tabular}[c]{@{}c@{}}Sparsity \\ Leveraged\end{tabular}}} & \multicolumn{3}{c|}{\multirow{2}{*}{\begin{tabular}[c]{@{}c@{}}Non-zero Data \\ Extraction\end{tabular}}} & \multirow{2}{*}{\begin{tabular}[c]{@{}c@{}}PE \\ Architecture\end{tabular}} & \multirow{3}{*}{\begin{tabular}[c]{@{}c@{}}Work \\ Synch-\\ ronization\end{tabular}} & \multirow{3}{*}{\begin{tabular}[c]{@{}c@{}}Freq.\\ (GHz)\end{tabular}} & \multirow{3}{*}{\begin{tabular}[c]{@{}c@{}}DRAM \\ BW\\ (GBPS)\end{tabular}} & \multicolumn{4}{c|}{Bit-width} \\ \cline{13-16} 
 &  &  & \multicolumn{2}{c|}{} & \multicolumn{3}{c|}{} &  &  &  &  & \multicolumn{2}{c|}{data} & \multicolumn{2}{c|}{metadata} \\ \cline{4-9} \cline{13-16} 
 &  &  & IA & W & Encoding & Discovery & Loc. & FU &  &  &  & IA / O & W & IA & W \\ \hline
A1 & EIE \cite{han2016eie} & GEMM & \multicolumn{2}{c|}{\begin{tabular}[c]{@{}c@{}}unstructured\end{tabular}} & CSR & \multirow{2}{*}{Indexing} & in-PE & Scalar & Prefetch & 0.8 & 256 & 16 & 4 & N/A & 4 \\ \cline{1-6} \cline{8-16} 
A2 & \begin{tabular}[c]{@{}c@{}}Cambricon-X\\ \cite{zhang2016cambricon}\end{tabular} & \multirow{2}{*}{\begin{tabular}[c]{@{}c@{}}CONV, \\ GEMM\end{tabular}} & dense & \makecell{unstru-\\ctured} & \begin{tabular}[c]{@{}c@{}}COO-\\ 1D\end{tabular} &  & \begin{tabular}[c]{@{}c@{}}central \\ (per PE)\end{tabular} & \multirow{2}{*}{\begin{tabular}[c]{@{}c@{}}Vector (16 \\ multipliers \&\\ adder tree)\end{tabular}} & \multirow{2}{*}{\begin{tabular}[c]{@{}c@{}}Every \\ Output \\ Activation\end{tabular}} & 1 & 256 & 16 & 16 & N/A & 8 \\ \cline{1-2} \cline{4-8} \cline{11-16} 
A3 & \begin{tabular}[c]{@{}c@{}}Cambricon-S\\ \cite{zhou2018cambricon}\end{tabular} &  & \makecell{unstru-\\ctured} & \begin{tabular}[c]{@{}c@{}}block-\\ sparse\end{tabular} & \multirow{2}{*}{Bitmap} & \multirow{2}{*}{\begin{tabular}[c]{@{}c@{}}Inter-\\ section\end{tabular}} & \begin{tabular}[c]{@{}c@{}}central, \\ shared\end{tabular} &  &  & 1 & 256 & 16 & 8 & 1 & 1 \\ \cline{1-5} \cline{8-16} 
A4 & \begin{tabular}[c]{@{}c@{}}ZENA-\\ IA-W \cite{kim2017zena}\end{tabular} & CONV & \multicolumn{2}{c|}{unstructured} &  &  & in-PE & Scalar & \begin{tabular}[c]{@{}c@{}}Intra-\\ Workgroup\end{tabular} & 0.2 & 12 & 16 & 16 & 1 & 1 \\ \hline
\end{tabular}%
\addtolength{\tabcolsep}{3pt}
\end{table*}

This section analyzes the sparsity of recent DNN models (for NLP and CV) and the acceleration that can be achieved with some of the popular accelerators-alike architectures.

\textbf{DNN models:} Table \ref{tab:sparsity-dnn-models} summarizes analyzed DNN models and their overall sparsity across all CONV, GEMM, and DW-CONV operations. For each of these DNN operations, $W$-sparsity was obtained from sparse DNN models (listed in the last column). $IA$-sparsity was obtained by performing inference with sample data (images and text sequences). $GOps$ corresponds to processing of a single image for CV models and sequences of 24
and 107 tokens for Transformer and BERT. 

\textbf{Accelerators:} Table \ref{tab:microarch-features-sparse-DNN-accelerators} summarizes analyzed accelerators and their sparsity-centered features. Their architectures targeted unstructured or block sparsity of activations and/or weights. Their features represent variations across data encoding, data extraction, vector processing, memory hierarchy, NoC, and load balancing.

\textbf{Methodology:} To determine the impact of sparsity on achievable acceleration, we performed a data-driven analysis of the execution latency. For each DNN layer, zeros (or blocks of zeros) were induced randomly according to the sparsity of its tensors. The overall execution time was determined from the latency of processing on functional units, data decoding, extraction of non-zeros, work synchronization, and off-chip memory transfers, which were calculated based on analytical modeling of the microarchitectural features. The speedups were calculated over oracle processing of dense tensors at the accelerator's peak utilization of computational resources and off-chip bandwidth. In this study, we do not consider the processing of DW-CONV on these accelerators, since they are often not pruned, and their execution needs to be group-wise, which is extremely inefficient. Such unsupported performance-critical operators were assumed to be processed with dense tensors at peak utilization of hardware resources.

\begin{figure}[t]
\centering
\centerline{\includegraphics[width=\linewidth]{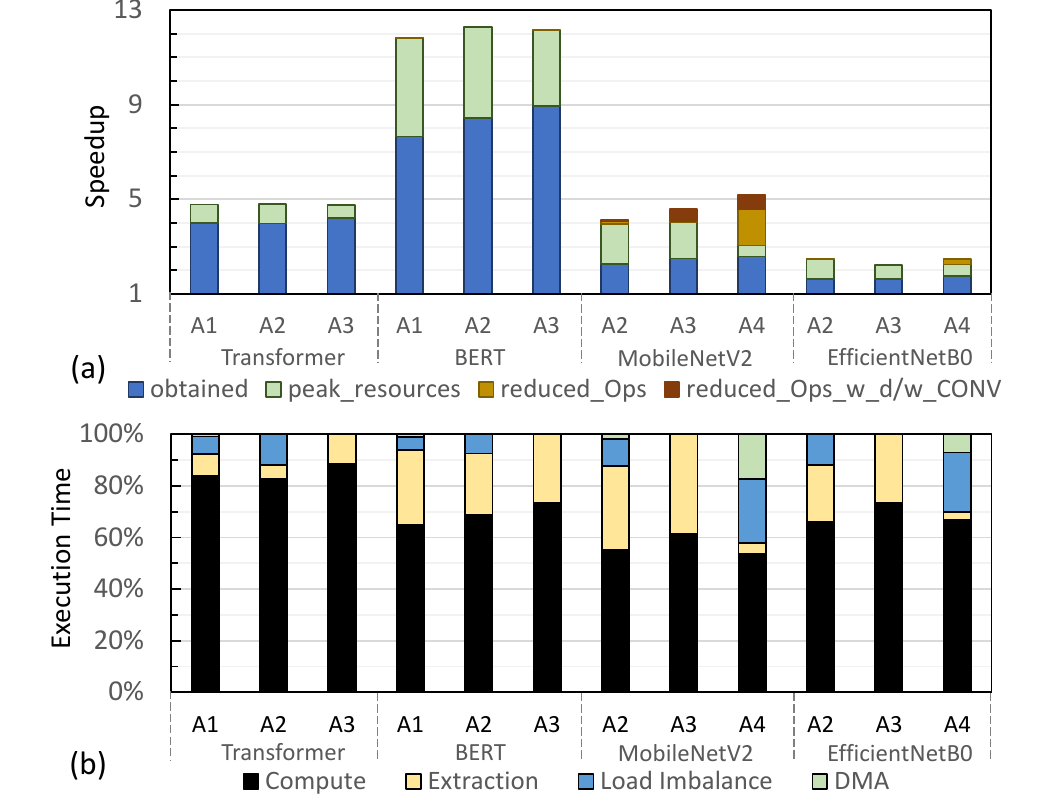}}
\caption{(a) Obtained speedups for accelerators listed in Table \ref{tab:microarch-features-sparse-DNN-accelerators}. (b) Analysis of execution time overheads for obtained accelerations.}
\label{fig::sparse-dnn-acceleration-analysis}
\end{figure}

\textbf{Analysis:} Fig. \ref{fig::sparse-dnn-acceleration-analysis}(a) shows speedups of accelerators for targeted DNN models, for leveraging the sparsity of supported DNN operators. It illustrates speedups for (i) reduction in the operations due to sparsity (desired), (ii) peak utilization of accelerator's computational resources and off-chip bandwidth while leveraging sparsity, over such oracle processing of dense tensors (potential), and (iii) actual processing on accelerator over oracle processing of dense tensors (obtained). For understanding implications of execution overheads including those incurred by metadata processing and load imbalance, Fig. \ref{fig::sparse-dnn-acceleration-analysis}(b) illustrates fractions for desired computation time and execution overheads in a stacked format. The overheads were extracted for layer-wise processing and then accumulated to determine the overall impact. Fractions include:
$\bullet$ Computation time: Minimum execution time required for processing at peak on accelerator's functional units.
$\bullet$ NZ extraction: Time required for decoding NZs from communicated operands and extracting matching operands for feeding the functional units. It also corresponds to balanced computations.
$\bullet$ Load imbalance: Time required for on-chip processing on the accelerator, considering the imbalanced computations subjected to the accelerator's work synchronization and work sharing schemes.
$\bullet$ DMA time: Time required for off-chip data communication via DMA transfers, in addition to on-chip processing.

Fig. \ref{fig::sparse-dnn-acceleration-analysis}(a) shows that accelerators efficiently exploited moderate sparsity. E.g., for 4.8$\times$ reductions in operations of Transformer due to $W$-sparsity, they achieved about 4$\times$--4.2$\times$ speedup. The exploitation of speedup lowers when activations are dense and weights are highly or hyper-sparse. This is because accelerators like EIE and Cambricon-X broadcast activations to PEs and extract matching pairs corresponding to NZ weights. So, communication of activations and extraction of matching NZ operands consume significant execution time, while there are fewer operations to feed the functional units (Fig. \ref{fig::sparse-dnn-acceleration-analysis}b). E.g., for BERT-base-uncased \cite{devlin2019bert} (92\% sparse weights \cite{huggingface}) on SQuAD \cite{rajpurkar2016squad}, they achieved about 7.7$\times$--8.9$\times$ speedup out of 12.2$\times$ speedup for processing at peak. Due to block-sparse weights, computations on PEs of Cambricon-S are always balanced. Therefore, it achieved higher speedups. By using blocks of 16$\times$16 or even 1$\times$16 (across input and output channels) for pruning, inducing similar sparsity is not possible sometimes. So, the reduction in operations and potential for the speedup was slightly lower for Cambricon-S (e.g., for EfficientNetB0). In general, due to high DRAM bandwidth, overheads incurred by DMA transfers were hidden (for Cambricon-X/S) or negligible for non-interleaved transfers (e.g., for EIE).

Fig. \ref{fig::sparse-dnn-acceleration-analysis}(a) also shows that Cambricon-S and ZENA-IA-W achieved higher speedups for CV models by leveraging unstructured sparsity of activations. High IA-sparsity amplified total sparsity during processing several layers (e.g., MobileNetV2), incurring considerable excess processing in data extraction for Cambricon-X/S and in load imbalance for ZENA-IA-W. With zero-aware static sorting of filters and dynamic load balance, ZENA \cite{kim2017zena} could overcome such imbalance. But, it would suffer through high on-chip communication time since it used only one shared bus for multicast via NoC and collecting outputs. We disregarded such communication overhead for ZENA-IA-W in this study, as most accelerators use separate NoCs or buses for alleviating communication overheads. Also, due to low DRAM bandwidth, overheads incurred by DMA transfers were higher for ZENA-IA-W, mainly for executing DW-CONVs with dense tensors.

\section{Encodings for Compressing Sparse Tensors}
\label{sec::sparse-data-coding}

\begin{figure}[t]
\centering
\centerline{\includegraphics[width=0.85\linewidth]{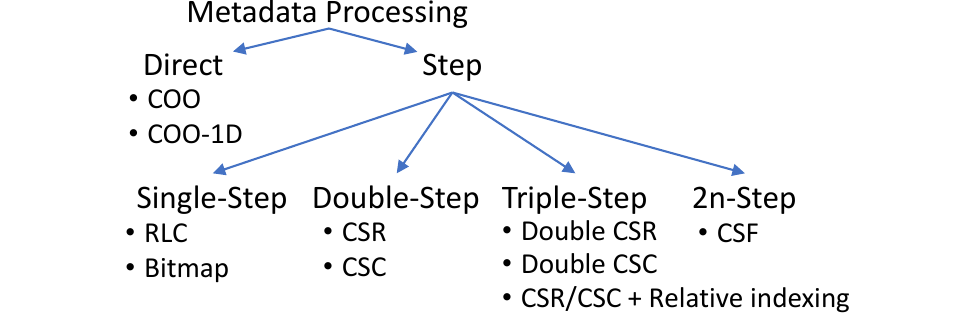}}
\caption{A taxonomy for the required processing on the metadata during data extraction when a sparse tensor is encoded using different formats.}
\label{fig::sparse-encoding-schemes-overview}
\end{figure}

\begin{figure*}[t]
\centering
\centerline{\includegraphics[width=0.9\linewidth]{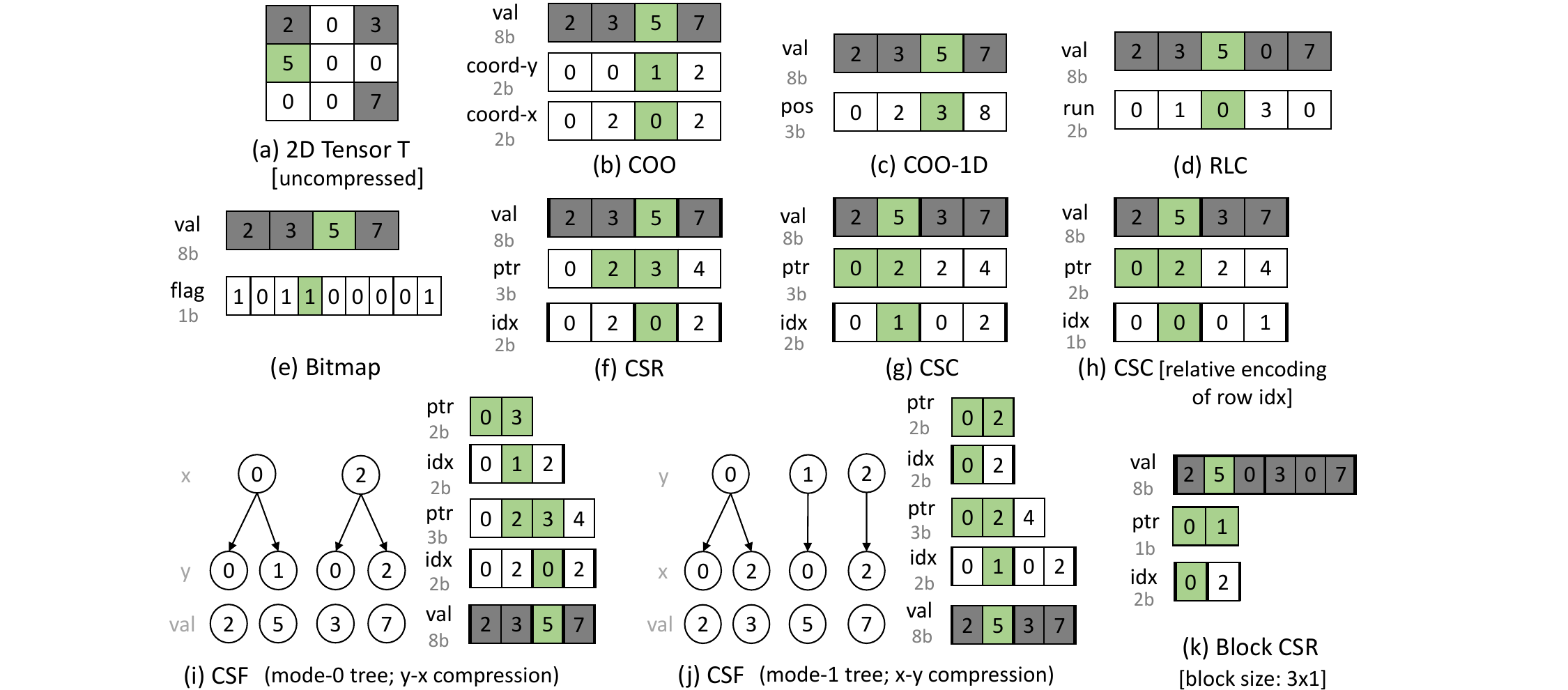}}
\caption{Encodings to store sparse tensors in different formats. Elements with green shade encode the same NZ element. (Figure inspired by \cite{chou2018format}.)}
\label{fig::sparse-data-encodings}
\end{figure*}

A sparse tensor is compressed with an encoding format. An encoded tensor contains actual \textit{data} (NZ values) and \textit{metadata} (information about positions of NZs). Later, metadata is used by an accelerator's data indexing logic to locate and extract NZs. This section discusses commonly used encodings through an example (Fig. \ref{fig::sparse-data-encodings}) and their implications on the storage and processing requirements. For different formats, Fig. \ref{fig::sparse-encoding-schemes-overview} introduces a taxonomy for processing metadata during data extraction, and Table \ref{tab:sparse-encodings-overhead} lists the corresponding storage overhead. Depending on the mapping of a layer onto the accelerator, tensors are divided into blocks (per PE-wise work) which are encoded separately. We refer to such processing as a \textit{group-wise encoding}, which is discussed later. Finally, this section briefly describes encoding on the fly and further opportunities.

\subsection{Encoding Formats and Implications}

\mysubsubsection{Coordinate (COO)} It stores absolute positions of NZs. As Fig. \ref{fig::sparse-data-encodings}(b) shows, all NZs of an uncompressed tensor $T$ are stored in a data vector $val$, and vectors $coord\_y$ and $coord\_x$ indicate the coordinates of each NZ value. So, COO is a natural way to express sparse tensors and is used commonly (e.g., in PyTorch). Formats adopted by FROSTT \cite{frosttTensorFormat} and matrix market \cite{MatrixMarket} closely resemble COO.

The COO format stores all coordinates in the uncompressed format. E.g., as Fig. \ref{fig::sparse-data-encodings}(b) shows, the metadata for values '2' and '3' (same row) or '2' and '5' (same column) are not compressed, i.e., duplicate values of row and column indices exist in coordinate vectors. So, the overhead of storing $n$ coordinates per NZ value is about $\sum_1^n { \lceil \log_2{d_i} \rceil}$ bits (vector $d$ contains the tensor's dimensions). It makes COO inefficient for storing tensors with low or moderate sparsity. 

\begin{figure}[t]
\centering
\centerline{\includegraphics[width=0.9\linewidth]{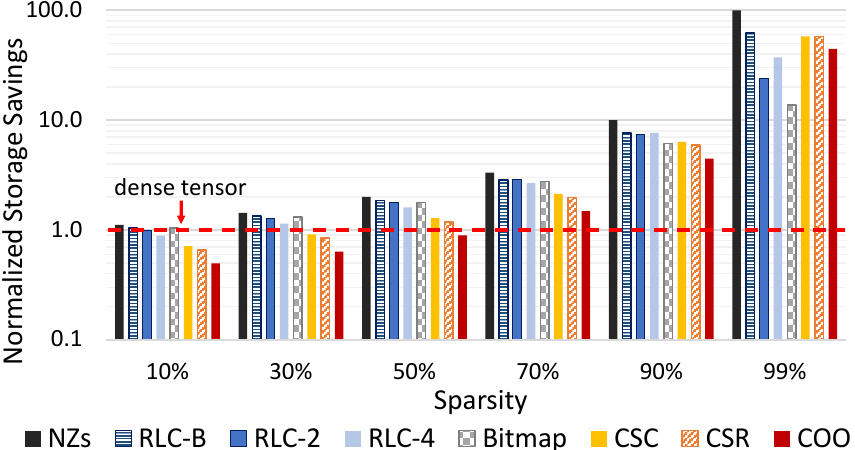}}
\caption{Storage benefits for encoding a sparse tensor (512$\times$2048 matrix with 16b elements) in different formats, normalized to the size of the fully dense tensor (Figure inspired by \cite{qin2020sigma}).}
\label{fig::sparse-data-encodings-overhead}
\end{figure}

Fig. \ref{fig::sparse-data-encodings-overhead} shows storage benefits for encoding 2 MB matrices of various sparsity in different formats. We calculated storage requirements with the analysis presented in Table \ref{tab:sparse-encodings-overhead} and normalized them to the matrix's size in dense format. We used the Scipy library \cite{jones2001scipy} to generate matrices of various sparsity and encode them in COO, CSR, and CSC. Fig. \ref{fig::sparse-data-encodings-overhead} shows that for a 2 MB matrix, COO achieves storage efficiency for 70\%+ sparsity. However, COO may yield simple indexing logic, as both the data and metadata can be directly extracted.

\mysubsubsection{COO-1D} For \emph{tile-wise processing} of an encoded tensor, accelerators often process only a block of NZs at a time, where block elements vary across only a single dimension. For example, Cnvlutin \cite{albericio2016cnvlutin} processes the input activations and weights across the channel direction.  Therefore, the data block is encoded with COO-1D, which is just like COO, but there is only one $pos$ vector for storing coordinates of NZs in the flattened block. For instance, if we flatten $T$ and consider it a block, then the value '5' is indexed by position '3'. 

\mysubsubsection{Run-Length Coding (RLC)} It compresses a sequence of values by replacing consecutive duplicate values with a single value and the number of repetitions (aka \textit{run}). For RLC-encoded sparse tensor, ``run'' indicates a total number of zeros before (after) an NZ. Fig. \ref{fig::sparse-data-encodings}(d) shows RLC encoding of $T$. Run values for '2' and '3' are '0' and '1', respectively. A few accelerators, including Eyeriss \cite{chen2016eyeriss}, encode both the NZs and runs altogether in the same vector $val$. For example, $T$ can be encoded as $val$: (0, 2, 1, 3, 0, 5, 4, 7).

RLC requires a \emph{step-processing} on metadata, as run-length needs to be calculated by accumulating runs and preceding \textit{number of NZs (NNZs)}, for determining the position of an NZ. The storage overhead for RLC-B is NNZ $\times$ B bits, where B is bit-width of the run. If a vector $d$ contains tensor dimensions, then B can be set as up to $\lceil \log_2{(\prod_1^n d_i)} \rceil$ bits for accommodating the number of leading zeros in a highly sparse tensor. When B is set lower, it cannot always capture the number of zeros as \textit{run}. Fig. \ref{fig::sparse-data-encodings}(d) shows RLC-2b encoding, where leading zeros before '7' are four. This cannot be expressed in 2 bits. As a work-around, padding zeros \cite{han2016eie} are inserted and treated as NZs. In this example, a padding zero is inserted between '5' and '7'; run values corresponding to the padding zero and '7' are '3' and '0', which contributes to the total run of four.

To accelerate CNNs with 30\%--90\% sparsity of tensors, designers have set B as two or four bits. In general, setting the B as $\lfloor \log_2{(\frac{sparsity}{density})} \rfloor + 1$ bits can effectively compress tensors and provide a feasible bit-width to indicate leading zeros. Here, $sparsity$ and $density$ are fractional numbers indicating the actual or anticipated number of zeros and non-zeros in the tensor, respectively. Thus, setting the B as 1, 1, 1, 2, 4, and 7 efficiently encodes tensors with sparsity of 10\%, 30\%, 50\%, 70\%, 90\%, and 99\%, which is depicted in Fig. \ref{fig::sparse-data-encodings-overhead}.

As RLC requires step-processing on metadata, the indexing logic needs an accumulator to determine the position of an NZ. When an encoded tensor is not processed block-wise but rather indexed by n-dimensions, the indexing logic may require performing division and modulo operations on the metadata. Alternatively, a multi-dimension representation can be used where $run$ for the coordinates of each dimension can be calculated separately and stored. The overall computational cost (arithmetic and logical operations realized in hardware) for such step-processing can be low. Therefore, several accelerator designs, including Eyeriss \cite{chen2016eyeriss} and SCNN \cite{parashar2017scnn}, used RLC or its variant. As run indicates repetition of a value, CompAct \cite{zhang2019compact} used an enhanced RLC format for encoding both the sparse and similar-value activations. 

\mysubsubsection{Bitmap} It stores all NZs in a tensor $val$ along with a tensor \textit{flag} which contains 1-bit flags for all elements of an uncompressed tensor $T$. As Fig. \ref{fig::sparse-data-encodings}(e) shows, a flag indicates whether an element is NZ or not. Storage overhead for the bitmap (aka bit-mask) is $\prod_1^n d_i$ bits (where vector $d$ stores $n$ dimensions of $T$) \cite{aimar2018nullhop}. Since bitmap stores metadata for all elements, it is effective for compressing the tensors of low or moderate sparsity. Like RLC, decoding or indexing  bitmap also requires step-processing. The indexing logic to locate an NZ typically consists of at least an adder and a comparator \cite{zhang2016cambricon}. Due to moderate storage overhead and low encoding/decoding cost, several accelerators used bitmap, including Cambricon-X \cite{zhang2016cambricon}, SparTen \cite{gondimalla2019sparten}, and SIGMA \cite{qin2020sigma}, as shown in Table \ref{tab:accel-sparse-encodings}.

\mysubsubsection{Compressed Sparse Row (CSR)} It compresses a matrix by processing each row as a sparse vector. In a CSR-coded tensor, an array $val$ contains all NZ values (ordered row-wise), and an array $idx$ stores their column indices \cite{saad1990sparskit}. Array $ptr$ stores information about total NZs in each row $i$, which is obtained by calculating $ptr[i+1]$ - $ptr[i]$. The last element of $ptr$ contains the total number of NZs in $T$. Row-wise compression enables random accesses to any row.

While COO redundantly stores row-coordinates of NZs in the same row, CSR compresses such metadata by storing NZs row-wise \cite{chou2018format}. For example, in Fig. \ref{fig::sparse-data-encodings}(b) (COO), \textit{coord-y} stores row indices '0' and '0' for NZs '2' and '3'. This redundancy is removed in the CSR coding of Fig. \ref{fig::sparse-data-encodings}(f), as $ptr$ stores only total NZs in each row. For compressing an M$\times$N matrix using CSR, the total storage overhead is {NNZ $\times$ $\lceil \log_2{N} \rceil$} (for $idx$) + {(M + 1) $\times$ $\lfloor \log_2{NNZ} + 1 \rfloor$} (for $ptr$). Due to high storage overhead (proportional to NNZs and size of the row), CSR coding is efficient at high sparsity \cite{zhang2016cambricon, qin2020sigma}, e.g., 90\% or higher (Fig. \ref{fig::sparse-data-encodings-overhead}). 

Decoding a CSR-encoded tensor can require a two-step processing of metadata. The first step locates NZs of a row by iterating over $ptr$, and the next step locates an NZ element in the NZs of the row through the column index. Accelerators efficiently process CSR-coded matrices row-wise such that $ptr$ is accessed once for fetching each row, and then the decoder iterates through $idx$ (to locate column positions).

CSR variants can improve efficiency further. For example, $ptr$ stores duplicate values when consecutive rows are zero. Doubly CSR (DCSR) \cite{buluc2008representation} eliminates this redundancy and achieves additional compression for hyper-sparse matrices. Block CSR (BCSR) \cite{im1998model} stores a block of elements in $val$, if the block contains at least one NZ. As Fig. \ref{fig::sparse-data-encodings}(k) shows, in BCSR, $idx$ indicates the column index of a block, and $ptr$ informs about the number of dense blocks located in the same rows. BCSR avoids storing blocks of all zeros and populates dense regions, and hence suitable for encoding block-sparse structured weight tensors. Thus, BCSR-coded tensors can be efficiently executed not only on conventional processors but also on hardware accelerators (with additional support for appropriately indexing dense regions, e.g., \cite{asgari2019eridanus}). 

\begin{table}[t]
\centering
\caption{Storage overhead for common encodings. Vector $d$ stores $n$ dimensions of a tensor that contains $NNZ$ non-zero elements.}
\label{tab:sparse-encodings-overhead}
\begin{tabular}{|c|c|}    \hline
Format & Storage Overhead (bits) \\ \hline
COO & $NNZ \times \sum_1^n { \lceil \log_2{d_i} \rceil}$ \\ \hline
COO-1D & $NNZ \times \lceil \log_2{\prod_1^n d_i} \rceil$ \\ \hline
RLC & $NNZ \times B$ \\ \hline
Bitmap & $\prod_1^n d_i$ \\ \hline
CSR & \makecell{$NNZ \times \lceil \log_2{d_1} \rceil$ + $(d_0 + 1) \times \lfloor \log_2{NNZ} + 1 \rfloor$} \\ \hline
CSC & \makecell{$NNZ \times \lceil \log_2{d_0} \rceil$ + $(d_1 + 1) \times \lfloor \log_2{NNZ} + 1 \rfloor$} \\ \hline
\end{tabular}
\end{table}

\mysubsubsection{Compressed Sparse Column (CSC)} CSC is similar to CSR, except that NZs are stored column-wise \cite{saad1990sparskit}. As Fig. \ref{fig::sparse-data-encodings}(g) shows, an array $val$ contains NZs (organized column-wise); $idx$ stores their row indices; $ptr$ informs about the total NZs in each column. The storage overhead and hardware costs for encoding/decoding tensors in CSC format are similar to those for CSR. Accelerators, including EIE \cite{han2016eie} and Sticker \cite{yuan2018sticker}, processed high-sparsity tensors with CSC format. 

For alleviating the high storage overhead of CSR or CSC formats due to storing $idx$ and $ptr$ arrays, a few accelerators further encode the metadata $idx$ or $ptr$. For example, EIE \cite{han2016eie} and EyerissV2 \cite{chen2019eyeriss} encode $idx$ in RLC such that elements in $idx$ indicate zeros between column indices of NZs (similar to $run$ in RLC for NZ values). Fig. \ref{fig::sparse-data-encodings}(h) shows CSC encoding with such an RLC-encoded row index array. Values '2' and '5' have column index '0' and '1', respectively, which can be encoded as '0' and '0' since there are no leading zeros before NZs '2' and '5'. Similarly, if the first column of $T$ is (0, 2, 0, 0, 5), then the row indices for '2' and '5' can be encoded as '1' and '2'. $ptr$ can also be encoded likewise (store NZs per column instead of a cumulative number). However, encoding positions relatively requires additional step-processing on the metadata. Therefore, decoding a CSR or CSC encoded matrix with RLC-encoded metadata can require triple-step processing on metadata (additional hardware cost).

\mysubsubsection{Compressed Sparse Fiber (CSF)} CSF \cite{smith2015tensor} provides a generalization of CSR for higher-order ($n$-dimensional) tensors by forming a tree (with $n$ levels). Nodes at level $l$ contain indices for $l$th mode (dimension) of an uncompressed tensor $T$. Path from a root to a leaf node encodes different coordinates of an NZ, which are stored in the nodes throughout the path; each leaf node stores an NZ value. So, the height of the tree is the total dimensions of $T$; the width is NNZs in $T$. 

Fig. \ref{fig::sparse-data-encodings}(i) illustrates a mode-0 tree and corresponding arrays of index pointers. Root nodes represent the major mode (0 or $y$), and their child nodes represent the consecutive dimension (1 or $x$). Like in CSR, $ptr$ informs about a group of indices corresponding to a dimension. For instance, $ptr$ array at the beginning informs that one group of three coordinates corresponds to the mode 0 ($idx$ stores coordinates). Similarly, the next $ptr$ array informs about three different groups of coordinates for the next mode (dimension 1). The corresponding $idx$ array stores the coordinates for mode 1, separated into three groups (marked by thick outer vertical borders). 

Layering the arrays of index pointers reduces duplication of indices \cite{tew2016investigation}. Each time when a node directs to children, it eliminates duplicating indices for the corresponding mode. Storage benefits increase with the increase in dimensions and redundancy among coordinates of NZs. The organization of the data also impacts storage efficiency. For example, Fig. \ref{fig::sparse-data-encodings}(j) shows another ordering, which eliminates storing redundant coordinates of column (mode 1), achieving fewer nodes. For an n-mode CSF tensor, the storage overhead corresponds to more than $NNZ + n - 1$ coordinates and typically much less than $n \times NNZ$ coordinates. Works \cite{tew2016investigation, smith2015tensor} provide further details about managing higher-order tensors with CSF format. Processing metadata at each dimension requires two-step processing (just like processing $ptr$ and $idx$ in CSR), thereby up to 2n-step processing for an n-dimensional tensor. So, accelerator designers may opt for CSF format when processing high-dimensional tensors with high sparsity.

\mysubsubsection{Huffman coding} It typically is applied for compressing sparse tensors once they are quantized using precision lowering or value sharing. After quantization, values of the reduced range appear with different frequencies and can be compressed further with Huffman encoding \cite{han2015deep}. For example, Deep Compression \cite{han2015deep} pruned and quantized weights of AlexNet \cite{krizhevsky2012imagenet} and VGG-16 \cite{simonyan2014very}, achieving 8b/5b indices with a codebook of 256/32 weights for CONV/FC layers. With Huffman encoding, it compressed the models further by 22\% and 36\% (total compression of 35$\times$ and 49$\times$).

\begin{table}[!t]
\centering

\caption{Commonly Used Sparsity Encodings by Accelerators}
\label{tab:accel-sparse-encodings}
\begin{tabular}{|c|m{6cm}|}
\hline
COO & \cite{yuan2018sticker, hegde2019extensor} \\ \hline
COO-1D &  \cite{struharik2018conna, albericio2016cnvlutin, zhang2019snap, kang2019accelerator,lee2018stitch, mcdanel2019full, lu2019efficient, gao2018deltarnn} \\ \hline
RLC &  \cite{chen2016eyeriss, zheng2018kernelxform, zhang2019compact, parashar2017scnn, li2019squeezeflow, lee20197, buckler2018eva2, wang2020spatten} \\ \hline
Bitmap &  \cite{zhang2016cambricon, zhang2019compact, judd2017cnvlutin2, qin2020sigma, gondimalla2019sparten, yuan2018sticker, moons201714, zhou2018cambricon, nvdla, choles2018parsecore, aimar2018nullhop, kim2017zena, liu2020systolic} \\ \hline
CSR & \cite{mishra2017fine} \\ \hline 
CSC & \cite{han2016eie, chen2019eyeriss, mishra2017fine, yavits2017accelerator, guo2017software, han2017ese, lu2018spwa} \\ \hline
CSF & \cite{hegde2019extensor} \\ \hline
\end{tabular}
\end{table}

\mysubsubsection{Encodings for tensors with structured sparsity}
Density-bounded blocks (Fig. \ref{fig::sparsity-structures}c) can be encoded similarly as blocks with unstructured sparsity, e.g., with bitmap \cite{liu2020systolic}, COO-1D \cite{kang2019accelerator}, or RLC. So, for the same sparsity and block size, the overhead is similar to tile-wise processing of a tensor with unstructured sparsity. It is usually low for small block sizes (e.g., 8$\times$1 \cite{liu2020systolic}, 1$\times$4 in NVIDIA A100 Tensor Core GPU \cite{ampere}), since the position of each NZ is indicated by a few bits. Coarse-grain block-sparse tensors (Fig. \ref{fig::sparsity-structures}b) can be encoded at block-granularity, which can significantly reduce the metadata size (almost eliminated for dimensional pruning \cite{ma2019non}). Cambricon-S \cite{zhou2018cambricon} used bitmap to indicate the presence of each 1$\times$16 dense block with a single bit. Similarly, ERIDANUS \cite{asgari2019eridanus} used few bytes to process each 8$\times$8 dense block on systolic arrays. Such encodings require indicating the position of a dense block across rows or columns and additional indices for higher dimensions that indicate dense blocks packed per dimension, e.g., in block CSR (Fig. \ref{fig::sparse-data-encodings}-k).

\mysubsubsection{Other formats} Various encoding formats have been proposed, which improve the compression or efficiently access sparse tensors during execution on CPUs/GPUs (for high-performance and scientific computing). It includes compressed sparse blocks (CSB) \cite{bulucc2009parallel}, libsvm \cite{chang2011libsvm}, ELLPACK \cite{kincaid1984itpack}, diagonal (DIA) \cite{saad2003iterative}, dynamic CSR \cite{king2016dynamic}, delta-coded CSR \cite{willcock2006accelerating}, and mode-generic and mode-specific formats \cite{baskaran2012efficient}. Prior works including \cite{chou2018format}, SPARSKIT \cite{saad1990sparskit}, and \cite{bader2008efficient, vuduc2003automatic, tew2016investigation, hong2019adaptive} surveyed them along with additional formats and discussed their implications. Different libraries that provide support for encoding the tensors and for sparse tensor computations on CPUs or GPUs include MATLAB tensor toolbox \cite{TTB_Software}, Intel MKL \cite{wang2014intel}, SciPy \cite{jones2001scipy}, and cuSPARSE \cite{nvidiasparse}.

\subsection{Group-wise Encoding} 

One way of processing sparse tensors is to encode the whole tensor. Then, the accelerator's data management logic extracts an appropriate tile  (optionally decodes it) and communicates to the PEs. In contrast, for group-wise encoding, tensor tiles are encoded separately, based on pre-determined per-PE work. Depending on the mapping, each tile is typically communicated to a unique PE (or a PE-group) during execution. Thus, the encoding considers the dataflow, i.e., mapping of the tensor computations onto PEs. It can make the decoding and data extraction easier, as each group corresponds to execution on a distinct PE (or a PE-group). EIE \cite{han2016eie}, Cambricon-X \cite{zhang2016cambricon}, and CompAct \cite{zhang2019compact} used group-wise encoding.

\subsection{On-the-fly Encoding} 

Accelerator designers often target only static sparsity of weights and encode them off-line, e.g., DNN inference accelerators, including EIE \cite{han2016eie}, Cambricon-X \cite{zhang2016cambricon}, and \cite{zheng2018kernelxform}. However, on-the-fly encoding is required for efficiently processing dynamically sparsified tensors (sparse activations in the inference and tensors in training the models). Therefore, accelerators, such as CompAct \cite{zhang2019compact}, SCNN \cite{parashar2017scnn}, NullHop \cite{aimar2018nullhop}, Cnvlutin \cite{albericio2016cnvlutin}, and Sticker \cite{yuan2019sticker} employ an on-the-fly encoder. Typically, before encoding a tensor, the data is re-organized as per requirements of the group-wise encoding and dataflow mechanism for processing the subsequent layer. So, on-the-fly encoding is often combined with assembling the outputs from PEs (section \ref{sec::post-process-encoding} provides further details).

\subsection{Optimization Opportunities}

\noindent \textit{(i) Tailoring encoding formats for sparsity levels and patterns:} Various layers of deep learning models exhibit a wide range of sparsity (\textit{inter-layer, intra-tensor sparsity variation}). Moreover, even within a DNN layer, sparsity among tensors can be different (\textit{intra-layer, inter-tensor sparsity variation}). Accelerators need to support such sparsity variations effectively without incurring significant overheads for storage, encoding, and indexing. When the sparsity range or pattern of multiple tensors is diverse, designers can opt for the separate encoding of different tensors (e.g., \cite{yuan2018sticker}). These different sparsity-encodings can be utilized for off-chip storage, zero-guarding the PEs, or reducing the latency of on-chip extraction to locate intersecting NZs. When different formats are used for performance gains, the accelerator should provide hardware logic for decoding different tensors that are stored in different formats (and support for any on-the-fly encoding). Such decoding logic may use existing data extraction mechanisms, but it will require separate/configurable decoding logic for supporting multiple formats.

\section{Extraction of Matching Data for Computations on Non-Zeros}
\label{sec::NZ-data-extraction}

\begin{table}[!t]
\centering

\caption{Classification of NZ Data Extraction Techniques}
\label{tab:data-extraction}

\begin{tabular}{|c|c|c|m{3cm}|}
\hline
\makecell{Target \\ Sparsity}    & \makecell{PE Arch-\\itecture}   & \makecell{Functional Unit \\ Operation} & \makecell{Accelerators} \\ \hline
\multirow{3}{*}{\makecell{One \\ Tensor}}          
& Scalar                    & MAC   & \cite{zheng2018kernelxform, aimar2018nullhop, mishra2017fine, han2017ese} \\ \cline{2-4}
& \multirow{2}{*}{\makecell{SIMD/\\Vector}}   & Sc-Vec-Mul   & \cite{albericio2016cnvlutin, kang2019accelerator, lee20197} \\ \cline{3-4} 
&                           & Vec-Vec-Mul   & \cite{zhang2016cambricon, kang2019accelerator, lu2018spwa} \\ \hline

\multirow{3}{*}{\makecell{Both \\ Tensors}} 
& Scalar                    & MAC   & \cite{kim2017zena, struharik2018conna, han2016eie, hegde2019extensor, gondimalla2019sparten, mishra2017fine, yavits2017accelerator} \\ \cline{2-4} 
& \multirow{2}{*}{\makecell{SIMD/\\Vector}}   & Sc-Vec-Mul   & \cite{chen2019eyeriss, judd2017cnvlutin2} \\ \cline{3-4} 
&                           & Vec-Vec-Mul   & \cite{zhang2019snap, zhou2018cambricon, qin2020sigma} \\ \hline
\end{tabular}

\begin{tabular}{ccc}
& & \\
\end{tabular}

\begin{tabular}{|c|m{5.6cm}|}
\hline
\makecell{Location of \\ Extraction Units} & \makecell{Accelerators} \\ \hline
\makecell{Centralized/\\Shared} & \cite{han2016eie, zhang2016cambricon, judd2017cnvlutin2, zhang2019snap, mishra2017fine, zhou2018cambricon, qin2020sigma, aimar2018nullhop, yavits2017accelerator, li2019squeezeflow, lee20197} \\ \hline
In-PE       & \cite{zheng2018kernelxform, kim2017zena, struharik2018conna, albericio2016cnvlutin, kang2019accelerator, chen2019eyeriss, gondimalla2019sparten, choles2018parsecore, han2017ese, han2016eie, hegde2019extensor, parashar2017scnn, lu2018spwa, venkatesh2017accelerating} \\ \hline
\end{tabular}
\end{table}

Tensors are typically stored in the compressed format in the accelerator's memory. Therefore, locations of NZs that need to be processed are determined from the metadata. Once a matching pair is extracted (elements of two tensors that need to be added or multiplied), a PE can proceed for computations. Identifying effective NZs is the primary step towards eliminating ineffectual computations due to the sparsity of weights and/or activations. 
This section describes different data extraction mechanisms (Table \ref{tab:data-extraction} provides a taxonomy), their management in PEs or centrally, and their trade-offs. Then, it discusses further acceleration opportunities to exploit various sparsity levels.

\subsection{Non-Zero Detection and Extraction Mechanisms}
\label{sec::data-extraction-HW}

\insight{A data extraction mechanism needs to feed functional units of PEs every cycle.} So, based on their processing of scalars or vectors of NZs, Table \ref{tab:data-extraction} categorizes extraction mechanisms for (i) MAC operation on scalars, (ii) scalar-vector multiplication, and (iii) vector-vector multiplication.

\begin{figure}[!t]
\centering
\centerline{\includegraphics[width=\linewidth]{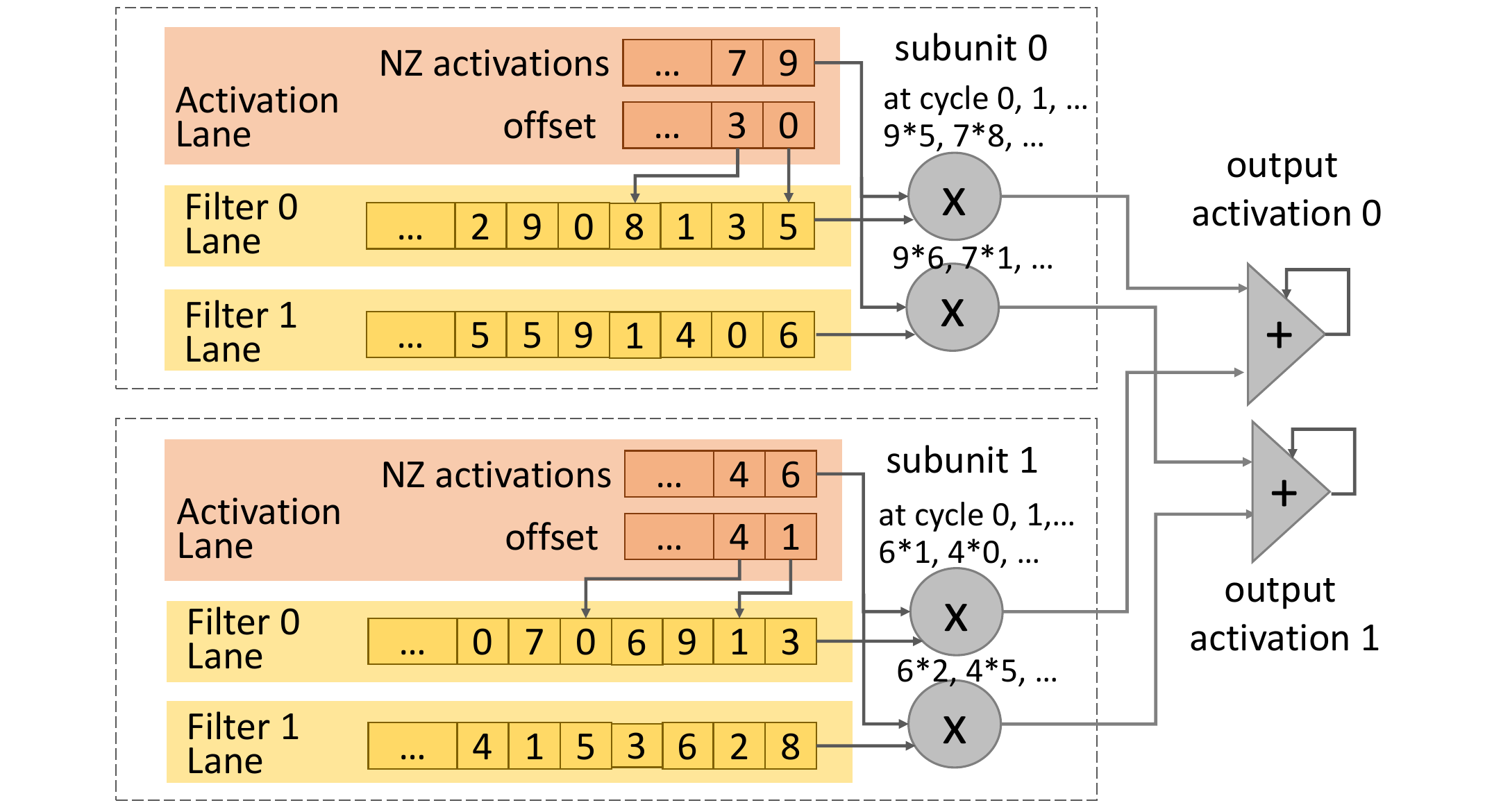}}
\caption{Data extraction in subunits of Cnvlutin PE (Figure adopted from \cite{albericio2016cnvlutin}).}
\label{fig::NZ-detection-cnvlutin}
\end{figure}

\mysubsubsection{Indexing dense tensors by indices of NZs of a sparse tensor}
Depending on sparsity, only one tensor may be treated as sparse and compressed (e.g., activations for Cnvlutin \cite{albericio2016cnvlutin} or weights for Cambricon-X \cite{zhang2016cambricon} and NVIDIA A100 \cite{ampere}). So, the position of an NZ can be used for indexing the other (i.e., dense) tensor to extract the corresponding value. 

\textit{MAC:} Consider the activation lane and filter lane 0 of subunit 0 in Fig. \ref{fig::NZ-detection-cnvlutin}, which can be visualized as processing on a scalar PE. For an NZ streaming from the activation lane, matching weight can be looked up and provided to the multiplier or MAC unit. For COO-1D encoded blocks, absolute positions of NZs can be obtained directly from metadata. Otherwise, absolute positions of NZs need to be computed explicitly by decoding metadata (e.g., bitmap or RLC) through simple combinational logic consisting of AND gates, multiplexers, and adders (e.g., in  \cite{zhang2016cambricon} and \cite{zheng2018kernelxform}). 

\textit{Sc-Vec Mul:} For SIMD processing, multiple arrays are indexed with the position of an NZ. Fig. \ref{fig::NZ-detection-cnvlutin} shows such mechanism used in Cnvlutin PEs \cite{albericio2016cnvlutin}. Each of 16 subunits in Cnvlutin PE featured an activation lane (streamed an input channel vector), 16 multipliers, and 16 filter lanes. A common NZ activation was fetched from the activation lane, and its position was used for looking up in all 16 filter lanes to obtain corresponding weights for multiplication.  

\begin{figure}[!t]
\centering
\centerline{\includegraphics[width=\linewidth]{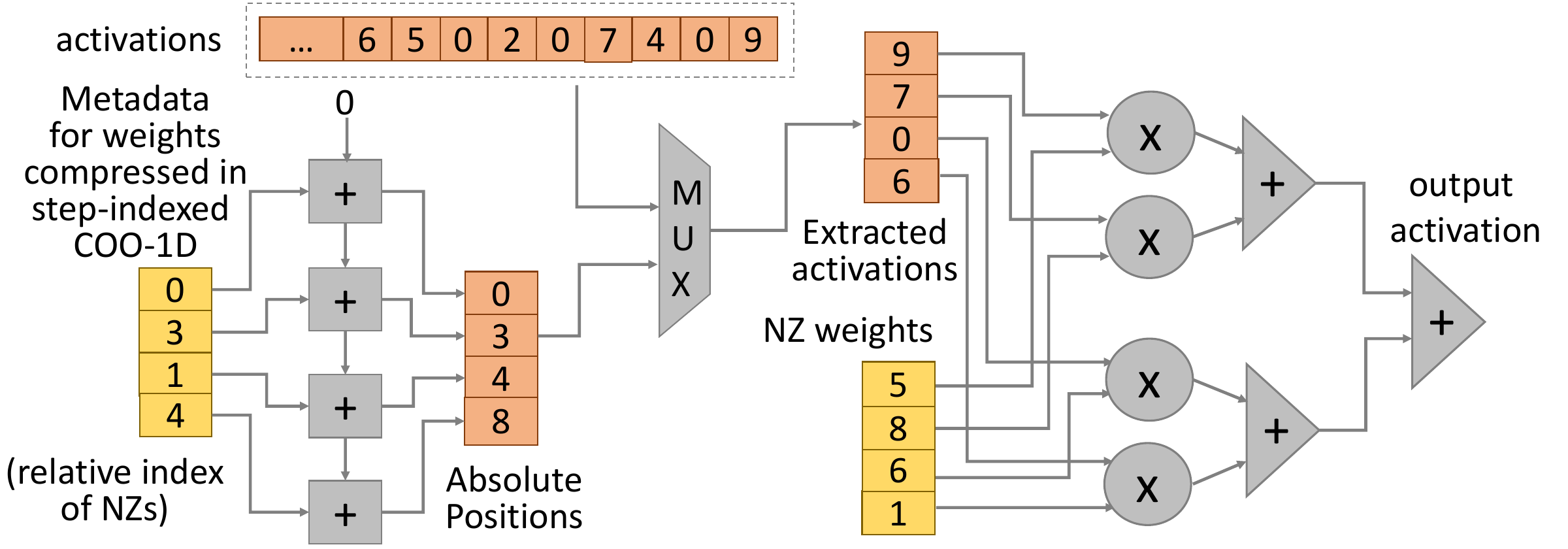}}
\caption{Data extraction via central indexing module in Cambricon-X \cite{zhang2016cambricon} accelerator. The indexing module decodes weights encoded in step-indexed COO-1D format to obtain the absolute positions of NZs. Then, it extracts the activations via a parallel look-up, which are later communicated to a PE via fat-tree NoC for a vector-vector multiplication. (Figure adopted from \cite{zhang2016cambricon}.)}
\label{fig::NZ-detection-cambriconX}
\end{figure}

\textit{Vec-Vec Mul:} PEs of some accelerators spatially process vectors at every cycle (e.g., with 16 multipliers and an adder tree in Cambricon-X). As Fig. \ref{fig::NZ-detection-cambriconX} illustrates, based on positions of NZs of a vector, a combinational logic with multiplexers can select matching data elements to feed the arithmetic units (e.g., in \cite{zhang2016cambricon, kang2019accelerator, ampere}). \insight{An associated challenge is overheads of parallel look-up}. To exploit high sparsity, larger multiplexers need to be used for indexing the dense tensor, as positions of scattered NZs are likely distant. With the search length set as 256 (supports 93.75\% sparsity for fetching 16 NZ elements), a central indexing module in Cambricon-X occupied about 31\% and 35\% of total on-chip area and power, respectively (exceeded total power of all 16 PEs) \cite{zhang2016cambricon}. 

\mysubsubsection{Compare metadata of sparse tensors for extracting matching pairs of NZs}
For effectual computations over multiple compressed tensors, the extraction logic determines pairs of NZs (intersections) by comparing indices either from metadata streams or in multi-stage indexing.

\textit{MAC:} Circuitry for extracting NZ scalars can consist of one or more comparators (or AND gates for comparing bitmaps) and an additional indexing logic (e.g., in ZENA \cite{kim2017zena} and SparTen \cite{gondimalla2019sparten}). The comparators match positions of NZs, and the indexing logic uses their outputs to extract the leading pair. Due to the diverse sparsity of tensors, positions of NZs may not match during comparison. Therefore, the detection logic uses several comparators to search within a large window, which usually can provide at least one pair at every cycle. Priority encoders provide the leading $n$-pairs for feeding $n$ computational units (n=1 for scalar PEs). The data extraction unit can use skip mechanisms (e.g., in ExTensor \cite{hegde2019extensor}) to quickly navigate through the lanes. 

Alternatively, multi-stage indexing logic is used for extracting the pair. The first stage obtains a position of an NZ from one tensor for indexing another tensor. Later stage checks if there is a corresponding NZ in another tensor and extracts it upon matching the positions. For example, in EIE \cite{han2016eie}, each PE loads an NZ activation from a queue; when it does not have any matching weights, it fetches the next activation from the queue in the next cycle. Depending on the sparsity level and pattern, the indexing-based design occasionally may not find the matching data, wasting the execution cycles, i.e., functional units in the pipeline are not utilized.  

\textit{Sc-Vec Mul:} PEs in EyerissV2 \cite{chen2019eyeriss} use multi-stage extraction. Each SIMD PE fetches an CSC-coded activation and its position, and checks positions of NZ weights. Upon match, it forwards the activation and weights to two MAC units. 

\begin{figure}[!t]
\centering
\centerline{\includegraphics[width=\linewidth]{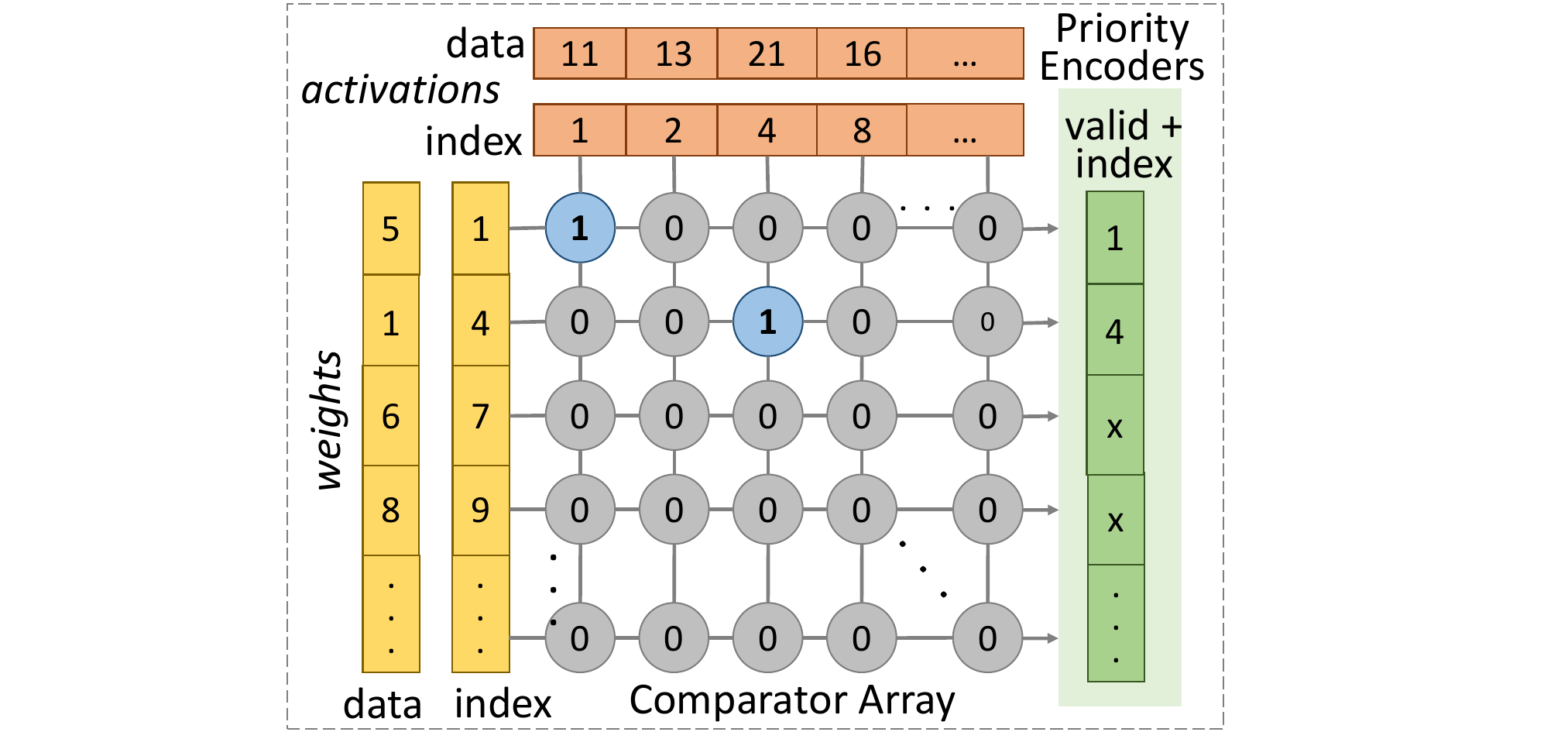}}
\caption{Associative index matching in SNAP (Figure adopted from \cite{zhang2019snap}).}
\label{fig::NZ-detection-SNAP}
\end{figure}

\textit{Vec-Vec Mul:} The data extraction logic to feed multiple arithmetic units of a vector PE requires multiple comparators followed by priority encoders or multiplexers. For example, in SNAP architecture \cite{zhang2019snap}, an associate index matching module (AIM, Fig. \ref{fig::NZ-detection-SNAP}) determines the positions of NZs in case of valid matches. Each PE of a row is interfaced with a shared AIM. Using comparison outcomes from AIM, a sequencer in each PE determines leading pairs of matching data, which are then fed to three multipliers within the PE. Cambricon-S \cite{zhou2018cambricon} uses similar extraction logic, but its comparator array is just ANDing of the bits due to bitmap encoding.    

\mysubsubsection{Eliminating extraction of intersecting NZs} Some accelerators do not require extracting unstructured NZs.   

\begin{figure}[b]
\centering
\centerline{\includegraphics[width=\linewidth]{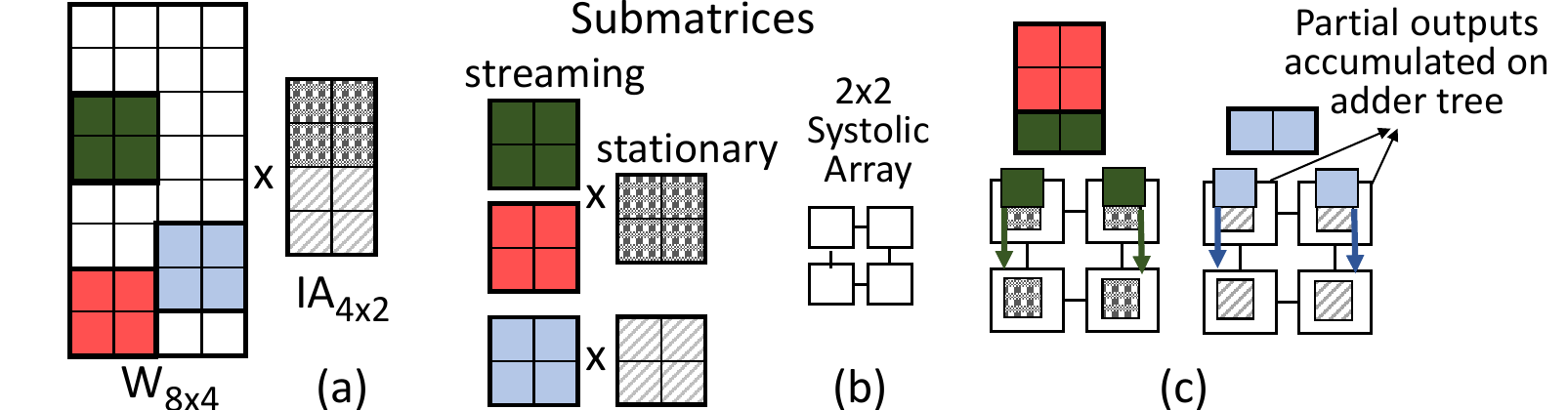}}
\caption{Computation of locally dense regions in ERIDANUS (Figure adopted from \cite{asgari2019eridanus}). (a) Matrix multiplication with block-sparse weights. (b) Sub-matrices for processing on a 2$\times$2 systolic array. (c) Multiplication of streaming blocks (NZs) with stationary data.}
\label{fig::orchestration-structured-sparsity}
\end{figure}

\textbf{Orchestrating structured computations:} A few techniques targeted high sparsity of single tensor (DNN weights). With data pruning or transformations, they achieved coarse-grain sparsity so that each PE can process a dense region of NZs. ERIDANUS \cite{asgari2019eridanus} proposed a pruning algorithm to cluster the weights (Fig. \ref{fig::orchestration-structured-sparsity}a). Blocks of NZ weights are streamed to PEs of systolic arrays for conventional processing (Fig. \ref{fig::orchestration-structured-sparsity}c). Corresponding activations are kept stationary. Partial products computed by each row of PEs are added on a separate adder tree. When block width for structured pruning can be set as the height/width of the systolic array, dot products can be accumulated linearly over the systolic array itself. Thus, \insight{structured sparsity allows executing denser blocks conventionally on accelerators, while requiring additional support to index and communicate the blocks.} Adaptive tiling \cite{kung2018adaptive} used a column-combining approach. For a sparse GEMM, NZ weights were statically combined such that each column of the systolic array could process multiple columns of input activations. Thus, it obviated the run-time data extraction and reduced total invocations of the systolic array by 2$\times$--3$\times$ for processing point-wise CONVs of MobileNet. 
CirCNN \cite{ding2017circnn} and C-LSTM \cite{wang2018c} proposed executing DNN operators as FFT (Fast Fourier Transform) on smaller block-circulant matrices.

\textbf{Coordinate computation unit:} SCNN \cite{parashar2017scnn} and SqueezeFlow \cite{li2019squeezeflow} perform unit-strided convolutions as a Cartesian product where all elements of two blocks of tensors should be multiplied together. Due to all-to-all multiplication, no special support is required for extracting matching pairs of NZs. However, index computation is still required to determine which partial-sums should be accumulated with partial products. This calculation is performed in a ``coordinate computation unit'' that processes metadata (indices of NZs) and determines indices of outputs. These approaches require conflict detection in hardware since it can't be pre-determined which accumulators would be accessed in any cycle. Since coordinate computation unit facilitates direct processing on compressed tensors, it may also be used for computing block-level indices for processing a \emph{coarse-grain block-sparse} tensor.

\subsection{Centralized vs. Distributed Management}

\mysubsubsection{Centralized} The data extraction unit can be either centralized (and shared among PEs) or within pipelines of PEs. Advantages of central mechanisms are: (i) PEs can be directly provided effective NZs for useful computations \cite{zhang2016cambricon}. It can also be used as a pre-processing unit for a PE-array that processes structured computations, e.g., systolic arrays or near-data accelerators. (ii) Centralized extraction in some architectures (e.g., Cambricon-X \cite{zhang2016cambricon}) duplicates hardware for concurrent extractions for PEs. However, the module can be time-shared by multiple PEs (e.g., in SNAP \cite{zhang2019snap}), which can reduce area and power. In fact, by leveraging structured $W$-sparsity, the module in Cambricon-S \emph{shares} extracted indices among \emph{all} PEs. (iii) Centralized logic extracts work for multiple PEs, and often it is coupled with a controller that allocates data to PEs. So, it can enable \emph{run-time load balancing}. However, a major challenge is to maintain spatial data reuse. This is because, the centralized unit mostly extracts data on a per-PE basis for communication to a unique PE. So, the common data for multiple PEs cannot be multi-cast. SNAP overcomes this limitation by sharing a module with a row of PEs and multicasting data to PEs. The multicast occurs first, followed by PEs communicating their metadata to the extraction unit. Then, extracted indices are streamed back to a PE, which uses them to obtain data from its local RF for computations.

\mysubsubsection{In-PE} PEs of several accelerators, such as Cnvlutin \cite{albericio2016cnvlutin}, ZENA \cite{kim2017zena}, and EyerissV2 \cite{chen2019eyeriss}, extract appropriate data.It allows a controller to multicast or broadcast tensor elements for spatial reuse. Then, in-PE logic extracts the data. However, challenges are: (i) in-PE logic may incur ineffectual cycles for extraction that cannot be hidden. (ii) employing inter-PE load-balancing in the hardware may be infeasible or costlier, as the actual work carried out by different PEs is unknown while offloading compressed tensors to PEs (until extraction in PE datapath).  

\begin{figure*}[!t]
\centering
\centerline{\includegraphics[width=\linewidth]{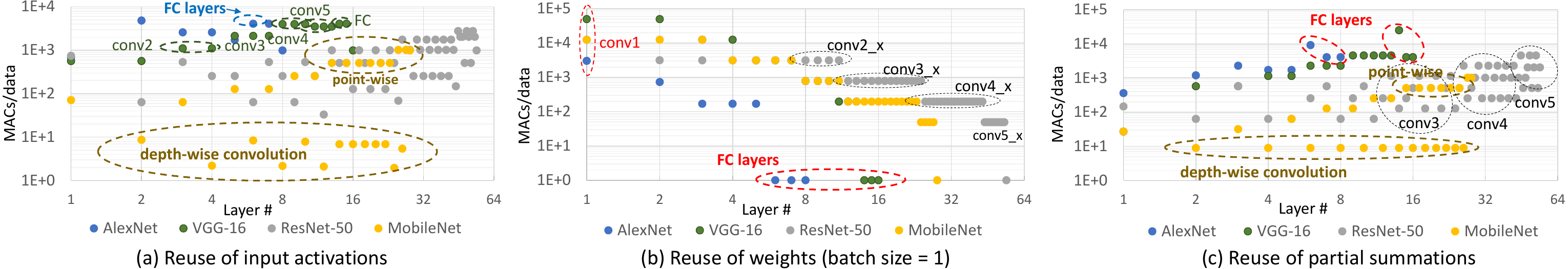}}
\caption{Data reuse opportunities for executing different CNN layers (dense tensors) on hardware accelerators (Figure inspired by \cite{chen2019eyeriss}).} 
\label{fig::analysis-data-reuse}
\end{figure*}

\subsection{Optimization Opportunities}

\textit{(i) Sparsity-adaptive low-cost data extraction mechanisms:} Encodings of sparse tensors are often selected with a focus on storage benefits. However, the computational overhead and hardware cost for encoding and decoding tensors should be also reduced, since they affect the performance and energy consumption. When the data extraction cannot feed $n$ pairs of NZs to $n$ computational units of a PE at every cycle, achieved speedup can be lower from the peak. Sustaining the acceleration across various sparsity of tensors can be challenging, as different extraction schemes may be cost-effective for only a certain sparsity range and patterns. For example, for similar sparsity, extraction logic with a few comparators may easily locate a pair of NZs. However, an indexing-based mechanism may be more effective, when one tensor is highly sparse and another is dense. Moreover, when positions of NZs in the two tensors are considerably distant (e.g., for diverse sparsity levels or for hyper-sparse tensors), the extraction logic needs to use several comparators or multiplexers for parallel lookup, so that it can extract at least one pair to feed each computational unit. Therefore, the extraction module needs to be \emph{configurable} or consist of (and select among) multiple mechanisms so that it can exploit a variety of sparsity at a modest hardware cost. 
For the latter, it can dynamically use partial features for desired sparsity levels/patterns (power-gated otherwise).

\textit{(ii) Tightening integration with load balance mechanism:} Central data extraction module can enable dynamic load balancing of work among PEs (e.g., data-driven dynamic work dispatch in GraphDynS \cite{yan2019alleviating}). As section \ref{sec::load-balance} discusses, the inter-PE imbalance can be severe due to the irregular distribution of NZs in tensor blocks that are allocated to PEs. Its mitigation by structuring the data may not always be possible (e.g., for activations/weights of some models or applications beyond deep learning). Consequently, accelerators may attain ineffective utilization of PEs and low speedup. Although some accelerators used hardware modules for dynamic balancing, further efficiency may be achieved by enhancing the centralized extraction module with additional low-cost logic. This is because it already keeps the track of the data provided to PEs, which can lead to information about the number of operations performed by different PEs.

\section{Memory Management of Compressed Tensors}
\label{sec::memory-management}

Accelerators contain \emph{multi-banked} scratchpads, which are usually shared among PEs. Either a scratchpad is \emph{unified} \cite{yang2018dnn}, or separate buffers store different tensors \cite{zhang2019compact, kim2017zena}. Their sizes vary from several tens of KBs \cite{zhou2018cambricon, chen2019eyeriss} to several MBs \cite{albericio2016cnvlutin, jouppi2017datacenter}. Effective management of shared and local memory highly reuses data and hides memory access latency behind computations on PEs. This section discusses how sparsity and reduced shapes of tensors lower reuse. 
However, compressed tensors help to achieve better speedups and energy efficiency, as more data fits in on-chip memory, reducing off-chip accesses. This section also describes how irregular accesses (e.g., arbitrating output activations) make management of the banks challenging. Then, it discusses reusing intermediate outputs via fused-layer executions and how sparsity affects it. 

\begin{figure*}[!t]
\centering
\centerline{\includegraphics[width=\linewidth]{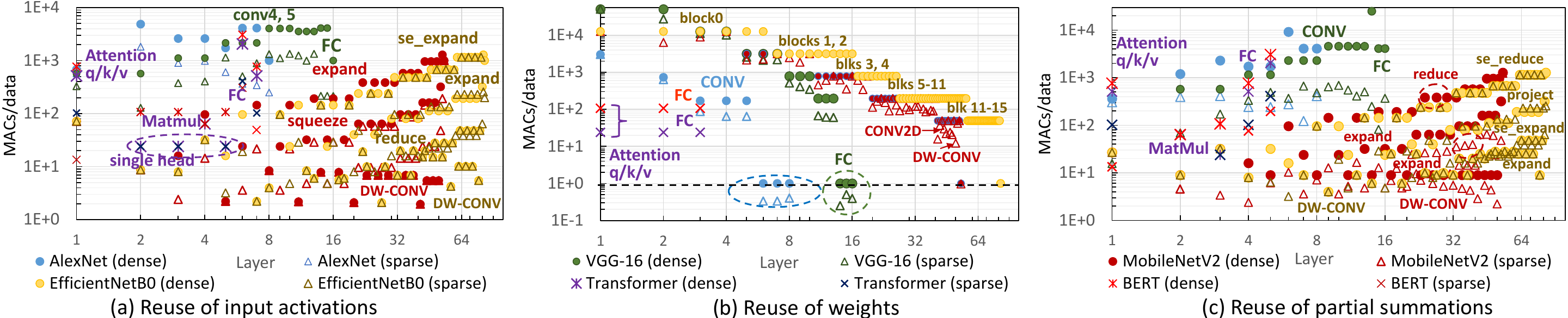}}
\caption{Impact of sparsity on data reuse opportunities for accelerating CNNs and NLP models.}
\label{fig::analysis-data-reuse-sparse}
\end{figure*}

\subsection{Leveraging Data Reuse Opportunities}
\label{sec::data-reuse}

\mysubsubsection{Reuse characteristics} Depending on the functionality of layers, there can be significant reuse of tensors. Figures \ref{fig::analysis-data-reuse} and \ref{fig::analysis-data-reuse-sparse} depict reuse opportunities for different layers (early CONV layers, later CONV layers, MLPs, DW-CONVs, PW-CONVs, expand or reduce layers, attention mechanism). For each tensor, the data reuse is calculated as the total number of MACs per data element. For better visualization, reuse factors and layers are plotted on a logarithmic scale. 
\textit{Input activations:} Reuse of input activations increases with going deeper in CNNs since the number of filters increases significantly. It is also high for 'expansion' layers in bottleneck blocks (Fig. \ref{fig::analysis-data-reuse-sparse}). DW-CONVs are an exception and present very low reuse as there is only one filter. 'Squeeze' or 'reduce' layers present moderate reuse for dense tensors. Reuse in FC layers or MLPs (e.g., in encoder/decoder layers of Transformers \cite{vaswani2017attention}) depends on the sizes of weight matrices (i.e., sizes of output tensors). 
\textit{Weights:} Since 2D feature maps in CNNs are usually much larger than 2D weights, weight reuse can be higher by an order of magnitude. With going deeper in CNNs, feature maps shrinks spatially, which lowers the reuse. There is no weight reuse for MLPs, but increasing the batch size linearly improves the weight reuse. Video processing applications use 3D CNNs (e.g., c3d \cite{tran2015c3d}), which can further increase the reuse opportunities \cite{hegde2018morph} for input activations and weights due to additional processing steps on consecutive frames. For NLP models such as Transformer \cite{vaswani2017attention} and BERT \cite{devlin2019bert}, Fig. \ref{fig::analysis-data-reuse-sparse} illustrates weight reuse for executing a sequence of 24 and 107 tokens, respectively. $MatMul$s in the attention-based calculation are shown for a single head.
\textit{Partial summations:} Input channels are increased as we go deeper into CNNs. Similarly, 'reduction' layers in bottleneck blocks involve more input channels. Both improve the reuse of partial summations. MLPs also usually provide high reuse due to larger input vectors. DW-CONVs show very low reuse because partial summations are not accumulated across input channels. 

\mysubsubsection{Impact of sparsity on reuse} Increase in sparsity can lead to lower reuse. To determine the impact of sparsity, we considered evaluations by Han et al. \cite{han2015learning} for pruned AlexNet and VGG-16 models. For recent DNNs like MobileNetV2 or BERT models, we considered sparse models as listed in Table \ref{tab:sparsity-dnn-models}. Then, we calculated the reuse as NZ MACs per NZ of a tensor. Fig. \ref{fig::analysis-data-reuse-sparse} plots the reuse opportunities for both dense and sparse tensors of CNNs and NLP models. Since execution in encoder/decoder modules of NLP models is repetitive, unique layers of a single module are only shown (sparsity averaged across all encoder/decoder modules). 
The figure shows that for sparse models, reuse characteristics are preserved, but the reuse factor decreases for almost all layers and tensors, as compared to processing dense tensors. Primarily, this is due to the reduced number of effectual MACs. For example, for MLPs without batching, weight reuse can drop below one. It means that even if a weight matrix consists of NZs, some of them are never used due to the unavailability of matching NZs in input activations. As an exception, reuse of weights remains the same, when activation sparsity is absent (e.g., EfficientNetB0 \cite{lu2019efficient}, BERT \cite{devlin2019bert}). Similarly, with dense weights, low or moderate reuse of activations remains the same for DW-CONV or 'excite' layers, respectively.

The reuse of partial summations also decreases since effectual MACs per partial summation decrease with sparsity. Note that each output activation element still needs to be populated or assembled before ReLU/encoding. Due to sparsity and fewer input channels, the reuse is low or moderate in 'expansion' layers. Similarly, small matrices in processing individual attention heads exhibit low reuse. The reuse remains high for 'reduce' layers in CNNs or query and value processing and FC layers in NLP models. To sum up, although sparsity reduces the reuse of tensors, there can be high data reuse for many layers (up to $1E+04$), which should be exploited for efficient accelerations.

\mysubsubsection{Temporally reusing data through shared on-chip memory} Like CPUs, accelerators have memory hierarchies because applications have different working set sizes. Data reuse can be leveraged \emph{temporally} (repeatedly accessing data from memory without accessing lower-level memory) and \emph{spatially} (providing the same data to multiple PEs without repeatedly accessing memory). After exploiting high temporal reuse, the highest energy is spent in upper buffers \cite{sze2017efficient, yang2018dnn}.  

\subsection{Hiding Miss Latency Behind Computations}

\mysubsubsection{Management of tiled data in double-buffered memory} On-chip buffers are typically not large enough to accommodate all tensors. Therefore, loops are tiled for reusing some tensors from buffers, while repeatedly accessing other tensors from the off-chip memory \cite{parashar2017scnn, chen2016eyeriss}. Since scratchpads are non-coherent and their management is software-directed, data is transferred by direct memory accesses (DMA) \cite{jouppi2017datacenter, dave2019dmazerunner}. PEs are kept engaged in useful computations by interleaving computations with memory accesses. Such an objective is usually achieved by double-buffering (aka ping-pong buffers) \cite{kim2017zena, zhou2018cambricon}. Loop optimization techniques, like loop tiling and ordering, can determine the sizes of tensor blocks to be managed in memories and sequence of memory accesses for high reuse and reduced data transfers \cite{dave2019dmazerunner, yang2018dnn}.

\mysubsubsection{Asynchronous communication} Some accelerators hide the latency of communicating data to the shared/local memory with an asynchronous mechanism that refills the memory after some data has been consumed (e.g., in Cambricon-X \cite{zhang2016cambricon}). For such execution, PEs and a DMA controller may simultaneously produce/consume data either through different banks or at the granularity of small blocks in the same bank. Similarly, when accessing shared memories via configurable communication networks \cite{chen2019eyeriss}, PEs can execute in a dataflow fashion and request partial refilling of their memory with new data. Such \insight{mechanisms for asynchronous communication and computations can alleviate work imbalance among PEs} that is caused by leveraging unstructured sparsity.
 
\mysubsubsection{Impact of sparsity on the latency of memory accesses and speedup} For memory-bounded execution (e.g., MLPs), even with effective prefetching, miss penalty may be significant. It restricts accelerators from achieving peak performance \cite{jouppi2017datacenter}. When tensors are sparse, the amount of data that needs to be transferred from off-chip reduces significantly, leading to substantial performance gains. For example, Cambricon-S reported up to 59.6$\times$ speedup of FC layers for hyper-sparse weights. However, higher $IA$-sparsity did not provide such gains (speedup saturated at about 14$\times$) since the latency of accessing weights dominated total execution time. For processing high sparsity (e.g., 90\%+) and low reuse, it becomes challenging to engage functional units into effectual computations. This is because, with \emph{low} arithmetic intensity, required data \emph{may not be prefetched} at available bandwidth.

\subsection{Management of Multi-Bank Memory}
\label{sec::memory-bank-management}

\mysubsubsection{Concurrent accesses to memory banks} While single-bank memory can be easier to manage, it is infeasible to provide multiple ports for the PE-array with just one bank \cite{kim2011high}. Moreover, multi-port unified memory consumes very high power and longer latency \cite{weste2015cmos}. 
So, on-chip memories are partitioned into smaller banks \cite{parashar2017scnn, chen2019eyeriss, shen2017maximizing}. For mapping a layer onto the accelerator, each bank is usually allocated to only one tensor (e.g., in EyerissV2 \cite{chen2019eyeriss}). Banked buffers provide multiple read and write ports, allowing simultaneous accesses to different tensors stored in different banks \cite{chen2016eyeriss, azizimazreah2019shortcut}. 
Sometimes, a data layout reorganization is required before loading into memory banks. Such transformation is done after loading it from DRAM or before writing outputs to DRAM, which consumes additional execution time and energy. For compressed tensors, such transformation can be done along with the data encoding \cite{zhang2019compact} at alleviated overheads.

\mysubsubsection{Arbitration and conflict management} Depending on the indexing logic and interconnect between memory and PEs, managing application data may require additional compilation support or hardware logic for data arbitration and conflict management \cite{parashar2017scnn, yuan2019sticker}. For regular memory accesses (e.g., dense or block-sparse data), allocation and accesses to banks can be determined for mappings of layers. However, computations on unstructured sparse data can lead to accessing \emph{arbitrary} banks and require special support. E.g., outputs from PEs may need to be written to different banks. Moreover, accelerators contain accumulator-buffers \cite{parashar2017scnn}, where PEs or their functional units are connected with memory banks via a crossbar. The crossbar arbitrates write-back of outputs to the appropriate bank \cite{parashar2017scnn, li2019squeezeflow}. Since these partial outcomes can correspond to non-contiguous elements in an output tensor, bank conflicts are possible during arbitration, i.e., multiple outputs need to be simultaneously handled by the same bank \cite{parashar2017scnn, yuan2019sticker}. To obviate conflicts, the buffer contains more banks (e.g., 2$\times$N banks for storing outputs from $N$ sources in SCNN \cite{parashar2017scnn}). It alleviates collisions in hashing irregular outputs into different memory banks. Consequently, the crossbar may require higher bandwidth and significant on-chip area (e.g., 21\% for a 16$\times$32 crossbar in each SCNN's PE). 

\subsection{Reusing Intermediate Tensors}

\mysubsubsection{Reusing intermediate tensors from large on-chip memory} Intermediate feature map in DNNs is an output of a layer that serves as input to later layers. It can be kept stationary and reused from on-chip memory to reduce off-chip traffic. Such reuse is amplified when input is the same for multiple layers due to residual connections \cite{he2016deep} or high cardinality (e.g., ResNeXt \cite{xie2017aggregated}). Leveraging it can be important for latency-bounded real-time applications. \insight{Sparsity-encoding and quantization significantly makes such reuse opportunities more feasible} due to reduced storage requirements. Accelerators with large memories (hundreds of KBs) such as SCNN \cite{parashar2017scnn} and Cnvlutin \cite{albericio2016cnvlutin}, can leverage such reuse. 

\mysubsubsection{Overcoming static bank assignment} Many accelerators process models layer-by-layer and do not leverage cross-layer reuse, i.e., write outputs for layer $L$ in DRAM and load them back later as inputs for layer $L+1$. It is more prevalent among accelerators with small memories. Moreover, bank assignment for each tensor is often fixed at design time \cite{azizimazreah2019shortcut}, which enforces write-back of outputs and reloading them later in other banks as inputs while processing next layers. Thus, in both cases, output activations are not reused on-chip, causing excessive off-chip memory traffic. To address this problem and exploit cross-layer reuse, shortcut-mining \cite{azizimazreah2019shortcut} used a flexible architecture with decoupled physical-logical buffers. 

For pre-known sparsity, prior techniques for statically determining the data allocation to memory banks may work well by estimating sizes of encoded tensors. However, for dynamic sparsity, conservative estimations may lead to inefficient utilization of banks, and efficient banking for non-conflicting accesses can also be challenging. 

\begin{figure}[!t]
\centering
\centerline{\includegraphics[width=\linewidth]{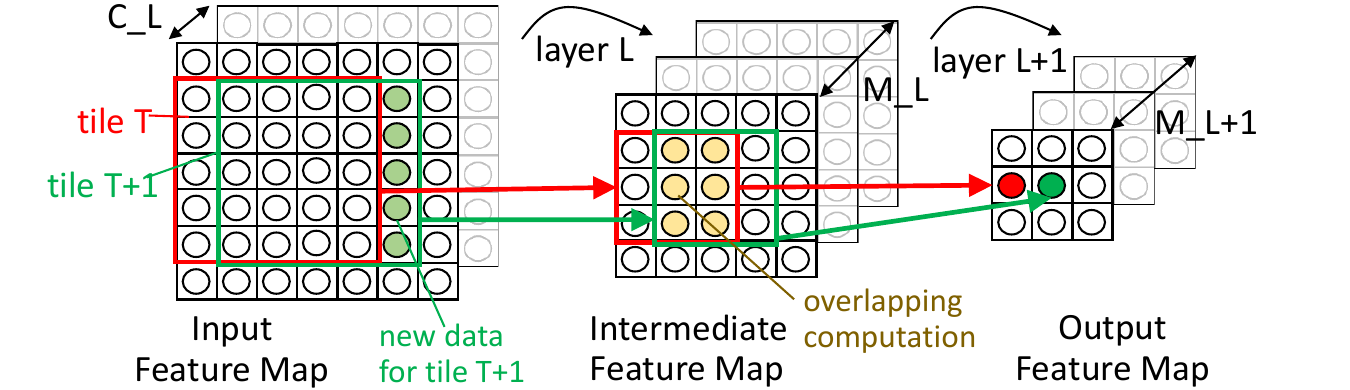}}
\caption{Fusing the execution of layers can significantly reuse intermediate activations \cite{alwani2016fused} (Figure adopted from \cite{alwani2016fused}).}
\label{fig::fused-layer-CNNs}
\end{figure}

\mysubsubsection{Fused-layer execution} Fused-layer CNNs \cite{alwani2016fused} leveraged cross-layer reuse by processing a small tile of activations such that outputs for few layers can be computed alongside while retaining the corresponding data in the on-chip memory. Fig. \ref{fig::fused-layer-CNNs} shows an example for processing an input tile of 5$\times$5 activations (C\_L input channels) for layer $L$ and finally obtaining 1$\times$1 output activations (M\_L+1 output channels) for layer $L+1$. Apart from reusing intermediate outputs for obtaining the output tile, corresponding tiles of intermediate activations and filters are maintained in the memory and reused partially for processing the next tiles (striding execution in the spatial direction). Alwani et al. \cite{alwani2016fused} reported reducing off-chip transfers of input feature maps by 28\% for the first two layers of AlexNet and by 95\% for  the first five layers of VGG-19. Since the cascading by storing all the filters and input channels (dense tensors) requires high memory, \cite{alwani2016fused} applied it to only early layers. However, encoded sparse tensors and further tiling across filters/channels allow fitting tensors for multiple layers in the small memory, making such reuse opportunities more feasible. The tile size and number of layers that can be fused are bounded by memory capacity. So, fusion parameters depend on the actual/anticipated sparsity levels. For efficient executions, fusion parameters need to be explored systematically with sparsity-aware dataflows.

\subsection{Techniques for Further Energy-Efficiency}

\mysubsubsection{Look-ahead snoozing} Depending on the sparsity, encoding of tensors, and mapping of the layer, several banks can be unused or inactive for certain time intervals. Accelerators achieve further energy efficiency by power gating unused or inactive banks. For example, look-ahead snoozing in CompAct \cite{zhang2019compact} targeted reducing the leakage power of large on-chip SRAMs. Each bank of its activation SRAM can be power-gated. Banks unutilized during the execution of a layer were put in the deep sleep mode (maximal savings in leakage power, while not preserving any data in unused banks). Further, the period of active cycles for each bank was determined based on the data movement schedule. Then, inactive banks were snoozed during execution (i.e., connecting to the data retention voltage for consuming lower leakage power). 

\mysubsubsection{Skipping memory hierarchy} Some layers do not provide significant reuse. Data reuse is also lowered due to sparsity and architectural choice for extracting or communicating NZs. Therefore, a few accelerators (e.g., EIE \cite{han2016eie} and Cambricon-X \cite{zhang2016cambricon}) obviate storing non-reusable data in the shared memory and directly feed it to appropriate PEs (weights for MLPs).  

\subsection{Optimization Opportunities}
 

\textit{(i) Managing both data and metadata in unified memory:} Accelerators often contain separate buffers for metadata (positions of NZs, indirection tables for shared values). Although such designs are easy to manage for processing tensors of some models encoded in a specific format, they may not work well across different levels of sparsity and value similarity, as storage requirements vary significantly. 
So, designers can explore unified memory architectures for managing both data and metadata (including memory partitioning and bank management) and their trade-offs. It can also be leveraged to tailor efficient designs for programming FPGAs.

\begin{figure}[!t]
\centering
\centerline{\includegraphics[width=\linewidth]{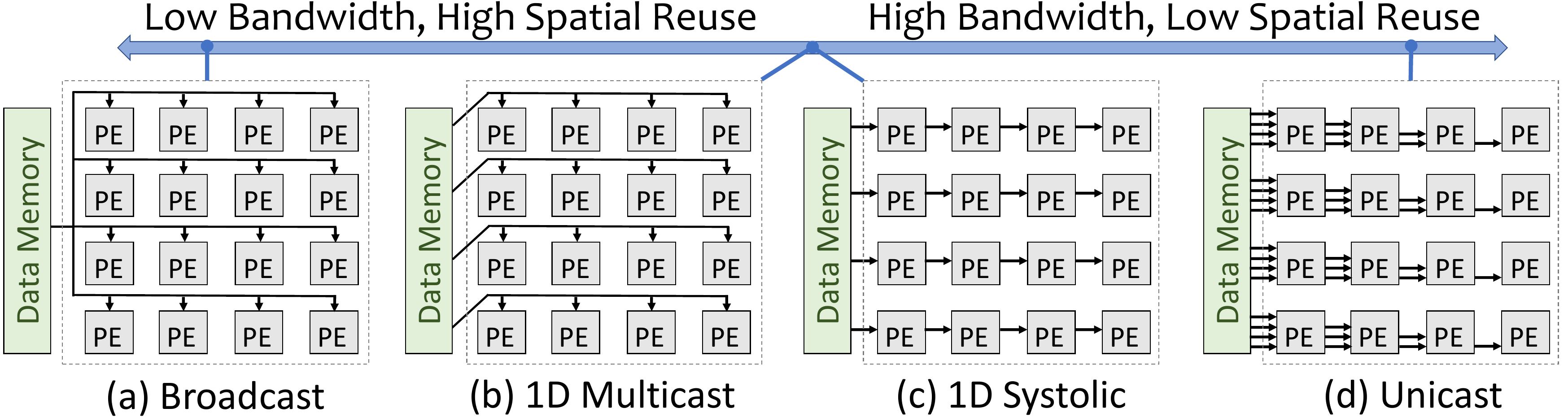}}
\caption{Common NoC designs (Figure adopted from \cite{chen2019eyeriss}).}
\label{fig::NoC-designs}
\end{figure}

\section{Interconnects for Distributing Non-Zeros and Reducing Partial Outputs}
\label{sec::comm-networks}

Network-on-chip (NoC) is required to efficiently distribute data to PEs, exchange data between PEs (for reducing partial outputs), and collect distinct outputs back from PEs. To process data-intensive ML models, accelerators employ multiple high-bandwidth interconnects for simultaneous communication of different tensors between PEs and buffers. At first, this section describes NoCs for the distribution of operands, which vary in terms of bandwidth and spatial reuse of the data. With efficient NoC design, PEs can be engaged in processing data from input FIFOs or local memory, which gets interleaved with communication of another set of data via NoC. This section also discusses \emph{configurable NoC} designs that can support various bandwidth requirements and spatial reuse opportunities due to variations in sparsity and tensor shapes. In processing sparse tensors, \insight{unstructured reduction of partial outputs among PEs can be challenging}. This section describes different mechanisms for \emph{accumulating} the outputs \emph{temporally} or \emph{spatially} at PE level and PE-array level. It also discusses configurable mechanisms for \emph{asymmetric} accumulation of \emph{variable-sized} partial outputs.

\subsection{Mechanisms for Distribution of Operands}

\begin{table}[!t]
\centering
\caption{NoC Designs for Distribution of Sparse Tensors}
\label{tab:NoC-data-distribution}
\addtolength{\tabcolsep}{-3pt}
\begin{tabular}{|c|c|m{5.7cm}|}
\hline
\multirow{5}{*}{\makecell{Topo-\\logy}}      
    & Unicast & \cite{zheng2018kernelxform, struharik2018conna, zhang2016cambricon, albericio2016cnvlutin, parashar2017scnn, li2019squeezeflow, hegde2019extensor, aimar2018nullhop, page2017sparcnet, lu2018spwa} \\ \cline{2-3} 
    & Multicast         & \cite{zhang2019snap, hegde2019extensor, aimar2018nullhop} \\ \cline{2-3} 
    & Broadcast         & \cite{zheng2018kernelxform, struharik2018conna, han2016eie, albericio2016cnvlutin, judd2017cnvlutin2, parashar2017scnn, zhang2019snap, lee20197, li2019squeezeflow, kim2017zena, zhou2018cambricon, gondimalla2019sparten, yuan2018sticker, han2017ese, page2017sparcnet, kang2019accelerator, lu2018spwa} \\ \cline{2-3} 
    & Mesh              & \cite{zhang2019compact, asgari2019eridanus, kung2018adaptive, venkatesh2017accelerating} \\ \cline{2-3}
    & Configurable      & \cite{kwon2017rethinking, chen2019eyeriss, qin2020sigma} \\ \hline
\multirow{2}{*}{\makecell{Spatial \\ Reuse}} 
    & Activations       & \cite{zheng2018kernelxform, han2016eie, albericio2016cnvlutin, judd2017cnvlutin2, struharik2018conna, zhou2018cambricon, gondimalla2019sparten, yuan2019sticker, kim2017zena, choles2018parsecore, aimar2018nullhop, han2017ese, page2017sparcnet, lee20197, zhang2019snap, kang2019accelerator, chen2019eyeriss, qin2020sigma, yuan2018sticker, lu2018spwa} \\ \cline{2-3} 
    & Weights           &\cite{parashar2017scnn, li2019squeezeflow, yuan2019sticker, kim2017zena, zhang2019snap, chen2019eyeriss, qin2020sigma, yuan2018sticker} \\ \hline 
\end{tabular}
\addtolength{\tabcolsep}{3pt}
\end{table}

Fig. \ref{fig::NoC-designs} shows some common NoC designs for distributing the operands, their bandwidth, and achievable spatial reuse \cite{chen2019eyeriss}. Data can be reused spatially by distributing it to multiple PEs or functional units. For layers with high reuse opportunities (Fig. \ref{fig::analysis-data-reuse}), it lowers communication and helps to hide the communication latency. Most accelerators leverage spatial reuse with multicast or broadcast NoC. They consist of configurable buses or trees that multicast the data to PEs (often in a single cycle) \cite{chen2016eyeriss, qin2020sigma}. In contrast, the mesh interconnect (e.g., in systolic arrays  \cite{jouppi2017datacenter}) or 1D buses communicate the data and reuse spatially with a store-and-forward mechanism. Low-reuse tensors are distributed with unicast NoCs. Table \ref{tab:NoC-data-distribution} lists common interconnect topologies used by previous accelerators for data distribution.

Communication requirements vary significantly depending on the sparsity of tensors, available reuse, and adopted dataflow mechanism. Prior work \cite{vainbrand2010network} provides a detailed analysis of different NoC topologies and \cite{kwon2017rethinking} characterizes the NoC bandwidth required for different dataflows. Similarly, analytical tools including \cite{kwon2019understanding} model implications of different dataflows on communication requirements and execution time. 

\mysubsubsection{Broadcast} Accelerators, including Cnvlutin \cite{albericio2016cnvlutin}, EIE \cite{han2016eie}, and Cambricon-S \cite{zhou2018cambricon}, use broadcast NoC to reuse activations for processing CONVs or MLPs. Similarly, in SCNN \cite{parashar2017scnn}, weights are broadcast to PEs for executing unit-strided convolutions with input stationary dataflow. For sparsity-encoded tensors, their NZs (and positions) can be broadcast for spatial reuse, as long as the NZs are indexed or extracted afterward (e.g., in-PE extraction in Cnvlutin and EIE). In Cambricon-S, positions of intersecting NZs are extracted centrally before broadcast, but due to structured sparsity, the same extracted positions are used by all PEs. So, NZ activations are broadcast to all PEs. 

\mysubsubsection{Multicast} Eyeriss \cite{chen2016eyeriss}, ZENA \cite{kim2017zena}, and SNAP \cite{zhang2019snap} use multicast NoC to reuse multiple operands spatially. For example, Eyeriss processed tensors with row-stationary dataflow where PEs of a row processed the same spatial rows of filters, and diagonal PEs processed the same spatial row of feature maps. Eyeriss facilitated such multicasting through its configurable NoCs, which consisted of row-wise and column-wise controllers for 2D PE-array. Each controller could be configured with a pre-determined tag value, which was compared with the row or column tag of a packet. Upon matching the tags, a row-wise controller forwarded the packet to associated column-wise controllers, and a column-wise controller forwarded it to the associated PE. Similarly, for processing bitmap-coded tensors in ZENA \cite{kim2017zena}, a block of activations was broadcast to a row of PEs, and a block of weights was multicast to PEs of the same column. 

\mysubsubsection{Mesh} A few accelerators, including Compact \cite{zhang2019compact}, ERIDANUS \cite{asgari2019eridanus}, and \cite{kung2018adaptive}, use systolic arrays with mesh interconnects. Since the same data is forwarded among PEs in the same row or column, such NoCs achieve the same amount of spatial reuse as multicast NoCs. But, for sparse tensors, efficient and correct processing becomes challenging. Hence, \emph{pre-processing} is needed to cluster appropriate NZs or index appropriate block of structured-sparse tensor before feeding PEs of the systolic array \cite{asgari2019eridanus, he2020sparse, qin2020sigma}.   

\mysubsubsection{Unicast} SCNN \cite{parashar2017scnn}, Cambricon-X \cite{zhang2016cambricon}, and SqueezeFlow \cite{li2019squeezeflow} use unicast NoC or point-to-point links. Such NoCs concurrently feed different elements to various PEs. They are used when spatial reuse of a tensor is infeasible (e.g., weights in MLPs, NZs are extracted beforehand, due to dataflow requirements) or outputs are collected simultaneously (section \ref{sec::write-back}). 
With high bandwidth, they reduce communication latency \cite{zhang2016cambricon}, but can incur high area and power. 

\mysubsubsection{Configurable}
\insight{Communication requirements vary with different dataflows} that are effective for only some DNN layers (section \ref{sec::dataflows} and Table \ref{fig::layer-characteristics}). Further, while communication may consist of gather, scatter, forward, or reduction patterns \cite{kwon2017rethinking, xu2020autodnnchip}, efficient execution may demand their combination or even non-uniform patterns including multi-hop communications among PEs \cite{yang2018dnn}. Therefore, configurable NoC designs are required, which can support various communication patterns that are amenable to different reuse and sparsity. Recent designs including EyerissV2 \cite{chen2019eyeriss}, microswitch-NoC \cite{kwon2017rethinking}, and SIGMA \cite{qin2020sigma} address some of these challenges.

\begin{figure}[!t]
\centering
\centerline{\includegraphics[width=\linewidth]{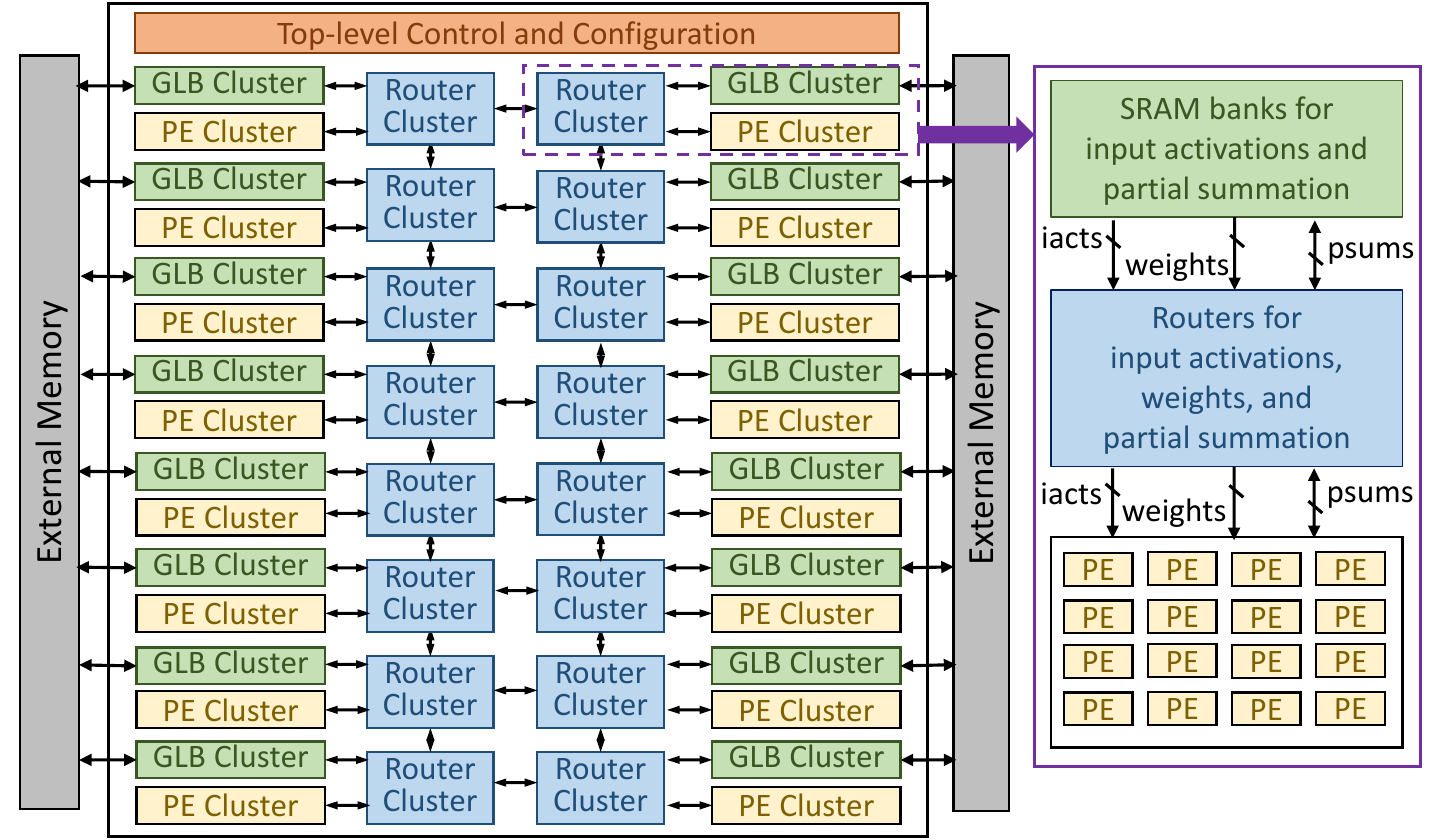}}
\caption{EyerissV2 accelerator architecture \cite{chen2019eyeriss} (Figure adopted from \cite{chen2019eyeriss}).}
\label{fig::architecture-EyerissV2}
\end{figure}

\begin{figure}[!t]
\centering
\centerline{\includegraphics[width=\linewidth]{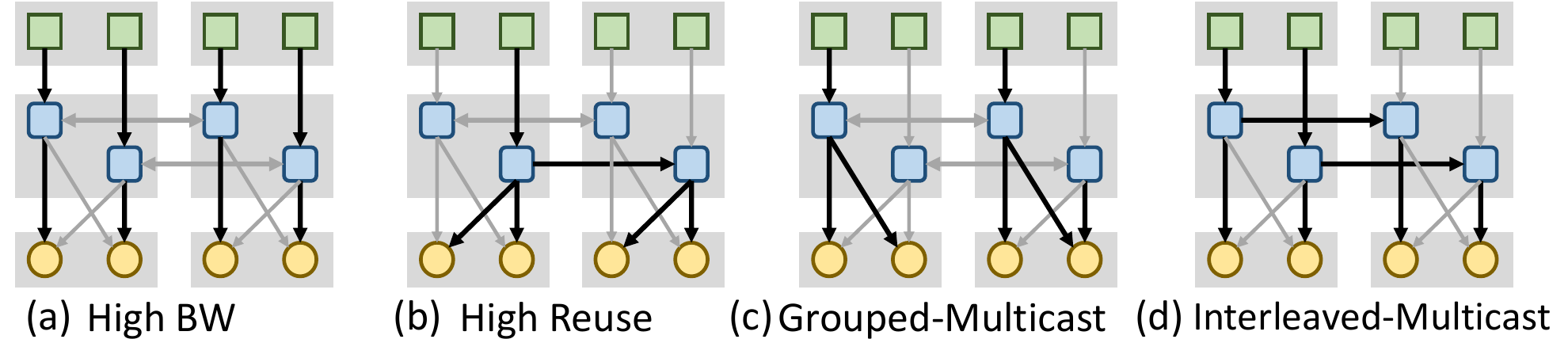}}
\caption{Different configuration modes of hierarchical mesh network in EyerissV2 architecture \cite{chen2019eyeriss} (Figure adopted from \cite{chen2019eyeriss}).}
\label{fig::NoC-Hierarchical-Mesh-EyerissV2}
\end{figure}

EyerissV2 \cite{chen2019eyeriss} uses a novel \emph{hierarchical-mesh} NoC, which is illustrated in Fig. \ref{fig::architecture-EyerissV2}. EyerissV2 contains 16 clusters (8$\times$2 array) of PEs and global buffers (GLBs). Each PE-cluster contains 3$\times$4 PEs, and each 12 kB GLB-cluster contains seven banks for input and output activations. At the top level, router clusters are connected through a 2D mesh, and they enable communication among different PE-clusters and GLB-clusters. For local communication among each PE-cluster and GLB-cluster, a router-cluster with ten routers is used. Each router connects PEs with a port of the GLB cluster for accessing GLB bank or off-chip memory (three, three, and four routers for managing input activations, weights, partial summations). Locally, an all-to-all NoC connects all PEs of a PE-cluster to the routers for each data type. As Fig. \ref{fig::NoC-Hierarchical-Mesh-EyerissV2}(a)--(d) illustrates, it facilitates multiple communication patterns including multicast, broadcast, and unicast of the tensors. 2D mesh topology enables inter-cluster communications, allowing an interleaved-multicast or broadcast to all clusters.

\begin{figure}[!b]
\centering
\centerline{\includegraphics[width=\linewidth]{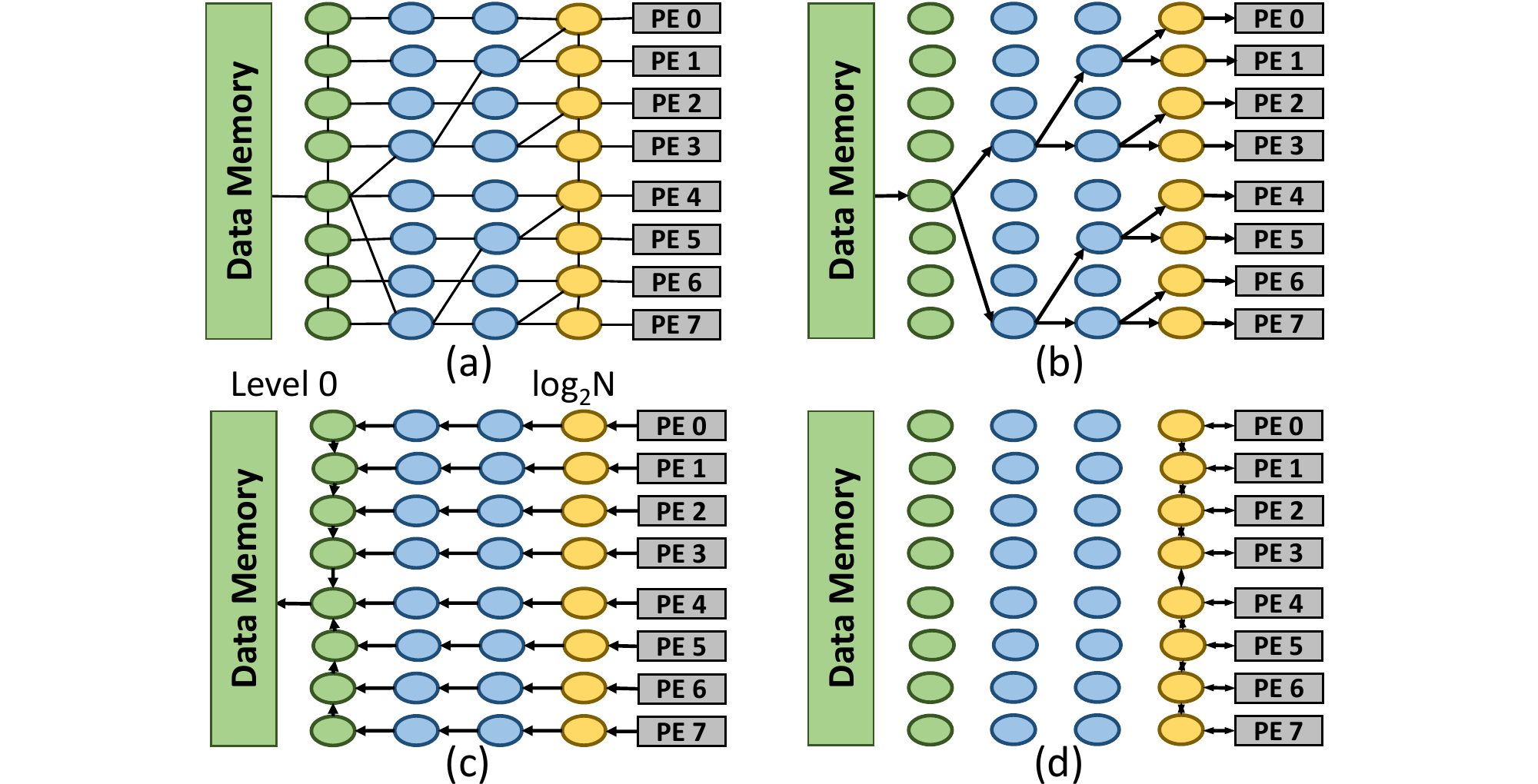}}
\caption{(a) Microswich network \cite{kwon2017rethinking}. NoC configurations: (b) multicast (c) gather (d) local communication. (Figure adopted from \cite{kwon2017rethinking}.)}
\label{fig::NoC-microswitches}
\end{figure}

For an N-PE accelerator, an array of microswitches (Fig. \ref{fig::NoC-microswitches}a) contains $\log_{2}N + 1$ levels with $N$ micro-switches at each level. Each microswitch contains a small combinational logic for configuration and up to two FIFOs for buffering the data during routing conflict. With small logic and storage, data traverses through several microswitches within each cycle \cite{kwon2017rethinking}. All microswitches contain \emph{gather} and \emph{scatter} units, and bottom microswitches (level $\log_{2}N$) also contain local units for inter-PE communication. In top microswitches (level 0), the scatter unit connects to memory banks, and the gather unit uses round-robin-based priority logic for arbitrating the incoming data in a pipelined manner. In middle microswitches, scatter units forward data to desired lower-level links, and gather units stream the data back. In bottom microswitches, scatter and gather units stream the data, and local units connect adjacent PEs. Fig. \ref{fig::NoC-microswitches}(b)--(d) shows how configurable microswitches can enable various communication patterns. 

\begin{figure}[!t]
\centering
\centerline{\includegraphics[width=0.95\linewidth]{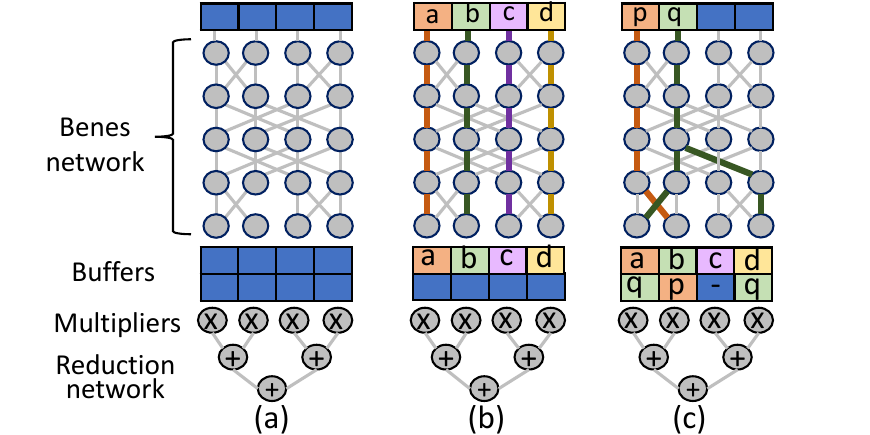}}
\caption{(a) Flexible dot product engine in SIGMA accelerator \cite{qin2020sigma} features a data distribution NoC with configurable switches interconnected via Benes topology. (b)--(c): Configuration of the interconnect facilitates different unicast and multicast communication patterns. (Figure adopted from \cite{qin2020sigma}.)}
\label{fig::NoC-benes-SIGMA}
\end{figure}

SIGMA \cite{qin2020sigma} used \emph{Benes topology} with \emph{configurable} switches (Fig. \ref{fig::NoC-benes-SIGMA}a). For N source and destinations, the interconnect contains $2\log_{2}N + 1$ levels, each with N number of 2$\times$2 switches. Each switch receives two control signals to determine whether to forward data vertically and/or diagonally. After combining communication requirements for distributing all elements to desired multipliers, switches can be configured to forward the data, as shown in Fig. \ref{fig::NoC-benes-SIGMA}(b)--(c).

\subsection{Mechanisms for Reduction of Partial Outputs}
\label{sec:reduction-partial-outputs}

Computation primitives of ML models require reduction (accumulation) of partial outputs from PEs or their functional units. It can be done temporally and/or spatially  (Table \ref{tab:reduction-mechanisms}).

\mysubsubsection{Temporal} All the reductions for computing an output scalar are performed on a \emph{single} PE (or a functional unit) during \emph{different} cycles. Accumulations are done temporally when different PEs compute distinct outputs, e.g., for output stationary dataflow. The temporal reduction makes the processing of sparse tensors \emph{simple} since PEs update partial outputs in their private memory or registers \emph{without communicating} to other PEs via an additional interconnect. Therefore, it is adopted by many accelerators, including EIE \cite{han2016eie}, ZENA \cite{kim2017zena}, and SparTen \cite{gondimalla2019sparten}. However, temporal accumulation requires \emph{indexing} the buffer for reading/writing partial outputs or accumulating computations in the output register (of MAC unit). So, it involves register/memory read and write operations, which consume higher energy than integer arithmetic \cite{sze2017efficient, yang2018dnn}. Besides, using local accumulator buffers for vector/SIMD functional units (e.g., in SCNN \cite{parashar2017scnn}) requires support for arbitration of partial outputs.

\mysubsubsection{Spatial} Partial outputs can be reduced spatially for an output scalar. It can be either done by functional units within a PE (e.g., adder-trees in Cambricon-X/S to sum up partial products) or inter-PE communication via a \emph{separate interconnect} (e.g., forwarding in systolic arrays). Inter-PE spatial reduction usually requires communication among neighboring PEs and is typically achieved through a mesh or similar topology \cite{kung2018adaptive, chen2016eyeriss}. Spatial reduction obviates buffer accesses and improves energy efficiency (e.g., by 2$\times$--3$\times$ \cite{lee2018stitch}, as compared to the temporal reduction on scalar PEs). These linear or tree-based reductions are typically symmetric. However, a major challenge is to enable \emph{asymmetric and asynchronous reductions} of a variable number of partial outputs, for adapting to high sparsity, tensor shape, or target functionality (e.g., DW-CONV). This is because, an efficient dataflow may require some of the interconnected functional units or PEs to process partial outputs for distinct output elements (e.g., different depth-wise groups); all partial outputs cannot be reduced altogether. Hence, configurable interconnects are needed. Otherwise, for high or hyper sparsity, functional units cannot be fed enough NZs and are poorly utilized. Note that structured sparsity can alleviate imbalance by inducing patterns such that all PEs process the same number of NZs. However, configurable mechanisms are still required to support different dataflows for the variations in functionalities or tensor shapes. 

\mysubsubsection{Spatio-temporal} Partial outputs can be reduced spatially and temporally and \emph{locally} (within PEs) and \emph{globally} (across PEs). Spatial and temporal reduction of outputs depends on the mapping of computation graph onto PEs \cite{dave2019dmazerunner}. In spatiotemporal reduction, different PEs or their functional units compute partial outputs at every cycle or a few, which are, at first, reduced spatially. The resultant partial output is then reduced temporally by updating the previous partial output in the memory. E.g., when data streams through PEs of a systolic array, there is an inter-PE spatial reduction of partial outputs (via PEs of each column). Then, the bottom PE-row provides the reduced partial outputs to accumulator buffers (CompAct \cite{zhang2019compact}, TPU \cite{jouppi2017datacenter}). PEs of SNAP \cite{zhang2019snap} perform spatiotemporal accumulation locally, where partial products are first spatially accumulated through a configurable adder-tree and then accumulated in PE's memory over time.    

\mysubsubsection{Temporo-spatial} In temporospatial reduction, PEs compute partial outputs and reduce them locally over time. Then, they are collected later and accumulated spatially via interconnect before further processing (e.g., write-back, encoding). For example, PEs of a cluster in EyerissV2 \cite{chen2019eyeriss} first locally accumulate partial summations. Then, partial outputs can be accumulated across vertically connected clusters. SCNN \cite{parashar2017scnn} PEs compute output tiles corresponding to distinct input feature maps stored in their buffers. Outputs are temporally reduced by indexing the accumulator buffers. Then, overlapping fractions of incomplete outputs are exchanged among neighboring PEs for reduction. SNAP \cite{zhang2019snap} also performs temporospatial reduction at PE-array (core) level. Its PEs accumulate outputs locally over time, which are reduced spatially across horizontal/diagonal PEs by a core-level reducer.

\begin{figure}[!t]
\centering
\centerline{\includegraphics[width=\linewidth]{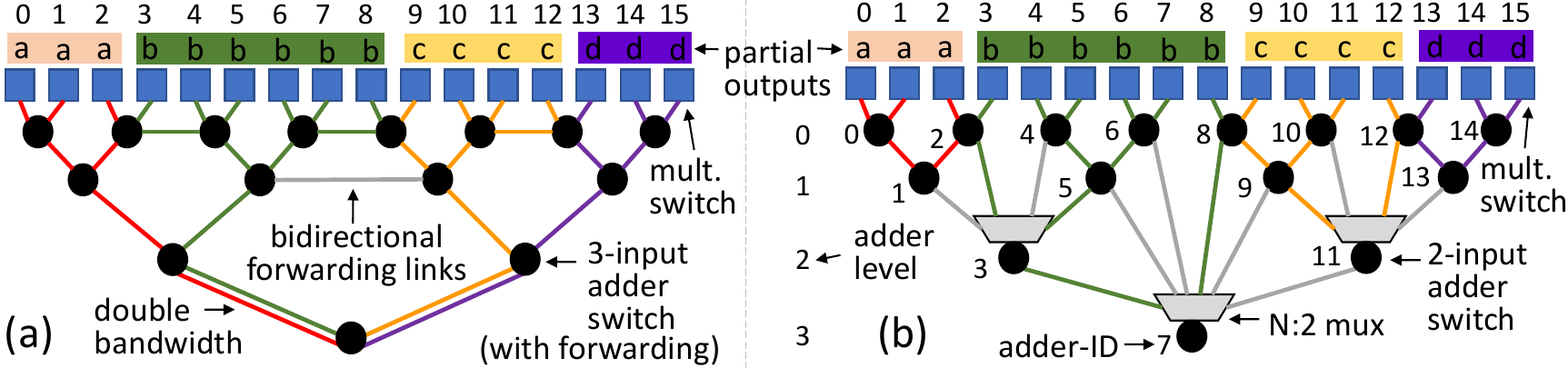}}
\caption{Configurable spatial reduction trees: (a) Augmented reduction tree in MAERI (Figure adopted from \cite{kwon2018maeri}.) (b) Forwarding adder network in SIGMA (Figure adopted from \cite{qin2020sigma}.)}
\label{fig::NoC-configurable-reduction}
\end{figure}

\mysubsubsection{Configurable} 
MAERI \cite{kwon2018maeri} and SIGMA \cite{qin2020sigma} employ configurable reduction trees for efficient and asymmetric spatial reduction of partial outputs. So, it can be useful for spatial processing of unstructured sparsity and variable-sized vectors for dot products. The augmented reduction tree in MAERI (Fig. \ref{fig::NoC-configurable-reduction}a) allows an asymmetric reduction of partial outputs with configurable adder switches and bidirectional forwarding links. Each 3-input adder switch can receive two partial outputs from the previous level and one via a forwarding link, and it can add and forward them. Plus, upper-levels of the tree (near root) have double bandwidth than lower-levels, allowing simultaneous collection of multiple reduced outputs. The forwarding adder network in SIGMA (Fig. \ref{fig::NoC-configurable-reduction}b) enables similar configurable reduction but at reduced area and power. Instead of 3-input adders, it uses 2-input adders and N:2 mux for selecting the inputs. Also, adders at the $0^{th}$ level allow bypassing of partial products to the next level.

\begin{table}[!t]
\centering

\caption{Mechanisms for Accumulations of Partial Outputs}
\label{tab:reduction-mechanisms}
\begin{tabular}{|c|m{5.1cm}|}
\hline
Temporal            & \cite{han2016eie, zheng2018kernelxform, struharik2018conna, gondimalla2019sparten, hegde2019extensor, choles2018parsecore, aimar2018nullhop, han2017ese, page2017sparcnet, whatmough2018dnn, lee20197, kim2017zena, lu2018spwa, venkatesh2017accelerating} \\ \hline
Spatial (intra-PE)  & \cite{zhang2016cambricon, zhou2018cambricon, qin2020sigma} \\ \hline 
Spatial (inter-PE)  & \cite{li2019squeezeflow, lee20197, kwon2018maeri, qin2020sigma} \\ \hline 
Spatio-temporal     & \cite{zhang2019compact, zhang2019snap, lee2018stitch} \\ \hline
Temporo-spatial     & \cite{chen2016eyeriss, chen2019eyeriss, parashar2017scnn, zhang2019snap} \\ \hline
Configurable        & \cite{kwon2018maeri, qin2020sigma, zhang2019snap} \\ \hline
\end{tabular}
\end{table} 

\subsection{Optimization Opportunities}

\textit{i) Low-cost flexible interconnects for accommodating spatial reuse opportunities, dynamic communication requirements, various sparsity, and different precision:} Variations in data reuse (Fig. \ref{fig::analysis-data-reuse-sparse}) are caused by the tensor size, functionality (e.g., stride, separable convolution), batch size, and sparsity of tensors. The communication  mechanism needs to leverage available reuse by supporting various multicast and unicast patterns \cite{kwon2017rethinking, chen2019eyeriss}. Moreover, the distribution, inter-PE communication, and collection of the outputs can be done \emph{asynchronously} and \emph{concurrently}. These require the interconnect switches to support \emph{dynamic} management (priority arbitration and congestion) at low cost. Furthermore, communication among distant PEs may be required (e.g., for store-and-forward or exchanging outputs during sparse computations). Finally, depending on sparsity and precision, the bit-width of the metadata and NZ value can differ significantly. Communicating different sizes of data and metadata can be facilitated by configurable interconnect buses and their interfacing with PEs and memory. For instance, in EyerissV2 \cite{chen2019eyeriss}, a 24-bit bus can supply PEs either three 8b uncompressed values or two pairs of 8b NZ and 4b metadata. Thus, configurable interconnect topologies should be explored for effectively serving various communication requirements. FPGAs can also be leveraged for designing accelerators with tailored interconnects.

\textit{ii) Programming of configurable interconnects and design exploration:} Configurable interconnections can support various communication patterns and dynamic data movement for sparse computations. But, compilation support is needed to program them as they often contain parameterized multi-level switches and switches with many-to-many links between source and destination (e.g., \cite{kwon2017rethinking, qin2020sigma}). Depending on the interconnect topology and optimized dataflow, the compiler may need to select efficient paths for distributing data from source to destination switches. Additionally, the underlying topology (e.g., lack of multi-hop connectivity) \emph{may not} support some dataflows (e.g., spatiotemporal accumulation of partial outputs from distant PEs in the absence of multi-hop connectivity). Further, a systematic methodology for mapping communication onto interconnect topology can enable design space exploration of interconnects needed for accelerating target ML models, allowing minimum overhead of \emph{run-time reconfiguration} of the interconnect to support various dataflows.

\section{PE Architecture Design}
\label{sec::PEArch}

\begin{figure}[!t]
\centering
\centerline{\includegraphics[width=0.9\linewidth]{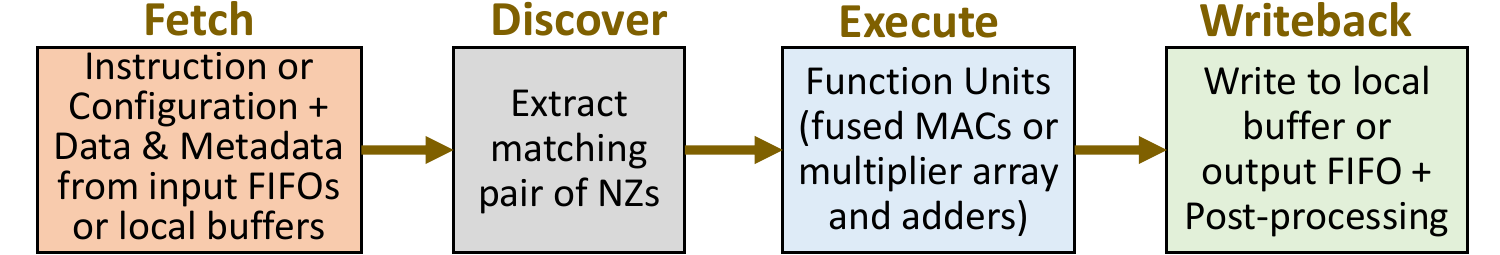}}
\caption{Overview of the PE pipeline for processing sparse and value-approximated tensors (Figure adopted from \cite{zhang2019snap}).} 
\label{fig::PE-pipeline}
\end{figure}

PE architecture consists of functional units, local memory (RFs or SRAMs), and local control (instruction buffer or finite state machine). Fig. \ref{fig::PE-pipeline} shows pipeline stages for processing sparse and value-approximated tensors. Depending upon PE's interface, it either gets data from the interconnect (typical) or directly accesses off-chip memory via DMA transfer. At every cycle or few, a PE (a) processes an instruction or events based on its state \cite{lu2017flexflow, yin2017thinker, yuan2019sticker}, (b) fetches data from local memory or interconnect, (c) computes tensor elements via functional unit, and (d) writes the intermediate result to local memory or interconnect. PEs may contain \emph{special-function} modules (e.g., for ReLU or sigmoid computations \cite{parashar2017scnn, han2017ese}). 

Processing compressed tensors can impose significant maneuvering efforts for PE design. For example, reusing tensors temporally through local memory (e.g., in EyerissV2 \cite{chen2019eyeriss}, SNAP \cite{zhang2019snap}) alleviates overheads of repeatedly accessing compressed tensors via memory hierarchy and decoding them. However, it requires communicating data to PEs before extracting NZs. Thus, the PE may require additional hardware for \emph{extracting} or \emph{correctly indexing} NZs (section \ref{sec::NZ-data-extraction}). Additionally, \insight{the selection of functional units is affected by the number of NZs that can be fed for various sparsity of tensors, support for mixed-precision, and functionality of the layers.} In such various scenarios, a single dataflow may not always be effective \cite{chen2019eyeriss, dave2020dmazerunner} and can lead to significant acceleration loss. So, PE datapath needs to be \emph{adaptive} for supporting multiple dataflows optimized for different layers and sparsity. Further, techniques for leveraging computation reuse due to \emph{value similarity} often require enhancements in the design. PEs may also post-process outputs or generate additional metadata for communicating outputs. So, an efficient pipeline needs to hide pre-processing and post-processing latency.       

\subsection{Functional Units}

\mysubsubsection{Scalar PEs} Table \ref{tab:PE-vectorization} lists accelerators based on their functional units for scalar, SIMD, or vector processing. Many architectures contain an array of scalar PEs; PE datapath contains a pipelined MAC unit (e.g., EIE \cite{han2016eie}, SparTen \cite{gondimalla2019sparten}). 

\mysubsubsection{SIMD/Vector PEs} PEs of Cnvlutin \cite{albericio2016cnvlutin} and Cambricon-S \cite{zhou2018cambricon} contain multiplier arrays and adder trees. By performing dot products at every cycle, they can deliver high throughput. Moreover, accumulation through adder-trees reuses data spatially, which lowers energy consumption (by 2$\times$--3$\times$ \cite{lee2018stitch}) as compared to temporal accumulation on scalar PEs by reading and writing partial summations via local memory. However, a major challenge is the inefficient utilization of multipliers and adders, which often leads to ineffectual computation cycles and acceleration loss. This is because, for high sparsity, enough NZs may not be extracted to feed all multipliers at every cycle. For example, a sensitivity analysis for Cambricon-X \cite{zhang2016cambricon} determined that, for hyper $W$-sparsity, it accelerated CONVs by about 8$\times$ (out of 16$\times$ peak speedup). The utilization may be improved by employing larger indexing or extraction modules (increased on-chip area and power). Alternatively, PEs can be designed with fewer multipliers to sustain the scalability and efficiency over a wide sparsity range.

\begin{table}[!t]
\centering

\caption{PE Architectures for Sparse Tensor Computations}
\label{tab:PE-vectorization}
\begin{tabular}{|c|m{6.6cm}|}
\hline 
Scalar & \cite{chen2016eyeriss, han2016eie, zheng2018kernelxform, struharik2018conna, kim2017zena, yin2017thinker, mishra2017fine, lee20197, moons201714, yuan2018sticker, li2019squeezeflow, gondimalla2019sparten, hegde2019extensor, yavits2017accelerator, aimar2018nullhop, han2017ese, page2017sparcnet, whatmough2018dnn} \\ \hline
\makecell{SIMD /\\Vector} & \cite{albericio2016cnvlutin, zhang2016cambricon, zhou2018cambricon, parashar2017scnn, zhang2019snap, kang2019accelerator, lee2018stitch, chen2019eyeriss, judd2017cnvlutin2, qin2020sigma, choles2018parsecore, lu2018spwa, gao2018deltarnn} \\ \hline 
\end{tabular}
\end{table}

While SIMD or vector PEs achieve spatial reuse, due to fixed designs, they are utilized poorly when executing some functionalities like DW-CONV. The efficiency of SIMD PEs is further affected by high sparsity, as functional units of the PE require synchronization, and there may not be enough effectual NZs to feed all of them. \emph{Configurable functional units} can overcome such limitations. For example, PEs of SNAP architecture \cite{zhang2019snap} use a configurable adder-tree. It processes inputs from three multipliers and computes different combinations of partial summations. With multiple adders and multiplexers, the PE can concurrently process different partial summations (vs. gather in adder-tree) without high-bandwidth crossbars. Such configurable designs can support different DNN operators (e.g., DW-CONVs).

\mysubsubsection{Multiplier-free PEs} Accelerators, such as ZENA \cite{kim2017zena} and \cite{zheng2018kernelxform}, use multiplier-free PEs for high energy efficiency. These PEs process tensors of very low-precision (binary or ternary values) or logarithmic quantization. So, they replace multipliers with \emph{simpler arithmetic} like 2's complement (inverters and adders or subtractors) \cite{zheng2018kernelxform, andri2017yodann} or bit-wise shift and additions \cite{tann2017hardware, lee2017lognet}. However, one challenge is to \emph{maintain} the accuracy of DNNs, as aggressive quantization often drops top-1 and top-5 accuracy, e.g., by 0.1\% \cite{tann2017hardware} -- 5\% \cite{lee2017lognet}. By trading off the flexibility with simple hardware, supporting various models can be challenging.

\begin{figure}[!b]
\centering
\centerline{\includegraphics[width=\linewidth]{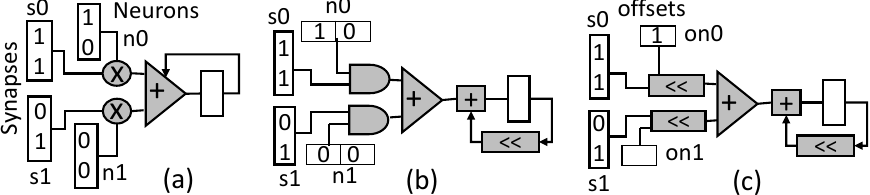}}
\caption{Bit-serial processing of sparse activations in Pragmatic \cite{albericio2017bit}. (a) Bit-parallel unit. (b) Bit-serial unit. (c) Bit-serial unit in Pragmatic for processing only essential bits. (Figure adopted from \cite{albericio2017bit}).}
\label{fig::bit-serial-pragmatic}
\end{figure}

\mysubsubsection{Bit-adaptive computing} Precision requirements for targeted accuracy can vary for different models \cite{sharma2018bit, albericio2017bit}, which can be supported by PEs with bit-adaptive computing. 

\textbf{Bit-serial computing:} Albericio et al. \cite{albericio2017bit} showed that \emph{zero bits in NZ activations} (8b or 16b precision) can be more than 50\% and proposed the Pragmatic accelerator to leverage sparsity of activation bits. Fig. \ref{fig::bit-serial-pragmatic}(b) shows the \emph{bit-serial} computation of an inner product with AND gates, adder tree, and bit-wise shift of partial output. AND gates are serially fed 1b activations (variable precision) and bit-parallel 16b weights (fixed precision). Fig. \ref{fig::bit-serial-pragmatic}(c) shows the processing of only NZ activations in Pragmatic (essential bits indicated by their positions). Laconic \cite{sharify2019laconic} achieved further accelerations by processing only NZ bits of both activations and weights. 

\textbf{Bit-composable computing:} Bit-fusion \cite{sharma2016high} employed fusion units consisting of an array of BitBricks. The fusion units can be configured for processing multiplications of 2b, 4b, 8b, or 16b operands. For processing NZs, PEs of CNN accelerator Envision \cite{moons201714} used a single-cycle N-subword-parallel multiplier, followed by an N$\times$48b/N reconfigurable adder. The subword-parallel design allowed the configuration of MAC units for processing the data of 4b, 8b, or 16b. SPU architecture \cite{dadu2019towards} employed DGRA, a decomposable CGRA, for efficiently processing stream-join accesses. The DGRA PE and interconnect switches enabled decomposing up to four 16b sub-word operands. DGRA also supported accessing sub-word data from the scratchpad. For DNN training with mixed-precision and sparse tensors, PEs of LNPU contained configurable MAC units that can process FP8 or FP16 tensors. Table \ref{tab:precision} lists precisions of sparse tensors that are supported by different accelerators. The precision indicates the bit-width of input operands (activations and weights). For MAC operations, accumulators usually produce high-precision output, which can be down-scaled or truncated afterward.

\begin{table}[!t]
\centering

\caption{Precision of Sparse Tensors Supported by Accelerators}
\label{tab:precision}
\begin{tabular}{|c|m{5.8cm}|}
\hline 
binary/ternary & \cite{zheng2018kernelxform, venkatesh2017accelerating} \\ \hline
int8        & \cite{zhang2019compact, gondimalla2019sparten} \\ \hline
int16       & \cite{struharik2018conna, chen2016eyeriss, han2016eie, zhang2016cambricon, kung2018adaptive, albericio2016cnvlutin, judd2017cnvlutin2, parashar2017scnn, zhang2019snap, zhou2018cambricon, li2019squeezeflow, aimar2018nullhop, han2017ese, whatmough2018dnn, lu2018spwa, gao2018deltarnn} \\ \hline
logarithmic & \cite{kim2017zena, mcdanel2019full} \\ \hline
bit-adaptive   & \cite{albericio2017bit, delmas2019bit, sharify2019laconic, moons201714, dadu2019towards} \\ \hline
FP8         & \cite{lee20197} \\ \hline
FP16        & \cite{lee20197, page2017sparcnet, venkatesh2017accelerating} \\ \hline
FP32        & \cite{mishra2017fine, qin2020sigma, venkatesh2017accelerating} \\ \hline
FP64        & \cite{yavits2017accelerator, hegde2019extensor} \\ \hline 
\end{tabular}
\end{table}

\mysubsubsection{Clock-gated PEs} PEs can be clock-gated when \emph{not used} for executing a layer and for ineffectual computations. For example, Eyeriss \cite{chen2016eyeriss}, Thinker \cite{yin2017thinker}, and Minerva \cite{reagen2016minerva} use zero detection for clock gating the datapath in the PE pipeline. PEs check whether the value being read is zero (or compare with a threshold, e.g., in MNNFast \cite{jang2019mnnfast}). Based on the comparator output, their clock gating logic prevent the MAC datapath from switching in the consecutive cycle, which reduces energy (e.g., it saved power consumption of Eyeriss PE by 45\%). Zero-skipping through flags in Envision \cite{moons201714} and Sticker \cite{yuan2019sticker} achieved similar savings.

\mysubsubsection{Optimization opportunities}

\textit{(i) Exploring efficient designs of functional units for various sparsity ranges/patterns and functionality:} Utilization of vector/SIMD units can drop significantly due to unstructured sparsity \cite{zhang2019snap} and functionality beyond structured dot products (e.g., DW-CONV). So, for exploring design hyperparameters such as the number of functional units, designers need to consider the impacts of sparsity, data extraction mechanism, required synchronization among computational units, and configurations required to support various functionalities. Moreover, for low sparsity, designs should deliver performance \emph{at par} with a sparsity-oblivious design. For example, for processing dense tensors, SCNN \cite{parashar2017scnn} achieved 79\% of the performance and consumed 33\% higher energy as compared to the baseline accelerator for processing dense tensors. So, designers may ensure that additional features for exploiting sparsity and configurable components do not increase the critical path latency and are power-gated if not used.

\subsection{Dataflow Mechanisms}
\label{sec::dataflows}

\mysubsubsection{Background} The efficiency of executing a layer onto hardware accelerator depends on the computational, communication, and memory access patterns, which are commonly referred to as dataflow mechanisms \cite{sze2017efficient, dave2019dmazerunner}. A dataflow refers to the spatiotemporal execution of a model layer (nested loop) on architectural resources \cite{yang2018dnn, dave2019dmazerunner}. Here, spatial execution corresponds to how PEs exploits parallelism in the computation graph and processes different subsets of tensors. Temporal execution drives the data accessed throughout memory hierarchy and data communication via interconnects. Thus, depending on the functionality and tensor dimensions, dataflow can significantly impact the utilization of resources, data reuse, and latency hiding for memory accesses and data communication, and consequently, the execution time and energy consumption \cite{sze2017efficient, chen2019eyeriss, parashar2019timeloop, dave2019dmazerunner, kwon2019understanding}. 

\begin{figure}[!t]
\centering
\centerline{\includegraphics[width=\linewidth]{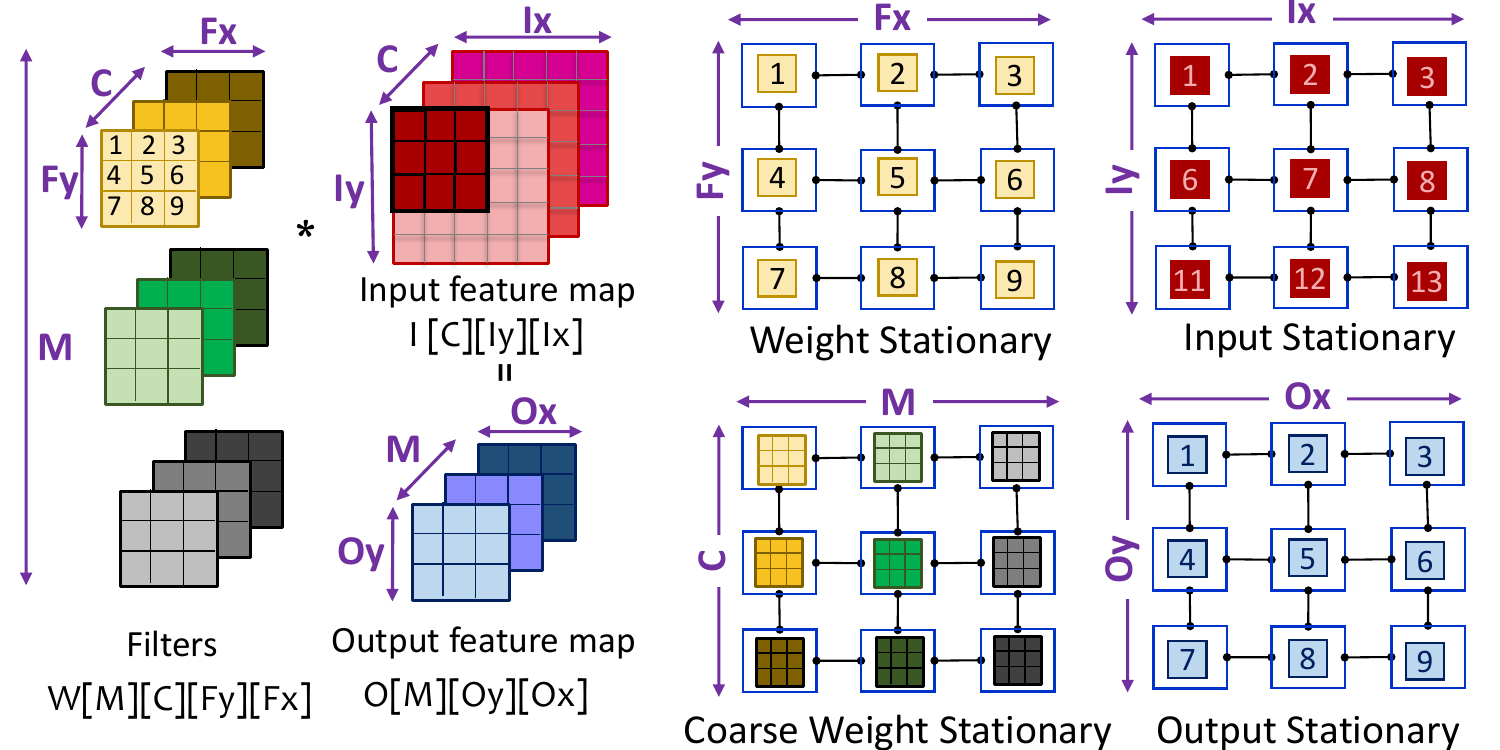}}
\caption{Commonly used dataflow mechanisms for executing convolution layers on hardware accelerators.}
\label{fig::dataflows}
\end{figure}

\begin{figure}[!t]
\centering
\centerline{\includegraphics[width=\linewidth]{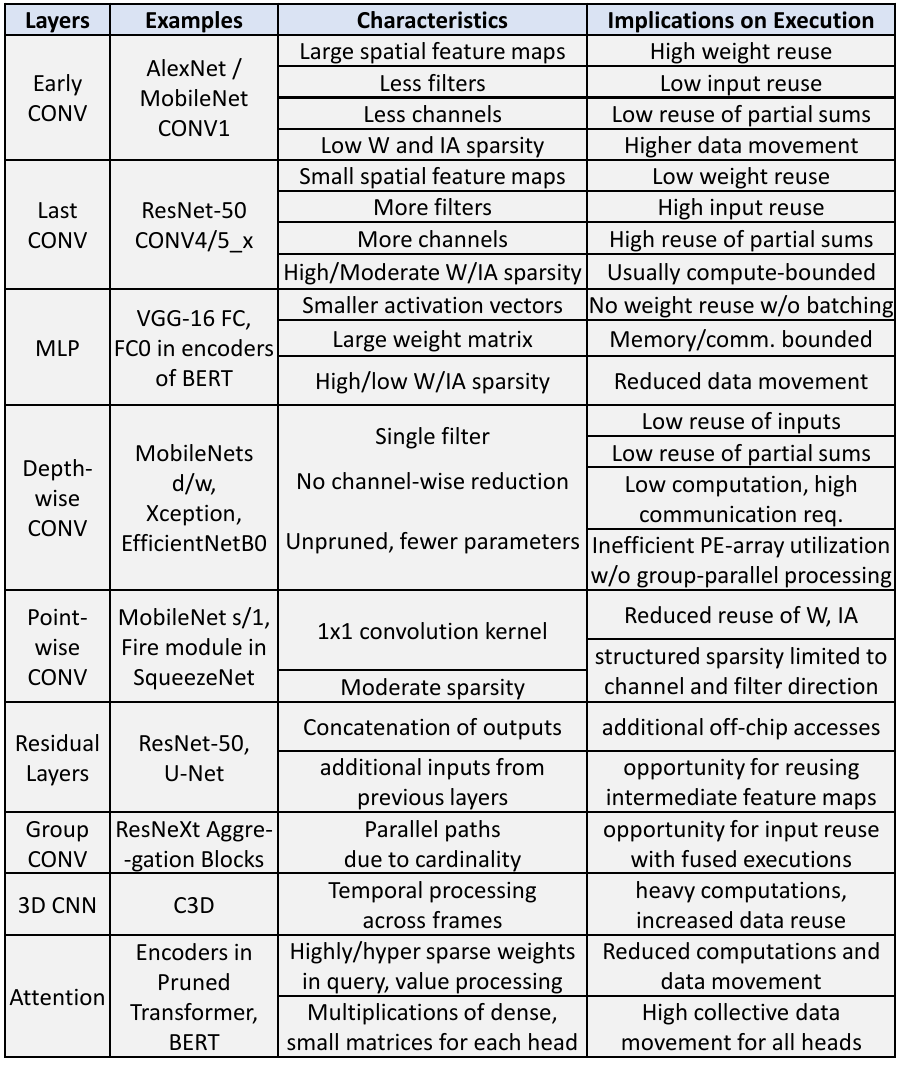}}
\caption{Characteristics of different DNN layers pertaining to hardware execution (Figure inspired by \cite{kwon2019understanding, chen2019eyeriss}).}
\label{fig::layer-characteristics}
\end{figure}

One way to classify dataflows is by what data is kept ``stationary'' in registers or local memory of PEs (and reused fully before eviction), while other data is being iterated over.
Some commonly used dataflow mechanisms are output stationary, weight stationary, input stationary, row stationary, and no local reuse. Fig. \ref{fig::dataflows} shows an example of convolution and the layout of the stationary data for mapping the convolution with these dataflows. In weight stationary dataflow, each weight (of a 2D filter) remains stationary on a unique PE, and reused many times, during processing activations (corresponding to the same input channel $C$). By processing a unique weight, each PE produces partial summations for output activations, which are communicated by PEs and accumulated before outputs are written back to the memory hierarchy. Thus, input and output activations are accessed from off-chip memory (via shared scratchpad and PE's local memory) several times, while weights are continuously reused. After reuse, a new set of weights is loaded from memory, and the execution repeats. Weight reuse is higher in processing larger feature maps (CNNs) and multi-folded for processing data in a batch (e.g., images for CNNs, tokens of sentences for NLP models). Fig. \ref{fig::layer-characteristics} lists such characteristics of different layers. 

Dataflows can be applied at a coarser level, where PEs process a data block or plane (1D/2D/3D). In a coarse weight stationary approach \cite{yang2018dnn}, each PE processes weights of an entire 2D filter (dimensions $C$ and/or $M$ are laid out spatially on PEs). Rows and columns of PEs process the data corresponding to unique input and output channels, respectively. So, activations need to be multicast to the PEs of a row, different weights need to be provided to each PE, and partial summations for output channels can be accumulated vertically \cite{yang2018dnn}. Similarly, in an input stationary dataflow, unique activations (or blocks of input feature maps) remain stationary and are reused. In an output stationary dataflow, each PE produces a unique output activation (corresponding to the same or different output channel) \cite{sze2017efficient}. By processing spatial data and input channels first, partial summations are accumulated and reused in the memory of each PE. With the temporal accumulation of outputs on PEs, the output stationary dataflow does not need to reduce partial outputs spatially by collecting them from appropriate PEs, which is otherwise challenging for unstructured sparse data (section \ref{sec:reduction-partial-outputs}). Therefore, many accelerators opt for such dataflow. In no local reuse dataflow, input operands are streamed to PEs, but they are not stored in PE's memory \cite{sze2017efficient, jouppi2017datacenter}. In row stationary dataflow, PEs of the same row process the same weights (a row of a filter), diagonal PEs process the same row of input activations, and partial summations for rows of the output feature map are accumulated through vertically connected PEs \cite{chen2016eyeriss}. 
Thus, different dataflows uniquely exploit the spatial parallelism and reuse of different tensors. 

\begin{figure}[!t]
\centering
\centerline{\includegraphics[width=\linewidth]{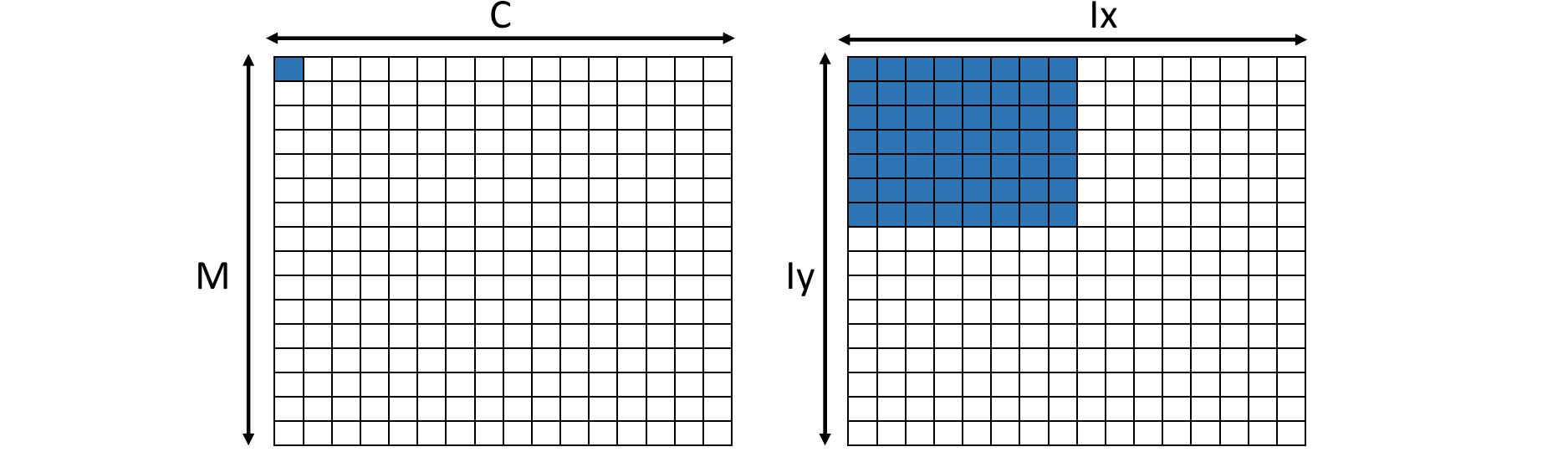}}
\caption{Low utilization of a 16$\times$16 PE-array in (a) coarse weight stationary dataflow when executing depth-wise layers and (b) input stationary dataflow for executing later layers of deep CNN models (Figure inspired by \cite{chen2019eyeriss}).}
\label{fig::dataflow-low-PE-utilization}
\end{figure}

\textit{Dataflow optimization:} As dimensions of tensors are often large, many ways exist for spatiotemporally executing a layer onto the computational and memory resources of an accelerator. Optimization of the dataflow is important as it can significantly impact the performance and energy consumption \cite{yang2018dnn, dave2019dmazerunner, kwon2019understanding}. For instance, mappings with similar performance can consume an order of magnitude higher energy \cite{parashar2019timeloop} or vice versa. Further, as Fig. \ref{fig::layer-characteristics} shows, reuse characteristics, tensor dimensions, functionality, and sparsity can vary significantly for different DNN layers. Hence, a single dataflow may not always be effective for acceleration. Fig. \ref{fig::dataflow-low-PE-utilization} provides two such examples that lead to low PE-array utilization. The coarse weight stationary dataflow processes different 2D filters on different PEs. So, it is inefficient for DW-CONV. Similarly, output-stationary or input-stationary dataflows can result in low utilization of PEs for processing later layers of deep CNNs. With the vast space of execution methods and the growing development of new models (with variations in tensor dimensions), it becomes hard for non-experts to figure out optimized execution methods and designs. Therefore, many optimization tools have been proposed recently including Timeloop \cite{parashar2019timeloop}, dMazeRunner \cite{dave2019dmazerunner}, MAESTRO \cite{kwon2019understanding}, and Interstellar \cite{yang2018dnn}. They analytically model the execution of accelerators to estimate execution metrics and evaluate a set of mappings from the pruned space of various dataflows. 

\mysubsubsection{Sparsity-aware dataflows} Dataflows for processing sparse tensors are typically similar to those for dense tensors while processing the data in compressed format. For correct functionality, dataflow executions are facilitated by extraction/orchestration of NZs, which is done either in PEs \cite{parashar2017scnn, chen2019eyeriss}, on a different module \cite{zhou2018cambricon}, or by a separate controller. For example, SCNN \cite{parashar2017scnn} used PT-IS-CP dataflow. It processed planar tiles of feature maps with input stationary dataflow. SCNN's PT-IS-CP-sparse dataflow extended the PT-IS-CP. It processed only NZ activations and weights in compressed format while accessing them from memory and performing computations. The coordinate computation module in each PE ensured that partial products generated by all-to-all multiplications of NZ inputs and weights were accumulated correctly and stored in appropriate buffers. Table \ref{tab:dataflows} lists sparsity-aware dataflow mechanisms used by accelerators.

\begin{table}[!t]
\centering

\caption{Dataflow Mechanisms of Accelerators}
\label{tab:dataflows}
\begin{tabular}{|c|m{4.6cm}|}
\hline 
Input Stationary            & \cite{parashar2017scnn, zheng2018kernelxform, qin2020sigma} \\ \hline 
Output Stationary           & \cite{han2016eie, zhang2016cambricon, albericio2016cnvlutin, judd2017cnvlutin2, kang2019accelerator, mishra2017fine, lee20197, moons201714, li2019squeezeflow, yuan2018sticker, gondimalla2019sparten, zhou2018cambricon, hegde2019extensor, choles2018parsecore,  han2017ese, page2017sparcnet, whatmough2018dnn, lu2018spwa, zhang2019snap, venkatesh2017accelerating} \\ \hline
Weight Stationary           & \cite{kung2018adaptive, qin2020sigma, asgari2019eridanus} \\ \hline
Coarse Weight Stationary    & \cite{albericio2016cnvlutin}, \cite{struharik2018conna} \\ \hline
Row Stationary              & \cite{chen2016eyeriss, chen2019eyeriss} \\ \hline
\end{tabular}
\end{table}

EyerissV2 \cite{chen2019eyeriss} used an enhanced row-stationary dataflow. By using statically known sparsity of weights, more NZ weights were allocated in local memories and global scratchpads. For example, each PE can store up to 192 NZ weights. Mappings of CONV and FC layers of AlexNet with row-stationary dataflow allocated 64--174 NZ weights, which corresponded to a total of 132--480 weights in the dense format. With in-PE data extraction logic, each PE only processed NZ values from CSC-encoded data. Thus, sparsity-aware dataflow can be optimized with the pre-known (or expected bounds of) sparsity and value similarity.            

\mysubsubsection{Optimization opportunities}

\textit{(i) Dataflow optimizations accounting for storage and computational overheads for metadata and codebooks:} Sparse and value-shared tensors are processed along with metadata (indicates positions of NZs) and codebook (common values shared among tensor elements), respectively. It requires additional processing, e.g., buffer management, communication via interconnects, and indexing the appropriate values. Depending on the dataflow, such processing can amplify the execution costs, which needs to be optimized. Existing tools for optimizing dataflows target dense tensor computations. Accelerators EIE, SCNN, and Cambricon-S process sparse tensor computations but with customized dataflows. Hence, frameworks for mapping and design explorations need to consider the sparsity and value similarity of tensors and their variations across layers/models. Such tools can include additional costs corresponding to storage, communication, and extraction in their analytical models. Explorations supporting multiple dataflows can help to achieve efficient designs for handling different functionality and variations in sparsity, shapes, and quantizations of tensors.

\textit{(ii) Sparsity-aware resource partitioning:} Acceleration of deep learning models is scaled by simultaneously processing multiple layers. It is done either by partitioning resources \cite{shen2017maximizing} of a scaled-up accelerator or on multiple accelerators (scale-out) by leveraging model- or data-parallelism \cite{song2019hypar}. Techniques for resource partitioning aim to highly reuse data from the on-chip memory of accelerators. It involves evaluating many-to-many mappings between layers and accelerators. Such optimizations can be crucial for several applications that require low latency, real-time processing, or high frame rates (e.g., processing the frames for multiple object detection models of an autonomous vehicle's perception system). Exploiting sparsity can provide further opportunities due to fewer computation, communication, and storage.

\subsection{Leveraging Value Similarity}

Several techniques have leveraged value similarity for accelerating DNNs by value sharing and computation reuse (Table \ref{tab:value-similarity}). Video frames exhibit high similarity spatially (among neighboring pixels) and temporally (over consecutive frames) \cite{riera2018computation, gonccalves2019aggressive}. After precision lowering, values of limited range repeat frequently \cite{han2015deep, hegde2018ucnn}, which are further compressed by maintaining a codebook of unique values \cite{han2016eie}. With repetition of values, computation (outputs) can be reused, either partially during processing a layer \cite{riera2018computation, hegde2018ucnn} or by skipping processing of a whole layer \cite{buckler2018eva2}. This subsection describes such techniques and corresponding hardware enhancements.

\begin{table}[!t]
\centering

\caption{Techniques for leveraging value similarity.}
\label{tab:value-similarity}
\begin{tabular}{|c|c|m{3.5cm}|}
\hline
\multirow{2}{*}{\makecell{Value sharing}}
    & Weights     & \cite{han2016eie, zhou2018cambricon, hegde2018ucnn, mahdiani2019deltann, yuan2018sticker}   \\ \cline{2-3} 
    & Activations & \cite{riera2018computation, mahmoud2018diffy}     \\ \hline
\multirow{2}{*}{\makecell{Computation reuse \\ and memoization}} 
    & Partial     & \cite{riera2018computation, hegde2018ucnn, mahdiani2019deltann, mahmoud2018diffy, silfa2019neuron, wang2019none, gao2018deltarnn}     \\ \cline{2-3} 
    & Full        & \cite{gonccalves2019aggressive, buckler2018eva2, zhu2018euphrates} \\ \hline
\multicolumn{2}{|c|}{Early termination of computations}  & \cite{akhlaghi2018snapea, zhu2018sparsenn, lee2018compend}   \\ \hline 
\end{tabular}
\end{table}

\mysubsubsection{Weight similarity} Prior studies have shown that weights can be approximated with a small set of values. 
Hegde et al. \cite{hegde2018ucnn} showed that for 8b weights of DNNs, each NZ value mostly repeated more than 10 times and even more than 100 times in later layers of AlexNet and ResNet-50 models. Han et al. \cite{han2015deep} pruned weights of DNNs with k-means clustering for value sharing. Shared unique values were represented with 4 or 5 bits without dropping classification accuracy. Local quantization (applying clustering separately over different sub-tensors) can achieve even smaller codebooks \cite{zhou2018cambricon}. Leveraging the weight similarity can compress pruned models further by up to an order of magnitude \cite{han2015deep, zhou2018cambricon}.

Value-shared weights are processed by augmenting the PE datapath with a weight decoder (e.g., in EIE \cite{han2016eie}). For processing NZ weights, the PE provides the encoded index of the weight to the decoder and obtains shared value.  
Depending on the lookup mechanism and total bits to be extracted at every cycle, the decoder can incur considerable area and power costs (e.g., for Cambricon-S \cite{zhou2018cambricon}, 32.56\% and 3.98\% of the total on-chip area and power, respectively). 

\mysubsubsection{Input similarity} Audio or video frames can contain high similarity spatially or temporally. This is because a speech signal can be quasi-stationary for a short interval. Also, successive executions of DNNs process overlapping windows of audio frames for context extraction \cite{riera2018computation}. Feature maps for CNNs exhibit high spatial correlation \cite{mahmoud2018diffy}. High input similarity enables only storing unique values and reusing computations by \emph{differential computing} over non-similar data. 

Riera et al. \cite{riera2018computation} showed that after uniform linear quantization of inputs of DNNs (e.g., C3D \cite{tran2015c3d}, EESEN \cite{miao2015eesen}, CNN for self-driving cars \cite{bojarski2016end}), about 61\% of input activations are the same as previous execution, and 66\% computations can be avoided. Their accelerator maintains centroids of quantized inputs and the index corresponding to each input element. Then, consecutive frames are processed layer-wise with differential computing. For example, for each activation of an FC layer (of a new frame), the accelerator calculates centroid and index, and then it compares calculated centroid to memoized centroid. If the difference is zero, then output from the previous execution is reused, and the next activation is processed. Otherwise, a new value of the index is updated in the buffer, and new values for output activations are computed by accumulating multiplications of weights with the difference.

\begin{figure}[!t]
\centering
\centerline{\includegraphics[width=\linewidth]{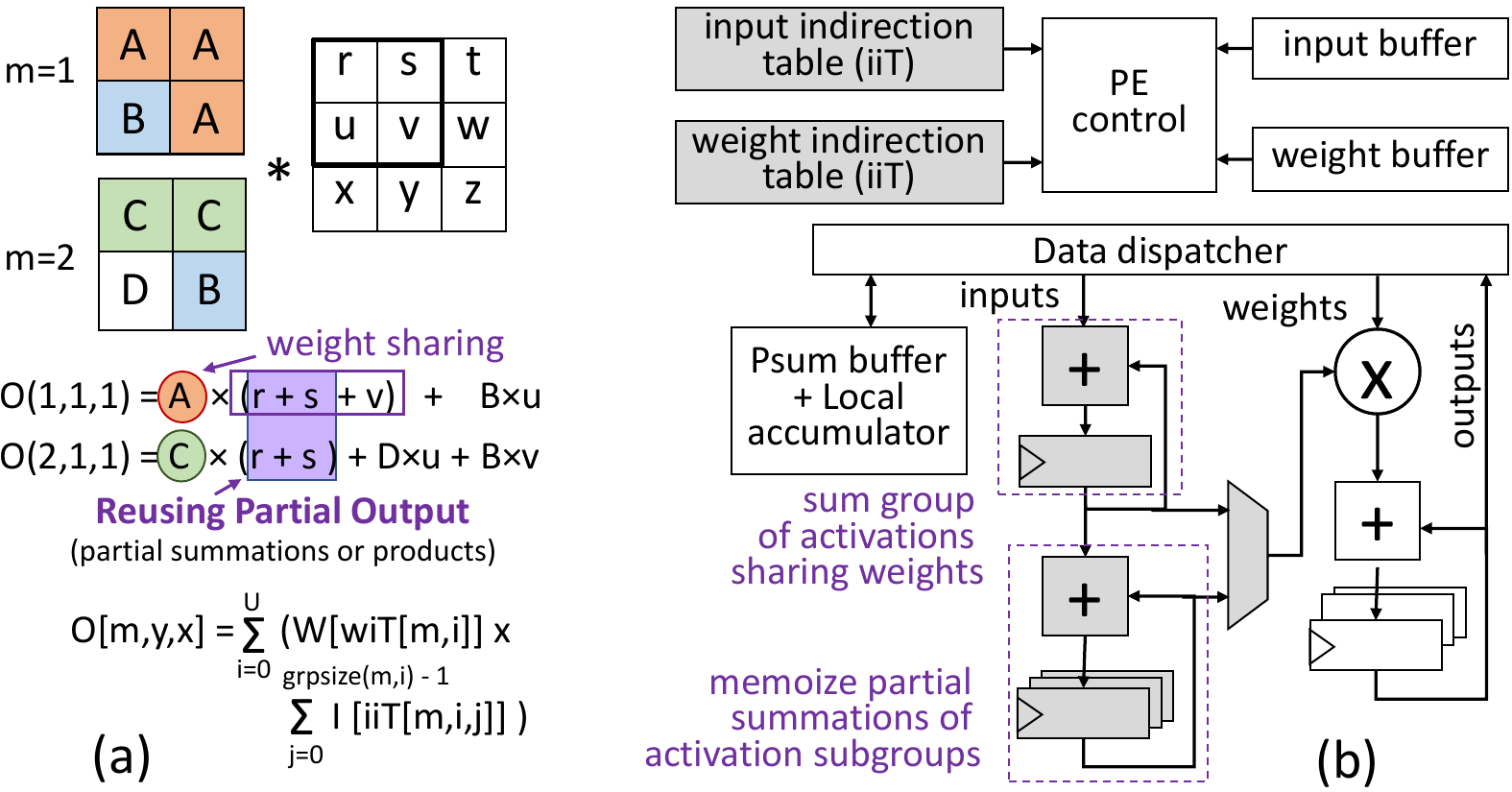}}
\caption{(a) Leveraging weight similarity and reuse of partial outputs \cite{hegde2018ucnn}. (b) Modifications in UCNN PE architecture (shaded blocks) for buffering indirection tables, partial summations of activation groups, and memoization of partial outputs. (Figure adopted from \cite{hegde2018ucnn}.)}
\label{fig::weight-similarity-UCNN}
\end{figure}

\mysubsubsection{Computation reuse (partial during processing a layer)} UCNN \cite{hegde2018ucnn} leverages the repetition of weights by forming activation groups (summations of activations) that share the same weight. It also reuses activation sub-groups, i.e., memoizes partial summations of activations that can repeatedly appear across different filters. Fig. \ref{fig::weight-similarity-UCNN}(a) illustrates an example. Weights $A$ and $C$ can be shared among corresponding activation groups. For producing activation groups, subgroups like (r+s) can be reused with memoization. So, an output activation is calculated by indexing a unique weight value and corresponding activation groups. Indirection tables provide indices of the unique weight and grouped activations. Fig. \ref{fig::weight-similarity-UCNN}(b) shows corresponding modifications in the PE datapath. UCNN reported up to 24\% area overhead for a PE and 1.8$\times$ speedup for CNNs as compared to execution on a baseline accelerator without exploiting weight repetition.   

Silfa et al. \cite{silfa2019neuron} showed that for RNNs (e.g., DeepSpeech2 \cite{amodei2016deep}, EESEN \cite{miao2015eesen}), the relative difference between the output activations over consecutive frames was about 23\%. Leveraging temporal similarity of outputs saved about 24\% computations with negligible accuracy loss. For predicting whether an output activation leads to a value similar to the previous output, their technique extended each RNN layer with a binary neural network (BNN). With BNN outputs correlating to actual outputs, execution of much smaller BNN layers led to an efficient prediction of the temporal output similarity.   

\mysubsubsection{Computation reuse (skip processing of entire layer)} A few techniques predict outputs based on previous computations and skip heavy computations of some layers. Gon{\c{c}}alves et al. \cite{gonccalves2019aggressive} showed that 18\%--81\% of computations in AlexNet CONV layers could be reused due to spatial (intra-frame) and temporal (inter-frame) redundancy of the inputs. They leveraged such reuse with memory look-ups and avoided executing CONVs. 
For YOLO-v3 \cite{redmon2018yolov3}, it processed only 22\%--32\% frames while incurring negligible accuracy loss.
Buckler et al. \cite{buckler2018eva2} proposed skipping heavy processing of some CNN layers for several frames (predicted) and executing precise computations periodically for remaining (key) frames. For predicted frames, their algorithm estimated motion in the input frame. It used results for incrementally updating the output saved from the last key frame. Unlike other value similarity techniques that incur changes in PE datapath, such techniques can be efficiently executed on a separate module (e.g., EVA$^2$ \cite{buckler2018eva2}) or co-processor, while other modules of the same or different accelerator process sparse tensors of DNN layers. EVA$^2$ identified 78\%--96\% of the frames for AlexNet and 40\%--71\% of the frames for Faster-RCNN as predicted frames while processing the YouTube-BoundingBoxes dataset \cite{real2017youtube}.

\mysubsubsection{Early termination of computations by predicting outputs} SnaPEA \cite{akhlaghi2018snapea}, SparseNN \cite{zhu2018sparsenn}, and CompEND \cite{lee2018compend} reduce ineffectual computations by early prediction of the usefulness of outputs. They check whether computations contribute to the effective inputs for the subsequent layer (e.g., ReLU or max-pooling). If not, their PEs terminate such computations. To reduce computations corresponding to output sparsity, \cite{akhlaghi2018snapea} statically re-ordered weights based on their signs. PEs of its SnaPEA architecture contained prediction activation units (with each MAC), which checked the sign-bit of the partial summation and raised a termination signal to notify the controller as the sign-bit was set. 

\mysubsubsection{Optimization opportunities} 

\textit{(i) Joint exploration of spatial and temporal similarity of inputs, weights, and outputs:} Depending on the model's configurations (depth, layer dimensions, cardinality) and domain-specific data, opportunities for value sharing and computation reuse (at both fine-grain and coarse-grain levels) in processing activations and weights can vary considerably. A joint exploration for different tensors can help to identify storage-Ops-accuracy trade-offs for efficient acceleration. 

\textit{(ii) Leveraging value similarity through separate processing:} Determining value similarity and leveraging computation reuse often demands modifications in PE-array, increasing the accelerator's area, latency, or power. Designers may obviate it by providing a separate and compact module for differential computing that handles necessary pre-processing or post-processing and can be interfaced with the PE-array and on-chip memory. Upon requirement, it can trigger execution on PE-array for structured computations. Further, algorithms expressing the functionality of ML layers/models may be defined in terms of \emph{differential computing} (i.e., execution is conditional to the input mismatch, reused otherwise). With efficient accelerator/model co-designs for differential computing of tensors, accelerators may attain structured effectual computations with fewer overheads of metadata or memoization.

\section{Load Balancing of Effectual Computations}
\label{sec::load-balance}

Depending on the distribution of zeros, the inter-PE or intra-PE imbalance can cause low utilization of PEs or their functional units, which increases execution time and energy consumption. This section summarizes sources of such imbalance, and then it discusses different software-directed techniques or hardware designs for balancing the computations. Table \ref{tab:load-balance} categorizes these techniques. Software-based techniques facilitate structured computations by forming local regions of dense elements, sorting the data by combining same-sparsity tensor blocks, or regularizing models with structured pruning. Although requiring low/no additional hardware, they are often limited to static $W$-sparsity. Accelerators dynamically balance computations by prefetching work in FIFOs or memory, obviating fine-grained synchronization of computations on PEs. Some accelerators achieve further run-time balance across PEs by a central hardware module for work sharing. 

\subsection{Sources and Impact of Imbalance}

\mysubsubsection{Inter-PE imbalance} Zeros in different tensors can be scattered, and their positions may not be determined statically (e.g., unstructured $IA$-sparsity). For most accelerators, work to be performed concurrently by PEs is fixed statically. Also, executions with conventional dataflows usually require synchronization among PEs (e.g., in SCNN \cite{parashar2017scnn}, Cnvlutin \cite{albericio2016cnvlutin}), which is achieved by barriers implemented in software via instructions or in hardware via PE architecture or controller logic. Consequently, computations per PE during each execution pass can vary drastically (inter-PE load imbalance). So, many PEs may finish their computations early, get stalled, and wait for the next set of data due to synchronized execution, while other PEs still process the previously allocated data. It increases execution time and energy consumption. Kim et al. \cite{kim2017zena} analyzed the distribution of NZ weights in AlexNet CONV3 filters and showed that in an execution pass, NZs processed by the leading and trailing PEs differed by up to 6.5$\times$. Similarly, up to 40\% cycles were idle for executions of PEs in SCNN architecture \cite{parashar2017scnn}. Sensitivity analysis for EIE showed that, without any load balance, about 47\% of the cycles were idle for the 64-PE accelerator \cite{han2016eie}.

\begin{table}[!t]
\centering

\caption{Classification of Load Balancing Techniques}
\label{tab:load-balance}
\begin{tabular}{|c|c|m{3.95cm}|}
\hline

\multirow{3}{*}{\makecell{Software \\ Directed}} 
& Data Clustering & \cite{asgari2019eridanus, kung2018adaptive, li2019squeezeflow, he2020sparse} \\  \cline{2-3} 
& Data Reorganization & \cite{kim2017zena, gondimalla2019sparten, lu2018spwa} \\ \cline{2-3}
& Model Regularization & \cite{zhou2018cambricon, kang2019accelerator, choles2018parsecore, han2017ese, ampere, liu2020systolic, shi2020csb} \\ \hline 

\multirow{2}{*}{\makecell{Hardware \\ Module}} 
& Work Prefetching & \cite{han2016eie, zhang2016cambricon, han2017ese} \\ \cline{2-3} 
& Work Sharing & \cite{lee20197, kim2017zena, shi2020csb, geng2020awb}
\\ \hline

\end{tabular}
\end{table}

\mysubsubsection{Intra-PE imbalance} For SIMD or vector PEs, intra-PE load imbalance can also contribute to a significant acceleration loss. With unstructured sparsity of one or more tensors, enough NZs may not be extracted to feed all the functional units within PEs, which causes intra-PE load imbalance. Sensitivity analysis for the SNAP accelerator showed that with moderate sparsity, utilization of multipliers falls below 80\% and up to 20\% for 90\% sparsity \cite{zhang2019snap}. Similarly, SCNN \cite{parashar2017scnn} reported below 80\% utilization of multipliers for all GoogLeNet \cite{szegedy2015going} layers and 20\% for the last two inception modules. Moreover, a few architectures use PE designs with multiple subunits in each PE. For SIMD processing, a subunit works in synchronization with other subunits of the same PE, e.g., in Cnvlutin \cite{albericio2016cnvlutin}, \cite{kang2019accelerator}, and \cite{judd2017cnvlutin2}. With unstructured sparsity, multipliers and accumulators in some subunits can often be idle, while trailing subunits process computations.

\subsection{Software Directed Load Balance}

\mysubsubsection{Clustering of NZs for populating dense data regions} As described in section \ref{sec::data-extraction-HW}, a few techniques targeted high $W$-sparsity. They used structured pruning or data combining approaches for clustering the tensor elements in locally dense regions that can be dispatched to PEs for processing in a conventional manner \cite{asgari2019eridanus, kung2018adaptive}. Thus, they achieve high PE utilization and lower invocations to accelerator resources. However, such techniques may not be effective when algorithms cannot generate or pack structured sparse data (e.g., dynamic unstructured sparsity).

Concise convolution rules (CCR) \cite{li2019squeezeflow} partitioned sparse convolutions into effective and ineffective sub-convolutions for processing locally dense regions of filters and input feature maps. It eliminated a majority of ineffectual computations and their storage (for VGG-16, achieving reduction of about 79\% and 51\%, respectively) \cite{li2019squeezeflow}. Sub-convolutions after CCR transformation were executed on the SqueezeFlow accelerator \cite{li2019squeezeflow}. However, with PEs performing only all-to-all multiplications, it may not support generic tensor operations; it can be challenging to extend CCR methodology for other algorithms.   

\begin{figure}[!t]
\centering
\centerline{\includegraphics[width=\linewidth]{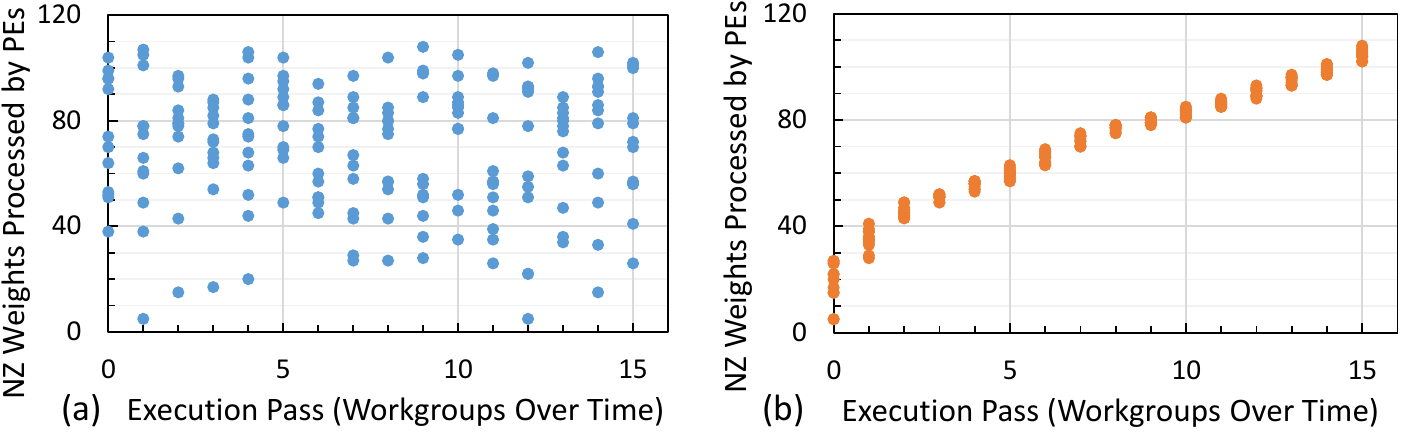}}
\caption{Distribution of NZ weights for executing CONV1 of AlexNet \cite{krizhevsky2014one} with coarse weight stationary dataflow on a 4$\times$3 PE-array. Distribution shows NZ weights in different workgroups (each workgroup contains NZs for 12 PEs): (a) without load balance (b) after sorting. (Figure inspired by \cite{kim2017zena}.)}
\label{fig::NZ-load-imbalance}
\end{figure}

\mysubsubsection{Data reorganization before work allocation} In ZENA \cite{kim2017zena}, each PE processed a different set of filters for processing a sub-workgroup. For balancing computations among these PEs, filters were sorted by sparsity and allocated to PEs such that all PEs executed filters of similar sparsity.

To determine the efficacy of such sorting, we considered AlexNet \cite{krizhevsky2014one} for ImageNet classification. We obtained the pruned model through the neural network distiller \cite{zmora2019neural} with a pruning algorithm similar to \cite{han2015learning}. For accelerating AlexNet \cite{krizhevsky2014one} CONV1 layer with coarse weight stationary dataflow, Fig. \ref{fig::NZ-load-imbalance} presents distributions of NZs in filters before and after reorganization. For processing 64 filters of size 3$\times$11$\times$11 on 4$\times$3 PEs, we consider execution through 16 different workgroups. Each workgroup contains NZ weights for concurrent processing of four filters and three channels on 12 PEs (up to 11$\times$11 on a PE). The next workgroup is initiated once all PEs entirely use previously allocated weights. Fig. \ref{fig::NZ-load-imbalance}(a) shows that before data re-organization, the total NZ weights allocated to PEs within workgroups differed by up to 21.4$\times$ (5 vs. 107 for 11$\times$11 filters) and 6.09$\times$ on average. Fig. \ref{fig::NZ-load-imbalance}(b) shows that after sorting the weights (both filter-wise and input channel-wise), it leads to an almost equal number of NZs for computations onto 12 PEs during each workgroup. The total allocated NZ weights differed by only 1.36$\times$. 

After static sorting, ZENA achieved about 20\%--32\% more acceleration for CONV layers of AlexNet and VGG-16 \cite{kim2017zena}. Depending on the sparsity, distribution of NZs, and achievable sorting granularity, the work allocated to PEs may differ considerably even after sorting. Moreover, such transformations are usually feasible only statically. So, ZENA also used dynamic work sharing, which we discuss in section \ref{sec::dynamic-load-balance}.

\mysubsubsection{Accelerator-aware regularization of the model} Recent accelerators, including Sparse Tensor Cores in NVIDIA Ampere architecture \cite{ampere}, \cite{kang2019accelerator}, and \cite{choles2018parsecore}, execute models pruned with $k$:$n$ block-sparsity (e.g., 2:4 sparsity supported by Ampere \cite{ampere}). Their PEs contain multiplexers that use indices of $k$ NZ weights to select $k$ out of $n$ dense activations. Then, functional units process extracted values. Like $k$:$n$ block-sparsity, ESE \cite{han2017ese} used a load-balance aware pruning for RNNs. It considered sub-matrices to be processed by different PEs and induced the same sparsity into all sub-matrices.

In some architectures, all PEs receive the same set of NZ activations. They process them with their unique weights and produce distinct output activations. One such architecture is Cambricon-S \cite{zhou2018cambricon} which used a coarse-grained pruning of weights. The block size for pruning depends on the total number of PEs (16). Over local regions, the pruning removed all connections between an input activation and all (16) output activations. So, when PEs processed output neurons, they processed the same number of NZ input activations and weights for computing MACs.

\subsection{Load Balancing with Hardware Structures}
\label{sec::dynamic-load-balance}

\mysubsubsection{Facilitating asynchronous computations by prefetching allocated work} One way to improve PE utilization (in the presence of load imbalance) is to prefetch the allocated work for PEs and avoid their fine-grain synchronization. So, even if there is a different amount of work (e.g., MACs per input activation), all the PEs may perform effectual computations at the same time (e.g., work on different activations). Thus, each PE can be engaged in performing some computations, before it runs out of the available data. This can be achieved by offloading more data into the FIFO or memory of each PE. For example, in EIE \cite{han2016eie}, activations are broadcast to FIFOs of all PEs. Once a PE finishes multiplying an activation to corresponding NZ weights or does not find any matching weights, it processes the next activation from its queue. FIFO size of 8 or higher ensured each PE almost having an NZ activation to process (during 87\%--91\% of computation cycles) and lowered idle time of PEs from about 47\% to 13\% \cite{han2016eie}. 

Cambricon-X \cite{zhang2016cambricon} allows asynchronous communication of weights to PEs. A centralized data extraction mechanism provides NZ activations to each PE via a unicast network, and compressed weights are loaded in the memory of each PE (2 KB). The memory access port is assigned to each PE for a short period, where it fetches several chunks of weights via DMA transfer. Depending on the prefetching interval and unstructured sparsity, each PE may asynchronously work on useful computations in most of the execution cycles. 

While asynchronous execution improves the utilization of PEs, the work allocated to PEs is still fixed. Plus, in-PE data fetching mechanisms may restrict PEs from finding the pending work in other PEs and sharing it. For highly imbalanced computations, straggling PEs can still be the bottleneck.

\mysubsubsection{Centralized load balance} In some accelerators, data is multicast to one or more rows (or columns) of PEs. A central logic processes the metadata (indices) of the tensor tiles to be distributed along with control signals from PE-rows and finds out work distribution. Then, it feeds the fast-acting rows/lanes of PEs and facilitates work sharing. For instance, ZENA \cite{kim2017zena} allocates work dynamically through down counters. Different PE-groups (e.g., PE-rows) process the same filters with different activation tiles. A central distribution mechanism contains down counters that store the number of remaining activation tiles for each PE-group. When a leading PE-group finishes its work (counter is zero), it obtains an activation tile from a straggling group (has the biggest count value) and then continues processing output activations. The work sharing improved acceleration by about 10\% for CONV layers of AlexNet and VGG-16 \cite{kim2017zena}. Memory port contention may occur when multiple leading groups simultaneously attempt to fetch the same set of input activation tiles. ZENA's execution mechanism overcomes this problem by reassigning only one activation tile at a time (to the leading group) and performing reassignments only during bus idle time.

\begin{figure}[!t]
\centering
\centerline{\includegraphics[width=\linewidth]{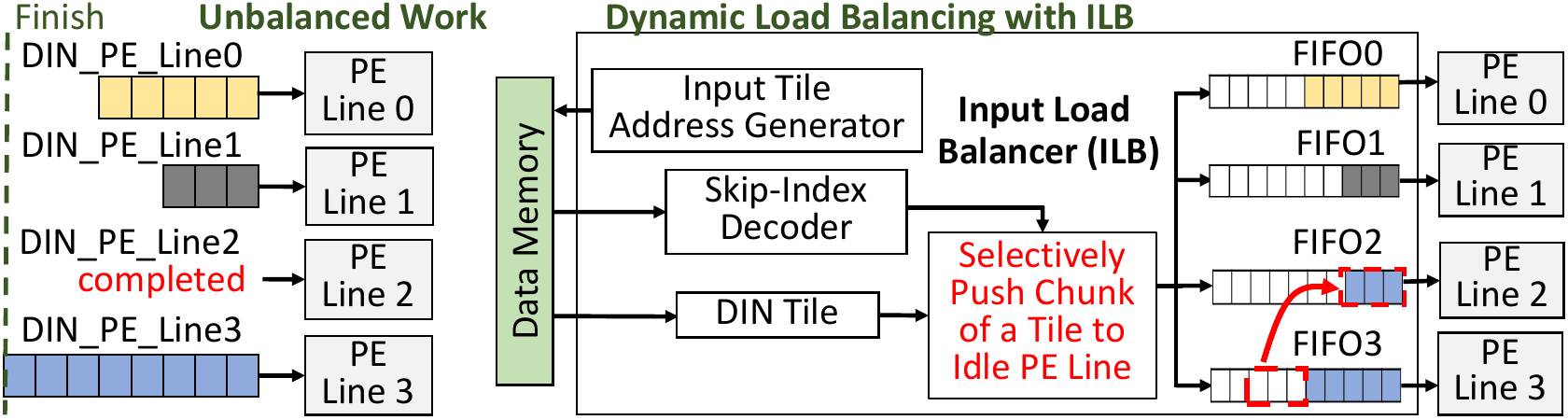}}
\caption{Load balance mechanism in LNPU \cite{lee20197} (Figure adopted from \cite{lee20197}).}
\label{fig::load-balance-LNPU}
\end{figure}

LNPU \cite{lee20197} uses an input load balancer (ILB) which is shared among PE-rows. As Fig. \ref{fig::load-balance-LNPU} shows, ILB contains address generator units to determine the indices of the compressed elements that need to be fetched. Once ILB fetches them, the skip-index decoder unit determines the appropriate indices for data extraction. It pushes them along with the NZ values into the FIFO of a PE-row. It also calculates bitmaps, which are used for pushing the data (indices and NZs) selectively into FIFOs of PE-rows at run time. Due to ILB, PE utilization in LNPU was increased by 2\%--26\% for 10\%--90\% sparsity of the inputs (activations or their gradients) \cite{lee20197}. Thus, centralized load balancing mechanisms can leverage the information about data allocation for PEs and provide equal work to PEs or feed the fast-acting PEs during run-time.  

\subsection{Optimization Opportunities} 

\textit{(i) Software-level or hardware/software/model co-design optimizations for low-cost load balance:} Most accelerators lack special support to balance computations among PEs, e.g., to avoid area and power overheads due to special hardware logic. (a) One technique is to reorganize the data \cite{kim2017zena, gondimalla2019sparten}. But, it can mostly exploit only static $W$-sparsity for inference at no/low hardware cost. So, we may require additional co-design optimizations for regularizing dynamic sparsity. (b) For pre-known or estimated sparsity, sparsity-aware mapping optimizations for accelerators may identify efficient dataflows that sustain high PE utilization. (c) When sparsity may be regularized at modest accuracy loss (e.g., for several DNNs), accelerator/model co-designs can induce the balance. It can be either done via structured pruning of activations or refactoring the operators (nonlinear activations, batch normalization \cite{gupta2019masr}, or quantization of outputs). Consequently, the co-designs may achieve structured computations over both activations and weights to an extent, leading to further accelerations.

\section{Write-Back and Post-processing}
\label{sec::post-processing}

Once PEs process allocated data, they write (partial) outputs back via interconnect. For unstructured sparsity, managing \textit{write-backs (WBs)} can be challenging because different PEs can produce outputs of different sizes at different times. Moreover, operations like ReLU, pooling, and batch-normalization need to be performed on outputs. They are usually not performance-critical like CONV or MLP. So, they can be either executed on PEs before WBs of outputs (in SCNN \cite{parashar2017scnn}, Cambricon-S \cite{zhou2018cambricon}, and EIE \cite{han2016eie}) or post-processed on central modules (in MAERI \cite{kwon2018maeri}, ZENA \cite{kim2017zena}, and SqueezeFlow \cite{li2019squeezeflow}). Central modules often assemble the outputs collected from PEs, transform data for the next layer, and encode sparse outputs on the fly. 

\subsection{Write-Back from PEs}
\label{sec::write-back}

\mysubsubsection{Simultaneous WB} Cambricon-X \cite{zhang2016cambricon} and SCNN \cite{parashar2017scnn} use fat-tree networks or point-to-point links, which allows simultaneous WBs from multiple PEs. Whenever ready, PEs can execute in a dataflow manner and immediately write outputs back after computations. This is important for processing unstructured sparsity because different PEs may process a different number of NZs and produce different amounts of output values for WB at different time intervals. With such high bandwidth, communication time can be reduced and interleaved with computations, which is important for processing models with low arithmetic intensity. These PEs write to a central module for post-processing (e.g., in Cambricon-X \cite{zhang2016cambricon}), the on-chip memory \cite{zhang2016cambricon}, or off-chip memory (e.g., in SCNN \cite{parashar2017scnn}). Although simultaneous WBs are faster, such a fat-tree network can incur considerable overhead due to increased bandwidth and inefficient bandwidth utilization in some scenarios. So, accelerator designs can instead use a common bus that is time-shared among multiple PEs; PEs can write the data back turn-wise or asynchronously.

\mysubsubsection{Sequential WB} PEs in several accelerator designs operate in a lock-stepped manner, where data blocks common to PEs are broadcast to them, and all PEs synchronize for processing the outputs (idle when done). Synchronized execution can allow WB in a specific sequence (e.g., a PE with the lowest PE-index writes the data first and so forth). It makes the programming of the accelerator easier. It also obviates overheads of specialized hardware/software support, which is required otherwise for asynchronous WB.  

\mysubsubsection{Asynchronous WB} With unstructured sparsity, PEs process a different amount of data and can asynchronously request WB during the execution. For facilitating such support, accelerator designs can employ additional hardware logic. For example, ZENA \cite{kim2017zena} used a common bus for multicasting blocks of filters and feature maps to PEs and collecting the output. Output buffers of PEs were flushed to the memory during the idle period of the bus, which avoided bus contention between broadcasting activations from memory and WB of partial summations. For prioritizing the requests from PEs to access the bus for WB, it determined the PE groups with a high number of pending output tiles. 

\subsection{Data Assembling}

PEs often process large output tiles. So, they perform fine-grained assembling of outputs locally. For example, SCNN \cite{parashar2017scnn} PEs use a coordinate computation unit that determines appropriate indices for arbitrating partial outputs to the local accumulator buffer. In other accelerators, PEs produce metadata and supplies it with outputs for correctly indexing the memory (e.g., in ZENA \cite{kim2017zena}) or assembling outputs on a central module (e.g., in Cambricon-X \cite{zhang2016cambricon}, CoNNA \cite{struharik2018conna}). The central module uses the metadata (e.g., output coordinates) from PEs or pre-known indices of PEs to assemble collected outputs before WB or post-processing. In some designs, data assembling is done by global accumulators that reduce partial summations and update outputs into appropriate memory banks (e.g., SNAP \cite{zhang2019snap}). The data assembling logic typically also handles data layout transformation (e.g., in \cite{zhang2019compact, struharik2018conna}), which is required for processing the subsequent layer. 

\subsection{Data Layout Transformations}

\begin{figure}[t]
\centering
\centerline{\includegraphics[width=\linewidth]{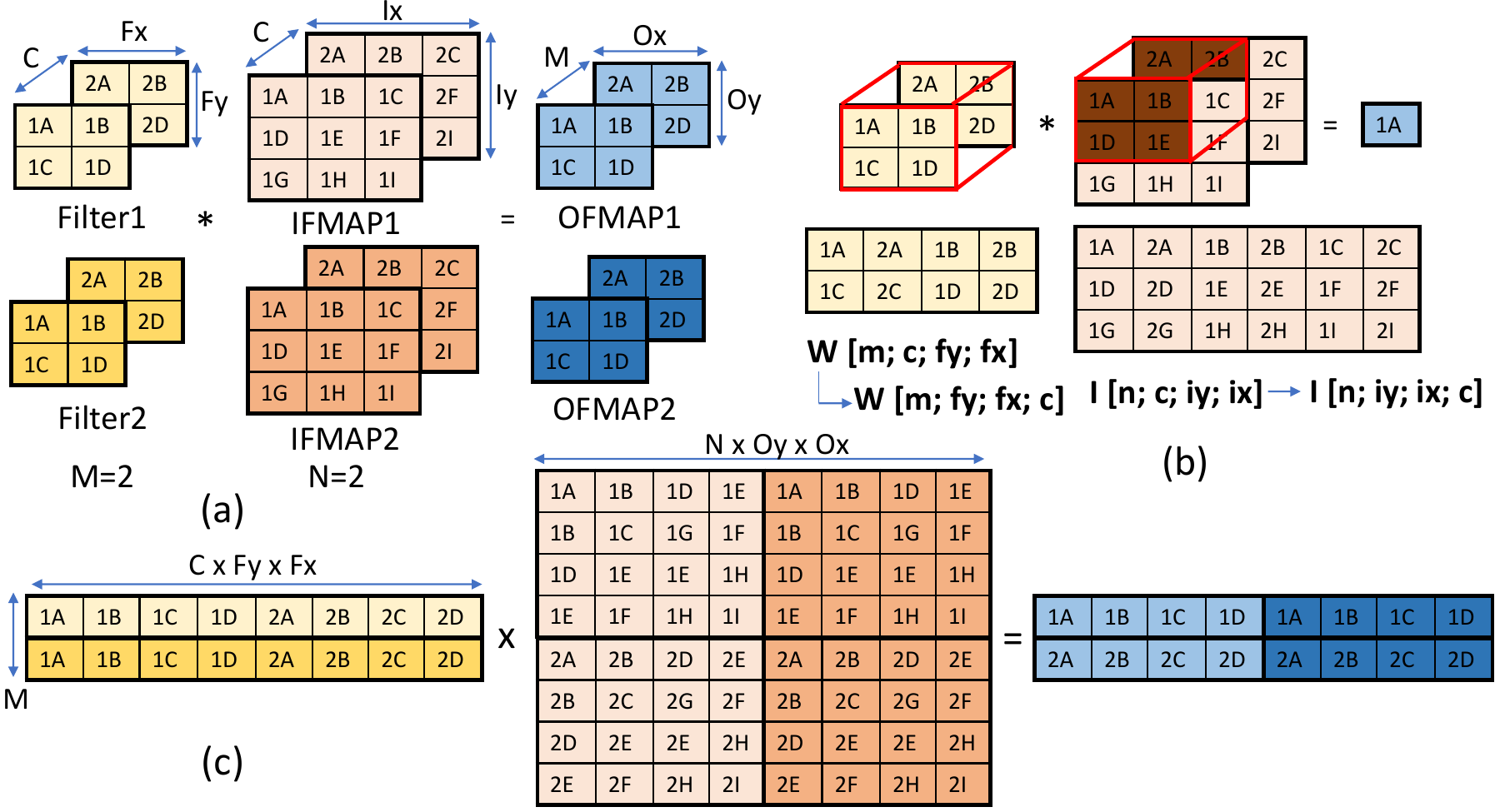}}
\caption{Data layout transformation for executing convolution. (a) Convolution of two 2$\times$3$\times$3 feature maps with two 2$\times$2$\times$2 filters. (b) Reorganizing data for striding execution. (c) Transforming feature maps into Toeplitz matrix.}
\label{fig::data-layout-xform}
\end{figure}

\mysubsubsection{Data reorganization} Accelerators are often designed for efficient vector or matrix multiplications. So, for processing convolutions, they (e.g., \cite{zhang2019compact, albericio2016cnvlutin}) require data layout in \textit{NHWC (channels-first)} format \cite{MKLMemoryFormats}, which is also used for processing on CPUs and GPUs. Fig. \ref{fig::data-layout-xform}(b) shows data reorganization for striding execution of the convolution of Fig. \ref{fig::data-layout-xform}(a). It shows iterative processing of the spatial data with channels-first processing. For example, an output activation $1A$ can be processed by fetching a block containing all channels of the first filter and ifmap. Vectors corresponding to channels can be processed iteratively. Sparse data blocks are also processed similarly but with computations on appropriate NZs.

\mysubsubsection{Transformations to Toeplitz matrix} Processing with NHWC format allows executing CONVs as iterative vector-vector multiplications, but it requires hardware support to fetch appropriate blocks. So, for processing CONVs as sparse GEMMs, a few accelerators, including ERIDANUS \cite{asgari2019eridanus} and \cite{kung2018adaptive}, transform sparse feature maps into Toeplitz matrix with \textit{im2col} transformation \cite{jia2014caffe}.Once transformed matrices are obtained and sparsity-encoded, accelerators compute sparse matrix multiplication. Fig. \ref{fig::data-layout-xform}(c) illustrates the transformation for tensors of Fig. \ref{fig::data-layout-xform}(a). It shows that neighborhood values for computing an output with a 2D CONV are combined as a vector. For multiple input channels of ifmaps or filters, corresponding elements are stacked in column-vectors or row-vectors. However, transforming ifmap into Toeplitz matrix duplicates neighborhood data and yields storage overhead (Fy$\times$Fx times higher memory for unit-strided CONV).

\subsection{On-the-fly Encoding}
\label{sec::post-process-encoding}

Several accelerators, such as SparTen \cite{gondimalla2019sparten}, SqueezeFlow \cite{li2019squeezeflow}, Eyeriss \cite{chen2016eyeriss}, and CompAct \cite{zhang2019compact}, use an encoding module. Such a module encodes blocks of output tensor on the fly and typically before WB. It reduces accesses to off-chip memory significantly \cite{chen2016eyeriss, li2019squeezeflow}. On-the-fly encoding allows efficient processing of dynamically sparsified tensors, i.e., sparse activations for DNN inference and tensors in the training of models. It typically consumes low on-chip area and power, e.g., 2.5\% and 3.06\% for the RLC encoder-decoder unit in SqueezeFlow \cite{li2019squeezeflow} and 0.3\% of the total on-chip area for the RLC unit in Eyeriss. The complexity of the hardware logic required for encoding depends on the coding format (section \ref{sec::sparse-data-coding}). For example, single-step processing for bitmap, RLC, or COO-1D incurs low overhead. A central bitmap-encoder in SparTen consisted of comparators (XNOR gates) for determining NZs and additional logic for shifting NZs to populate data vector. The encoding overhead may be lowered for block-sparse tensors, which requires indicating only positions of blocks of NZs.       

Sticker \cite{yuan2018sticker} facilitates sparsity-aware encoding. It uses three modes to encode DNN tensors of high, medium, or low sparsity with COO, bitmap, and dense format. The three modes are controlled by two threshold values. Since weights can be processed offline for DNN inference, they are pre-encoded in appropriate formats. To encode activations online, Sticker uses a sparsity adaptor module. . It consists of a sparsity detector, a four kB buffer, an encoder, and a controller. Sparsity detector contains counters that count zeros in activations of consecutive 16 channels. After the detector processes output activations (obtained after ReLU), they are stored in the buffer. Then, the controller determines the encoding mode with which the encoder can encode the data in the buffer.

\section{Compiler Support}
\label{sec::software-support}

This section provides an overview of the compiler support for sparse deep learning accelerators. It focuses on:
\begin{itemize}[leftmargin=*]
    \item Intermediate representations (IRs). They determine what type of code the compiler can support and what kind of compiler transformations it can perform.
    \item Support for sparse tensors. This subsection discusses compilation challenges in supporting sparse deep learning and compilers developed to overcome these challenges.
    \item Compiler optimizations. This subsection provides an overview of state-of-the-art techniques that allow the compiler to apply advanced optimizations and generate the most efficient code from high-level neural network descriptions.
    \item Accelerator ISAs and code generation. This subsection focuses on accelerator ISAs (e.g., instruction set for high-level tensor operations) and the compiler support for machine code generation for accelerators.
\end{itemize}

\subsection{Intermediate Representations}

IR determines which types of code can be represented by the compiler, whether it can support sparse tensor computations, the types of code transformations that can be done, and even the scalability of the compiler. 

\mysubsubsection{Need for high-level representations} A common example of low-level IR is LLVM IR which is well suited for low-level code optimizations such as register allocation but not for many high-level code optimizations needed for optimizing sparse deep learning. This is mainly because low-level IRs do not preserve information about loop structures and data layouts, and reconstructing such information is not trivial \cite{baghdadi2019tiramisu}. That is why many deep learning compilers such as TVM \cite{chen2018tvm}, Tiramisu \cite{baghdadi2019tiramisu}, and Halide \cite{halide_2012} apply many code optimizations on a high-level IR (an IR that has loops and represents multi-dimensional tensors). This is also one of the motivations for creating MLIR \cite{lattner2020mlir}, which serves as a high-level IR for low-level compilers like LLVM.

\mysubsubsection{Mathematical abstractions of code} While previous IRs have focused on representing program statements and program structure, many compilers use an additional mathematical representation (abstraction) to represent iteration domains\footnote{The iteration domain of loop iterators in a loop is all possible values that loop iterators can take.} and array accesses of statements. These mathematical representations are usually used in conjunction with the IR to simplify iteration domain and array access transformations. This subsection presents two major families of mathematical representations and compares their strengths and weaknesses. 

\myparagraph{2.A. Polyhedral representation.}
It is a unified mathematical representation for the iteration domains of statements, code transformations, and dependencies. It relies on two main concepts: \emph{integer sets} and \emph{maps}. \emph{Integer sets} represent iteration domains. \emph{Maps} are used for representing memory accesses and transforming iteration domains and memory accesses.

An integer set is a set of integer tuples described using affine constraints.  An example of a set of integer tuples is \polyc{$\{(1,1); (2,1); (1,2); (2,2); (1,3); (2,3)\}$}
Instead of listing all the tuples, we can describe the set by using affine constraints over loop iterators and symbolic constants as follows: \polyc{$\{S(i,j): 1 \leq i \leq 2 \wedge 1 \leq j \leq 3\}$} where $i$ and $j$ are the dimensions of the tuples in the set.

A map is a relation between two integer sets.  For example,
\polyc{$\{S1(i,j) \rightarrow S2(i+1,j+1) : 1 \leq i \leq 2 \wedge 1 \leq j \leq 3\}$}
is a map between tuples in the set S1 and tuples in the set S2. More details about the polyhedral model and formal definitions can be found in \cite{polyhedral, verdoolaege2010isl, pencil__paper}.

\myparagraph{Polyhedral compilers:}
Notable polyhedral compilers for deep learning include Tiramisu \cite{baghdadi2019tiramisu}, Tensor Comprehensions \cite{vasilache2018tensor}, Diesel \cite{elango2018diesel}, and TensorFlow XLA \cite{leary2017xla} (through affine MLIR dialect \cite{lattner2020mlir}). General-purpose compilers that support deep learning include PENCIL \cite{pencil__paper}, Pluto \cite{bondhugulapractical2008}, Polly \cite{polly}, PolyMage \cite{mullapudi2015polymage}, AlphaZ \cite{yuki2012alphaz}, and CHiLL \cite{chill}.

\myparagraph{Strengths of the polyhedral representation:}
\begin{itemize}[leftmargin=*]
    \item Unified representation: It eliminates friction within compiler IRs and greatly simplifies design of code transformations.
    \item Instance-wise representation: The representation granularity is instances of statement executions where each instance is a single execution of a statement during one loop iteration. Instance-wise representation includes iteration domains, data dependencies, data accesses, and code transformations, which allows the compiler to have a precise representation. 
    \item Support for the whole class of affine transformations: It allows applying any affine transformation on iteration domain and data accesses. An example of a complex affine transformation is  iteration space skewing, which allows extracting parallelism from multi-layer recurrent neural networks to increase hardware occupancy.
    \item Non-rectangular iteration domains: The representation allows compilers to naturally express non-rectangular iteration domains (i.e., iteration domains with an affine conditional).
\end{itemize}

\myparagraph{Weaknesses of the polyhedral representation:}
\begin{itemize}[leftmargin=*]
    \item Limited support for non-affine code: The polyhedral model mainly represents code and transformations using sets and maps described using affine constraints. So, it does not naturally support code that leads to non-affine constraints. This includes code with non-affine loop bounds, non-affine array accesses, and non-affine conditional. While the classical polyhedral model does not support non-affine constraints, recent work has extended the polyhedral representation to support non-affine array accesses, non-affine loop bounds, non-affine conditionals \cite{Benabderrahmane}, and parametric tiling \cite{hartono2009parametric}. The efficiency of these techniques has been demonstrated in practice by PENCIL \cite{pencil} and Tiramisu \cite{baghdadi2019tiramisu}.
    \item Slower compilation: While polyhedral operations are precise, they are computationally expensive. So, polyhedral compilers are slower than non-polyhedral compilers. Recent techniques reduce the number of statements by clustering groups of statements into macro-statements and scheduling macro-statements instead of individual statements \cite{48842}, reducing the compilation time notably. \\
\end{itemize} 

\myparagraph{2.B Non-polyhedral representation.} A common non-polyhedral representation used in deep learning compilers is \emph{interval}-based representation. It uses intervals and interval arithmetic to represent iteration domain and code transformations, respectively. Using intervals, N-dimensional loops are represented with N-dimensional boxes, e.g., iteration domain of a loop nest can be represented as: $(i,j) \in$ ([0, N],[2, M-2]). 

\myparagraph{Non-polyhedral DNN compilers:}
Their examples include TVM \cite{chen2018tvm}, Halide \cite{halide_2012}, DLVM \cite{wei2017dlvm}, and Latte \cite{truong2016latte}.

\myparagraph{Strengths of {interval}-based representations:}
\begin{itemize}[leftmargin=*]
    \item Better support for non-affine code: Non-polyhedral compilers can naturally support non-affine code transformations such as parametric tiling (loop tiling with parametric tile size). This is because the interval-based representation does not rely on using affine sets and affine relations to represent the code or dependencies. However, non-polyhedral compilers also have limited support for non-affine code (e.g., indirect memory accesses) and code transformations.
    \item Faster compilation: Operations on intervals are computationally less expensive than polyhedral equivalent operations on sets of integer points, which yields faster compilation.
\end{itemize}

\myparagraph{Weaknesses of {interval}-based representations:}
\begin{itemize}[leftmargin=*]
    \item Limited expressiveness: Interval-based non-polyhedral compilers cannot naturally represent non-rectangular iteration spaces (e.g., when bounds of loop iterators depend on a condition). It is also hard to perform certain complex affine transformations such as iteration space skewing. 

    \item Lack of support for programs with cyclic data-flow graphs: To simplify checking the legality of a schedule, many interval-based compilers assume that the program has an acyclic dataflow graph. This prevents users from expressing many programs with cyclic dataflow. For example, when a value produced by a loop is read by another loop, Halide \cite{halide_2012} does not allow fusion of the two loops (with \texttt{compute\_with} command). While it avoids illegal fusion, it prevents legal loop fusions in common cases. Polyhedral compilers avoid these over-conservative constraints by using dependence analysis \cite{feautrier1991dataflow} to check for the correctness of code transformations, which enables more schedules. While interval-based compilers can also implement non-polyhedral dependence analysis (by computing dependence distance vectors \cite{wolf1992improving}), it is not as precise as polyhedral dependence analysis \cite{feautrier1991dataflow}.
\end{itemize}

\subsection{Support for Sparse Tensors}

\mysubsubsection{Challenges in supporting sparse tensors}
While compiler support is needed in general for targeting ML hardware accelerators with diverse features, sparse tensor computations with various dataflows especially need further support. The code for manipulating sparse tensors exhibits \insight{non-static loop bounds, non-static array accesses, and conditionals}, which are difficult to analyze at compile time. The following pseudo-code shows one example of a direct convolution with sparse tensors (bounds of $j$ and accesses of $in$ are non-static).

\begin{lstlisting}
for each output channel c_o
  for j in (W.row_ptr[c_o], W.row_ptr[c_o + 1])
  {
    coeff = W.value[j]
    offset = W.col_idx[j]
    for y in (0, out_H)
      for x in (0, out_W)
        out[c_o][y][x] += coeff*in[y*out_W+x+offset]
  }
\end{lstlisting}

\mysubsubsection{DNN compilers supporting sparsity}
Their examples include Tiramisu \cite{baghdadi2019tiramisu}, Acorns \cite{acorns}, and Taichi \cite{hu2019taichi}.

Tiramisu supports $W$-sparsity by extending the polyhedral model in a way similar to \cite{Benabderrahmane}. For example, a non-affine conditional is transformed into a predicate that is attached to computation. The list of accesses of the computation is the union of the accesses of the computation in the two branches of the conditional, which is an over-approximation. During code generation, a pre-processing step inserts the conditional back into generated code. Non-static loop bounds and tensor accesses are represented as parameters in the polyhedral model. Statements that define those parameters are inserted just before the original statements that have non-static code. These techniques introduce approximations in the compiler. Their efficiency was demonstrated by \cite{Benabderrahmane} and confirmed by PENCIL \cite{pencil} and Tiramisu \cite{baghdadi2019tiramisu}.

Acorns \cite{acorns} optimizes the CNNs with \textit{IA}-sparsity. It fuses operators in a computation graph of a deep CNN, followed by sparse layout conversion (which ensures that dense/sparse tensors produced by each operator are compatible with the next operation), followed by code optimization and code generation. Acorns introduces a data layout for exploiting the sparsity structure of input data in certain domains (face detection, LiDAR, etc.) where only certain data regions are NZs. For code optimization and generation, the compiler processes a set of template codes for CNN operators (e.g., convolution, pooling) and applies optimizations such as loop tiling, vectorization, and weight packing. It does not implement advanced loop-nest optimizations like iteration space skewing.

TACO \cite{kjolstad2017tensor} uses a specific representation (iteration graphs) to generate code for sparse tensor operations and uses a scheduling language to guide the code optimization.

\subsection{Compiler Optimizations}

To generate efficient code for NN operators, a compiler has to apply a large set of complex code optimizations. It includes operator fusion; multi-level tiling and register blocking which improve data reuse; loop reordering, array packing \cite{goto2008anatomy} and data prefetching; loop skewing which enables the extraction of wavefront parallelism from multi-layer RNNs; parallelization; loop unrolling; vectorization; full/partial tile separation; tuning optimization parameters to the target architecture (e.g., tile sizes or loop unrolling factors). There are two major families of optimizing compilers: compilers that allow semi-automatic code optimization and fully automatic compilers.

\mysubsubsection{Compilers with semi-automatic code optimization (scheduling languages)}
The main idea in these compilers is to separate the algorithm from optimizations. A program, in this case, has two parts: The first part specifies the algorithm without specifying how it is optimized. The second part specifies how the algorithm is optimized (transformed). This is done through a set of high-level scheduling commands for common optimizations. Halide \cite{halide_2012}, Tiramisu \cite{baghdadi2019tiramisu}, and TVM \cite{chen2018tvm} are examples of compilers that allow semi-automatic optimization. The main advantage of this approach is it allows a user to have full control over how code should be optimized. This is important because fully automatic optimization techniques do not always succeed in providing the best performance.

Semi-automatic deep learning compilers usually provide a library of highly optimized deep learning operators. The compiler then only needs to decide automatically whether to apply certain optimizations such as operator fusion. All other optimizations are encoded manually in the library using scheduling commands. This minimizes the number of decisions that the compiler needs to make and thus guarantees the best possible performance. Note that semi-automatic compilers usually also have automatic optimization modules, but such modules can be disabled if necessary.

\mysubsubsection{Fully automatic compilers} Tensor Comprehensions \cite{vasilache2018tensor} and Diesel \cite{elango2018diesel} are examples of fully automatic compilers for deep learning. Other examples of fully automatic compilers include PENCIL \cite{pencil,pencil__paper}, Pluto \cite{bondhugulapractical2008}, and Polly \cite{polly}. All of them use Pluto \cite{bondhugulapractical2008} algorithm to automatically optimize code (choosing the schedule of statements). The main idea of Pluto algorithm is to use integer linear programming to model the problem of automatic code optimization where constraints are dependencies of the program and the objective function is the minimization of the distance between producer and consumer statements. Other compilers such as PolyMage \cite{mullapudi2015polymage} use a custom algorithm for automatic optimization.

All these compilers do not have a scheduling language and therefore do not allow the user to have fine-grain control over optimizations. Although fully automatic compilers provide productivity, they may not always obtain the best performance. Performance can be sub-optimal because they do not have a precise cost model to decide which optimizations are profitable. For instance, the Pluto \cite{bondhugulapractical2008} algorithm does not consider the redundant computations, data layout, or the complexity of the control flow of generated code.

\myparagraph{Cost models for automatic code optimization:} The goal of an automatic code optimization pass in a compiler is to find the best combination of code optimizations that minimizes the execution time. This can be modeled as a search problem where the search space is a set of combinations of code optimizations. Then, the compiler needs a search technique and a cost model to evaluate each combination. Classical compilers use hand-tuned cost models \cite{trifunovic2009polyhedral}, while others use machine learning to build cost models \cite{agakov2006using}. Both of these models do not precisely capture hardware complexity (different memory hierarchies, out-of-order execution, hardware prefetching, communication latency, etc.). Instead, state-of-the-art models are built using deep learning for better accuracy \cite{adams2019learning, mendis2019ithemal}. For example, Ithemal \cite{mendis2019ithemal} is a cost model that predicts the throughput of a basic block of x86 instructions and gets less than half the error of state-of-the-art hand-tuned models (llvm-mca in LLVM \cite{lattner2004llvm} and Intel’s IACA).

\subsection{Accelerator ISAs and Code Generation}

Accelerators, such as Cambricon-X \cite{zhang2016cambricon}, Scaledeep \cite{venkataramani2017scaledeep}, Thinker \cite{yin2017thinker}, and DnnWeaver \cite{sharma2016high}, expose a high-level ISA where some instructions perform tensor operations (e.g., dot product, matrix-matrix multiplication, convolution, pooling, and sigmoid). They simplify the compiler's  mission because it can invoke high-level operations instead of generating and optimizing a low level-code. However, it still has to manage data copies automatically. This subsection describes such high-level ISAs used by accelerators and machine code generation.

\mysubsubsection{Instruction sets} For tensor computations on hardware accelerators, ISAs typically feature instructions for arithmetic, logical, and data transfer operations with matrix, vector, and scalar data. Layers of ML models feature loops iterating thousands of times; dynamic instances of repetitive instructions can significantly increase the bandwidth requirements for delivering them to PEs at each cycle and the energy consumption. To mitigate such overheads, accelerators are designed with an array of vector or SIMD PEs. It allows PEs to process a single instruction for performing multiple computations on the blocks of tensors. Alternatively, PEs contain additional control logic such that they process an instruction once, but repeatedly perform the sequence of execution for a certain interval. 

Cambricon ISA for machine learning \cite{chen2019instruction} contains instructions for matrix and vector processing with arithmetic and logic operations, control (conditional branch and jump), and data transfer. Each operand of the instruction is either an immediate value or one of the 64 32b general-purpose registers. The registers are used for temporarily storing scalars or register-indirect addressing of the on-chip scratchpad memory. The tensor blocks are communicated between computational units from the on-chip scratchpad that is transparent to the compiler and programmers. The instructions support commonly used primitives in various ML models, e.g., multiplication, addition, subtraction, and division operations on matrices and vectors. It also supports max-pooling with a vector-greater-than-merge instruction and provides dedicated instruction for random vector generation with uniform distribution of values within [0, 1]. For supporting weight update during the training of DNNs, Cambricon provides additional instructions such as outer product, scalar-matrix multiplication, and matrix-matrix addition. However, it lacks support for managing data in the local memory of PEs and configuring NoC for communication in various dataflows. Moreover, it does not provide specific instructions for handling sparsity, e.g., predicated execution of encoded sparse data. 

The instruction set for Sticker \cite{yuan2019sticker} consists of instructions for high-level operations. For processing each layer, one of the instructions is executed only once. It configures instruction registers and common control signals that correspond to the sparsity levels and tensor dimensions. Then, at a certain time interval, a dynamic 32b instruction executes for computing convolution over data blocks on PE-array. Meanwhile, the accelerator controller distributes the next instruction, if there is no collision between the current and the next instruction. It allows hiding the execution of other dynamic instructions including the write-back and encoding of outputs and transferring data between on-chip and off-chip memory.

\mysubsubsection{Finite state machines (FSMs)} Some accelerators use FSMs for PE executions. The parameters of FSMs or PE's control logic correspond to tensor shapes and target functionality, and they are configured once (e.g., through bit-streams \cite{chen2016eyeriss}) before executing a model or a layer. Accelerator controllers (which usually initiate the data movement between on-chip and off-chip memory and configure PEs and NoC) can also contain FSMs. For example, in Thinker architecture \cite{yin2017thinker}, a finite-state controller is used for configuring the accelerator at three levels, i.e., PE-array level, model layer level, and PE level. Configuration word for PE-array level handles partitioning of the PE-array, and it points to the memory address of configurations for model layers. Each configuration word for a layer contains information about tensor dimensions and their memory addresses. Lastly, layer configurations for PEs correspond to PE functionality and the interval (loop iterations) of computations and idle time.

\mysubsubsection{Library support and code generation} The instructions for cycle-level executions or primitives are usually obtained offline. Accelerator system designers often provide users a template library that defines high-level primitives such as model layers or low-level primitives such as vector/matrix operations. Using these primitives, users can construct the model of their interest. Then, the low-level code is obtained automatically by the compiler or using the pre-defined optimized code \cite{venkataramani2017scaledeep, gopinath2019compiling}. For example, Zhang et al. \cite{zhang2016cambricon} programmed Cambricon-X accelerator with a set of library functions (written in C/C++) for primitives like convolution and matrix/vector multiplication and addition. Chen et al. \cite{chen2019instruction} proposed a programming framework consisting of assembly language, an assembler, and run-time support for executing ML models with their Cambricon ISA. For executing common layers, it also replaced the primitives with pre-defined code blocks. 

TVM \cite{chen2018tvm} supports defining custom back-ends for accelerators, which was demonstrated using a vanilla accelerator with a matrix-multiply engine. For executing primitives on accelerators, TVM enables Tensorization \cite{chen2018tvm}, i.e.,  decoupling the target hardware intrinsic from the schedule while mapping ML operators. To demonstrate code generation for the vanilla accelerator, TVM enabled a driver library and runtime support that constructs the instructions and offloads them to the accelerator. Its code generation module translated the program into appropriate function calls of the runtime API. Moreau et al. \cite{moreau2018hardware} leveraged the TVM stack and proposed a JIT compiler and a runtime system to generate code for a programmable VTA accelerator. 

It is important that the accelerator can support multiple front-ends corresponding to different ML frameworks such as TensorFlow \cite{abadi2016tensorflow}, PyTorch \cite{paszke2019pytorch}, and MXNet \cite{chen2015mxnet}. Integration of the programming, compilation, and runtime environment with the common frameworks for ML application development is necessary for supporting different compact ML models. Leveraging the existing system stack (e.g., TVM) can provide such opportunities to accelerator system developers. Note that although TVM supports defining custom accelerator back-ends and can lower optimized mappings to accelerator-specific code, it currently does not provide support for sparse tensors.

\section{Trends and Future Directions}
\label{sec::future-directions}

\subsection{Hardware/Software/Model Co-designs} 

\mysubsubsection{Hardware-aware compression techniques} The framework for exploring efficient model compression (either of quantization, pruning, and size reduction) should be aware of hardware features and provide directed search accordingly. For example, bit-widths of tensors that can be efficiently processed by different hardware platforms vary considerably (e.g., from multiples of 8-bits to arbitrary bit-widths). Accelerators typically support only uniform widths of tensors (activations and weights), and many accelerators do not support value sharing. Also, when hardware only supports widths that are multiple of 4 or 8 bits, quantization with other bit-widths requires zero paddings, which incurs inefficient storage and processing. Instead, the compression algorithm can opt for improving the accuracy, increasing sparsity, or trading off the bit-widths among layers for achieving higher compression and acceleration. Similarly, depending on the hardware support for fine-grained or block-sparsity, hardware-aware pruning can better achieve the compression objectives (model exploration time, performance, and energy efficiency, while meeting target accuracy). Efficiency can be further improved when compression techniques leverage execution models of hardware accelerators (e.g., energy-aware pruning \cite{yang2017designing}). Relatively simple logic modules of hardware accelerators have enabled recent techniques to estimate execution metrics through analytical cost models. Accommodating such cost models (including for different sparsity levels/patterns, precisions) enables the compression algorithms to select effective pruning ratios/structures, tensor shapes, and tensor precisions, which can help to achieve desired accelerations.

\begin{figure}[!t]
\centering
\centerline{\includegraphics[width=0.75\linewidth]{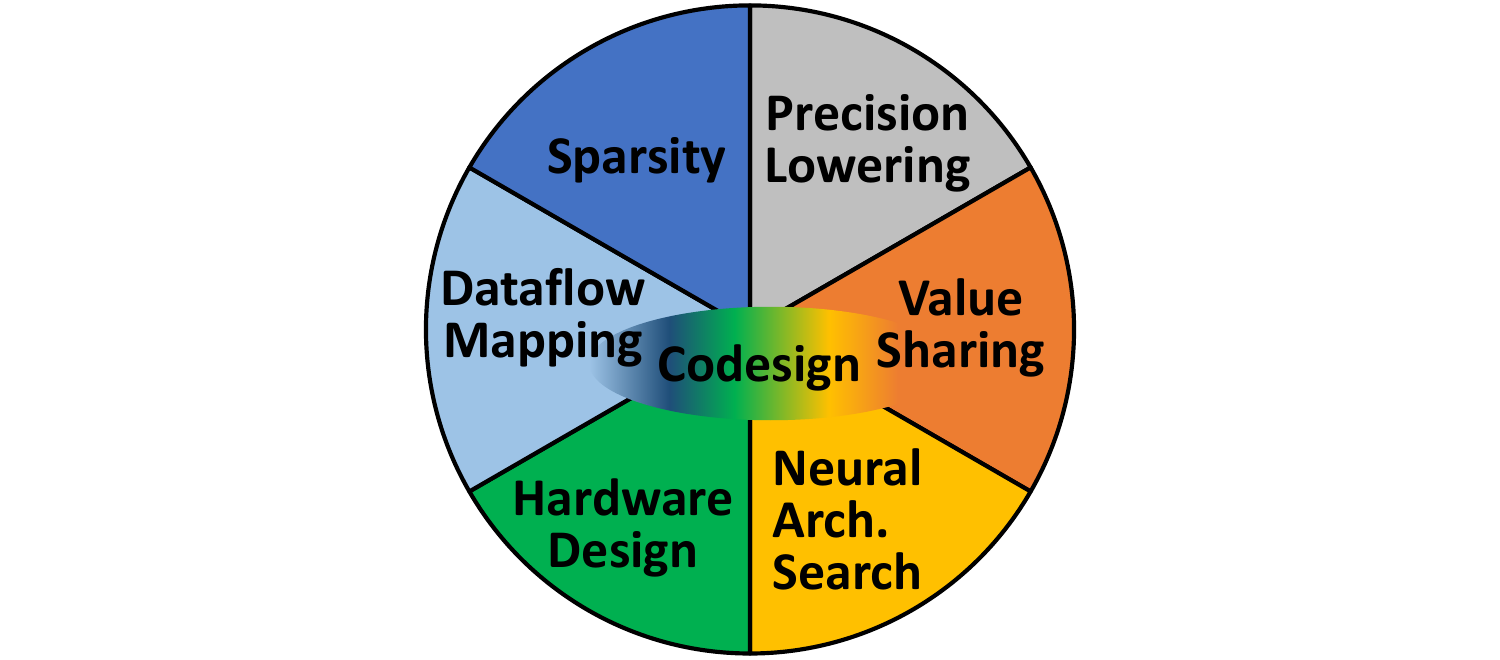}}
\caption{Co-designs can enable efficient accelerations of compact models.}
\label{fig::codesign}
\end{figure}

\mysubsubsection{Joint and automated exploration of sparsity, precision, and value similarity} Recent compression techniques typically employ structured or fine-grained data pruning during training with a fixed precision of tensors. Techniques for adaptive quantization often do not explore pruning. Joint explorations of pruning and quantization may achieve high compression due to the interplay of these compression mechanisms. For instance, quantization can increase sparsity considerably \cite{aimar2018nullhop}, as more values can be represented as zero after compressing the range \cite{han2020survey}. Likewise, pruning may reduce bit-widths further since fewer non-zero values in the pruned model may be expressed with a much lower numeric range and precision. Moreover, such compression techniques do not leverage temporal and spatial value similarity in inputs, outputs, or weights. So, joint exploration algorithms may be developed that use multiple compression strategies during training and automatically explore combinations that compress the model further. Recent techniques for automated explorations include CLIP-Q \cite{tung2018clip}, \cite{srivastava2019joint}, and \cite{yang2019automatic}. Exploring a wide range of compression combinations during the training may not be feasible. Therefore, model designers may reduce the space of compression choices by limiting effective options before beginning resource-extensive training, and if required, further limiting the search space by quick evaluations with a pre-trained model and fine-tuning. 

Compression benefits achieved through joint explorations need to be translated into efficient hardware accelerations. So, the exploration heuristic should not preclude experts from expressing a directed search for hardware-friendly executions, e.g., specifying pruning with 1D or $k$:$n$ block-sparsity, constraints for bit-widths, tolerable accuracy loss, etc. Moreover, the heuristic should also provide automated optimization/exploration of hyperparameters (including using cost models of accelerators). This is because the compression algorithm needs to adjust the strategy of pruning or quantization and its hyperparameters. For instance, the pruning algorithm needs to find out the pruning ratio for each iteration (epoch); pruning mechanism (which values to prune, e.g., below a certain threshold); pruning pattern (fine-grain, block size); bit-widths of tensors (quantization). All such hyperparameters or strategies need to be adjusted automatically (to an extent) such that the memory footprint or computations are greatly reduced, with no or tolerable accuracy loss.

\mysubsubsection{Value-aware neural architecture search (NAS) and accelerator/model co-designs} Techniques for NAS or AutoML can automatically obtain efficient models that surpass the accuracy of models devised by human developers. However, there remains scope for considerably improving NAS for obtaining highly compact models. Recent techniques \cite{kwon2018co, marculescu2018hardware, zhang2019neural, abdelfattah2020best} have explored accelerator/model co-designs that support quantized models and layers of different shapes. However, the efficiency can be further amplified by including the sparsity and adaptive bit-widths of model layers and analytically considering their implications on hardware accelerators. 

A major challenge faced by the model search techniques and automated accelerator/model co-designs is the vast search space. As Fig. \ref{fig::codesign} shows, explorations can be performed for (i) ML models (i.e., NAS) \cite{han2020survey}, (ii) compression strategies (e.g., automated pruning and quantization) \cite{he2018amc}, (iii) mappings of models on accelerators \cite{parashar2019timeloop, dave2020dmazerunner}, and (iv) specifications of hardware accelerators \cite{kwon2019understanding, dave2020dmazerunner}. The explorations of (i) and (ii) directly impact compression and accuracy, while search optimizations for (iii) and (iv) affect the performance and energy-efficiency of the accelerator for given models. Among these exploration spaces, NAS can be significantly time-consuming (several GPU days \cite{han2020survey}), followed by automated model compression (e.g., \cite{he2018amc}). Therefore, the resultant joint space for value-aware NAS and accelerator/model co-designs is many-folded. So, it may require notable efforts for developing automated exploration of co-designs that can obtain extremely efficient and accelerator-friendly compact models.

\mysubsubsection{Facilitating structured computations of sparse tensors} Designers may opt for accelerators that are effective for structured computations of dense tensors, e.g., systolic arrays (as near-data accelerators or coupled to processor cores) and in-memory processing with resistive crossbars. While sparsity or size reduction of tensors may need to be leveraged, significant design modifications are often infeasible due to design requirements (area/power budgets) or the increase in complexity of the system stack. So, techniques for pre-processing can be developed, which can arrange structured dense regions for feeding underlying engines or achieve structured data through sparsification/reorganization at almost no accuracy loss. Such pre-processing can be done on additional hardware modules or the host processor that handles the non-performance-critical tasks. Such disjoint mechanisms can obviate heavy design modifications in systolic arrays (e.g., \cite{kung2018adaptive}) or in-memory/near-data processing engines (e.g., ReCom \cite{ji2018recom}, SNNrram \cite{wang2018snrram}) while leveraging various sparsity and value similarity opportunities across different models.

\subsection{Design Tools and Frameworks}

\mysubsubsection{Framework for analyzing performance gains of accelerators due to sparsity} Given that tensors of several ML models are sparse, it is important to design accelerator systems that can exploit performance gains for multiple models through low-cost hardware modules and enhanced software support. As we discussed in sections \ref{sec::sparse-data-coding}--\ref{sec::software-support}, each enhancement presents multiple implementation choices at the hardware or software level. Although crafting a cycle-level simulation infrastructure for such a wide design space may be infeasible, a data-driven quantitative model can be significantly helpful for design explorations. It can process the actual data (or discover distributions of zeros), provide high-level modeling of common choices, and estimate the performance gains for each combination of the implementation choices. For newer models or functionality, hardware designers can run through a set of implementation choices in an early design phase. They can explore the implications of sparsity for the desired choice of encoding, data extraction logic, functional units, NoC, load balancing, and dataflow mechanism. 

\mysubsubsection{Accelerator design frameworks for compact models} Several frameworks for developing and simulating FPGA or ASIC based accelerators have recently been proposed, including DNNWeaver \cite{sharma2016high}, DNNBuilder \cite{zhang2018dnnbuilder}, T2S-Tensor \cite{srivastava2019t2s}, and HeteroCL \cite{lai2019heterocl} for FPGAs and NVDLA \cite{nvdla}, VTA \cite{moreau2018hardware}, MAGNet \cite{venkatesan2019magnet}, MAERI \cite{kwon2018maeri}, and AutoDNNChip \cite{xu2020autodnnchip} for specialized accelerators. Similarly, hardware construction languages or representations such as Chisel \cite{bachrach2012chisel} and $\mu$IR \cite{sharifian2019muir} enable expressing microarchitectural features through high-level primitives. Such infrastructures are key for the community since they can serve as a good learning resource for training the new professionals and provide a kick-starter baseline for developing new design features. 

However, most frameworks support designs for dense tensors of fixed bit-widths and lack support for sparsity-tailoring features. Such frameworks can provide some pre-built modules for encoding/extracting sparse data (e.g., with common formats like  RLC, bitmap, or for block-sparsity), dynamic load balancing or data reorganization, configurable functional units, and configurable interconnects for sparse and bit-adaptive computing, etc. Even with limited features, they may serve as reusable logic that can be leveraged by designers for quick prototyping and design explorations. Further, abstractions and specifications for constructing sparsity-tailored hardware and dataflows can enable automated and efficient design explorations and easier programming of accelerators.

\subsection{Accelerating Training of ML Models} While there have been significant advances in performing inference on hardware accelerators, efficient training of the models on hardware accelerators has received relatively little attention. Training has been done in high-performance computing environments containing CPU and GPU platforms and recently on FPGAs and TPU accelerators. Hardware accelerators can offer significant benefits to the model training in both edge and datacenter-scale computing environments, and they can notably improve performance and energy efficiency, respectively. In particular, they are promising for enabling online learning on edge devices through compact models.

Accelerators, such as \cite{fleischer2018scalable}, ScaleDeep \cite{venkataramani2017scaledeep}, and HyPar \cite{song2019hypar}, have been proposed for efficiently training the models. However, they either do not leverage sparsity, or may not be efficiently utilized for irregular-shaped tensors, or lack support for various precisions of weights, activations, gradients, and weight updates. This presents further opportunities for performance gains and energy efficiency. Additionally, designers can leverage cross-layer optimizations (e.g., by reusing the data of gradients during back-propagation) and support mixed-precision of tensors during the training of compact models.

\subsection{Applying Techniques for Sparsity to Other Domains} In this work, we considered a wide variety of techniques that leverage sparsity for the machine learning domain, which represents an enormous research effort. Many other domains face similar challenges in exploiting sparsity, and accelerators have been proposed for some of the more processing-intensive domains; this includes graph processing \cite{graphicionado, tesseract}, database operations \cite{q100}, genomics \cite{darwin, genax}, and compression \cite{fowers2015comrpession}. In some cases, computation primitives even extend across domains. For example, finding intersecting non-zeros is analogous to joins in a database context \cite{dadu2019towards}. Applying the lessons learned from extensive research on sparsity in an ML context can likely speed innovation in a broader context.

\section{Related Work}
\label{sec::related-works}

\textbf{Deep learning models and their applications:} Surveys \cite{pouyanfar2018survey, alom2019state} described different deep learning models along with different frameworks and datasets. Gu et al. \cite{gu2018recent} discussed applications of CNNs in computer vision and language processing. Recent surveys have also discussed applications of deep learning in medical image analysis \cite{litjens2017survey}, biomedical applications \cite{wei2020review}, wireless and networking \cite{zhang2019deep}, and embedded systems \cite{rezk2020recurrent}. Elsken et al. \cite{elsken2019neural} surveyed techniques for neural architecture search.

\textbf{Compact models:} Cheng et al. \cite{cheng2018model} surveyed techniques for parameter pruning and low-rank factorization. Wang et al. \cite{wang2019deep} surveyed techniques for pruning, precision lowering, weight sharing, low-rank factorization, and knowledge distillation. Deng et al. \cite{han2020survey} described techniques to obtain compact models including sparsification, quantization, tensor decomposition, and joint-way compression. 

\textbf{Hardware accelerators for dense ML models:} Shawahna et al. \cite{shawahna2018fpga} surveyed FPGA accelerators for processing dense tensor computations of deep learning applications. Venieris et al. \cite{venieris2018toolflows} discussed different CNN-to-FPGA toolchains and described their hardware architectures, design space exploration techniques, and support for different precisions of tensors. They also compared execution metrics of the designs obtained with various toolchains and those with the previously proposed FPGA accelerators for CNNs. Sze et al. \cite{sze2017efficient} presented a survey about efficiently executing DNNs on hardware accelerators. It described different DNNs, different compression techniques for compact models, and optimized dataflows for spatial architectures. Reuther et al. \cite{reuther2019survey} benchmarked executions of different ML accelerators. Li et al. \cite{li2020deep} discussed different ML frameworks and compilers for deep learning models.

\textbf{Hardware accelerators for compact ML models:} Mittal \cite{mittal2018survey} surveyed executing compact models, including BNNs, on FPGAs. It also discussed processing convolutions with the Winograd algorithm and executions on multiple FPGAs. Deng et al. \cite{han2020survey} surveyed hardware accelerators that support bit-adaptive computing and the data extraction modules for leveraging the sparsity of inputs, weights, or outputs. Du et al. \cite{du2020self} recently proposed MinMaxNN system for dynamically switching NN models. They surveyed techniques for designing self-aware NN systems (which can continuously sense information from the environment and dynamically react), including leveraging sparsity and tensor quantization. Wang et al. \cite{wang2019deep} surveyed hardware implementations for processing tensors of lower precisions (binary, ternary, and logarithmic quantizations). Ignatov et al. \cite{ignatov2019ai} benchmarked executions of quantized deep learning models on mobile AI accelerators.

In contrast to the above surveys, this work highlights sources of sparsity and size reduction of tensors in ML models and challenges in efficiently executing them on hardware accelerators. Then, it surveys and discusses the corresponding hardware and software support, including encodings and extraction of sparse data, sparsity-aware dataflows, memory management and on-chip communication of sparse tensors while leveraging data reuse, load balancing of computations, and compiler support. It also discusses techniques for computations of mixed-precision and value-shared sparse tensors. 

\section{Summary}

For efficient and hardware-friendly processing, compact deep learning models have been designed. They consume less storage and computations and consist of tensors with considerable sparsity, asymmetric shapes, and variable precisions. While these compact models can be accelerated on hardware accelerators efficiently, it requires special hardware and software support. We have highlighted challenges in efficiently accelerating their sparse and irregular tensor computations. Leveraging sparsity, especially unstructured, requires a significant redesign to store, extract, communicate, compute, and load-balance only non-zeros. Moreover, the sparsity levels and patterns due to various sources lead to unique challenges and solutions in hardware/software/model co-designs. 

In this article, we have discussed how exploiting sparsity effectively depends on tailoring the data encoding and extraction, dataflow, memory bank structure, interconnect design, and write-back mechanisms. We provided an overview of corresponding enhancements in accelerator designs and their effectiveness in exploiting sparsity. Categorization of different techniques informs how they leveraged structured or unstructured sparsity of weight or activations during learning or inference of ML models (Tables \ref{tab:overview-sparse-accel}, \ref{tab:target-sparsity}). For recent DNNs, we analyzed achievable accelerations for a few popular accelerators (section \ref{sec::sparse-dnn-acceleration-case-study}). The analysis showed that accelerators exploit moderate sparsity and achieve high speedups as sparsity increases. However, exploiting high or hyper sparsity can further provide considerable opportunities, which would also need efficient mechanisms for data extraction and load balancing. Also, configurable architectures for NoCs, functional units, and buffers are required for catering to various functionalities and metadata management.

Our analysis of sparsity-encodings describes their storage efficiency for various sparsity and the decoding requirements. While bitmaps and RLC/CSC formats are commonly used for moderate and high sparsity, respectively, storage efficiency can be improved with block-sparse tensors (especially at hyper sparsity). We have introduced a taxonomy for non-zero extraction techniques that are used for feeding the functional units of PEs. Existing data extraction mechanisms (e.g., in EIE \cite{han2016eie}, EyerissV2 \cite{chen2019eyeriss}, Cambricon-X/S \cite{zhang2016cambricon, zhou2018cambricon}) exploit moderate sparsity. But, they may not extract enough NZs at high or hyper sparsity of large tensors (e.g., sparse BERT \cite{huggingface}), achieving lower speedups. We also discuss how block-sparsity can simplify data extraction and facilitate balanced computations. For exploiting diverse sparsity across tensors of different models, designers can explore multiple or configurable mechanisms for decoding and extraction of non-zeros. 

Data reuse opportunities in processing common DNNs vary significantly, and sparsity lowers the reuse due to fewer effectual computations. However, compressed tensors allow to fit and reuse more data in on-chip memory, which reduces accesses to off-chip memory and overall latency. We have discussed techniques for memory bank management to support unstructured accesses for sparse computations and hiding the memory access latency. At high or hyper sparsity, execution may become \emph{bandwidth-bounded}, as enough data may not be prefetched always. Hence, techniques for efficient data management (e.g., cross-layer, on-chip data reuse) and exploiting high bandwidths need to be explored. Different accelerator designs have used various interconnects for the distribution of operands, reduction of partial outputs, and collecting the outputs. They vary in terms of the bandwidth requirement and exploiting spatial data reuse. \emph{Configurable interconnects} (e.g., in EyerissV2 \cite{chen2019eyeriss}, SIGMA \cite{qin2020sigma}) are required for accelerating different DNNs of diverse sparsity, functionality, and tensor shapes, since they can support a mix of communication patterns. They are important for enabling \emph{asymmetric} spatial accumulations of partial outputs (for sparse tensor computations) and concurrent spatial processing of different groups, e.g., for DW-CONV.  

Processing compressed tensors can impose significant maneuvering efforts in the PE architecture design. We discuss further opportunities including configurable designs of functional units for efficient vector processing and flexible sparsity-aware dataflows for high utilization across variations in sparsity and functionality of different layers. We also surveyed techniques for approximated computing through multiplier-free PEs and leveraging temporal and spatial similarity of values, which improve execution efficiency further. Sparse tensor computations over different PEs can be highly imbalanced. We have surveyed different techniques that sustain the acceleration by balancing the work through hardware modules for asynchronous computations or work sharing (e.g., EIE \cite{han2016eie}, ZENA \cite{kim2017zena}). Software-directed regularization such as structured sparsity eliminates load imbalance, e.g., in leveraging weight/activation sparsity for Cambricon-S \cite{zhou2018cambricon} and 50\% weight sparsity for NVIDIA A100 \cite{ampere}. Techniques including data transformations and refactoring of DNN operators may achieve low-cost load balance, including for dynamic sparsity. We have also surveyed mechanisms for asynchronous write-backs of outputs and sparsity-aware encodings on the fly. Compilation for the accelerators requires the ability to efficiently express sparsity in intermediate representations, flexibly apply different compiler optimizations, and emit efficient accelerator-specific code. The survey has discussed techniques that can enable such support and open challenges. 

Accelerator/model co-designs can efficiently leverage various precision and value similarity of different tensors and induce sparsity for accelerator-friendly executions. Automated and joint explorations of accelerator-aware compression algorithms can advance acceleration opportunities further. We have highlighted future directions for such co-designs and the system stack development (section \ref{sec::future-directions}). In individual sections, we have also discussed further opportunities for tailoring different hardware or software enhancements for sparsity. While our discussions focused on leveraging sparsity for ML models, exploiting diverse sparsity can also aid the efficient processing of applications of other domains \cite{duff1977survey, hegde2019extensor}.

In conclusion, while different accelerators and compression algorithms have been proposed for efficiently processing compact ML models, it remains an active research frontier. In particular, hardware/software/model co-designs and automated and joint explorations of tensor sparsity, adaptive quantization, shape reductions, and dataflow will likely provide further opportunities for innovations across the system. With a boost in energy-efficient accelerations of the learning and inference at the cloud and edge, they can be anticipated to further improve the intelligence of various systems or applications. 

\newpage
\appendix[Hardware Accelerators Can \\ Exploit Sparsity Better]
Exploiting acceleration opportunities due to sparsity (especially unstructured) is relatively hard for execution on CPUs and GPUs \cite{zhang2016cambricon, zhou2018cambricon, qin2020sigma}. The performance of ML models can even degrade, as compared to the execution with dense data (e.g., for a GEMM, when unstructured $W$-sparsity is below 70\% \cite{gale2020sparse}). For executing AlexNet layers on GPUs, \cite{wen2016learning} analyzed speedup for processing CSR-encoded matrices with cuSPARSE and dense matrices with cuBLAS. Their experiments showed obtaining \emph{limited} speedups (below 1.4$\times$) or even slowdowns for high sparsity. This is because unstructured sparsity may yield poor data locality for scattered effectual NZs. Plus, it is challenging to skip ineffectual computations and equally distribute the work among multiple threads or computational units of processor cores. Zhang et al. \cite{zhang2016cambricon} analyzed performance benefits of executing sparse models (LeNet, AlexNet, and VGG-16) on CPU (with sparse BLAS) and GPU (with cuSPARSE) platforms, as compared to processing dense models (with Caffe \cite{jia2014caffe}). For average sparsity of 90.94\%, they reported geomean speedup of only 23.34\% for GPU and 110\% more time on CPU. In their sparsity-sensitivity analysis, CPU and GPU showed marginal speedup only at moderate or high sparsity due to non-trivial costs of sparse data processing. But, for Cambricon-X  \cite{zhang2016cambricon}, performance gains were reported for 5\% or more sparsity due to its design tailored for sparse tensor computations. For hyper sparsity, it achieved high speedups (e.g., 15.5$\times$ for CONV and 48.5$\times$ for FC layer), as compared to executing dense tensors \cite{zhang2016cambricon}. Thus, with special support for sparse and irregular tensor computations, hardware accelerators can achieve notable speedups and efficiency. 

\section*{Acknowledgment}
This work was partially supported by funding from NSF grant CCF 1723476 - NSF/Intel joint research center for Computer Assisted Programming for Heterogeneous Architectures (CAPA). The authors thank anonymous reviewers for providing valuable feedback.

\ifCLASSOPTIONcaptionsoff
  \newpage
\fi
\bibliographystyle{IEEEtran}
\bibliography{ref.bib}

\begin{thebibliography}{100}
\providecommand{\url}[1]{#1}
\csname url@samestyle\endcsname
\providecommand{\newblock}{\relax}
\providecommand{\bibinfo}[2]{#2}
\providecommand{\BIBentrySTDinterwordspacing}{\spaceskip=0pt\relax}
\providecommand{\BIBentryALTinterwordstretchfactor}{4}
\providecommand{\BIBentryALTinterwordspacing}{\spaceskip=\fontdimen2\font plus
\BIBentryALTinterwordstretchfactor\fontdimen3\font minus
  \fontdimen4\font\relax}
\providecommand{\BIBforeignlanguage}[2]{{%
\expandafter\ifx\csname l@#1\endcsname\relax
\typeout{** WARNING: IEEEtran.bst: No hyphenation pattern has been}%
\typeout{** loaded for the language `#1'. Using the pattern for}%
\typeout{** the default language instead.}%
\else
\language=\csname l@#1\endcsname
\fi
#2}}
\providecommand{\BIBdecl}{\relax}
\BIBdecl

\bibitem{krizhevsky2012imagenet}
A.~Krizhevsky, I.~Sutskever, and G.~E. Hinton, ``Imagenet classification with
  deep convolutional neural networks,'' in \emph{Advances in neural information
  processing systems}, 2012, pp. 1097--1105.

\bibitem{he2016deep}
K.~He, X.~Zhang, S.~Ren, and J.~Sun, ``Deep residual learning for image
  recognition,'' in \emph{Proceedings of the IEEE conference on computer vision
  and pattern recognition}, 2016, pp. 770--778.

\bibitem{tan2019efficientnet}
M.~Tan and Q.~Le, ``Efficientnet: Rethinking model scaling for convolutional
  neural networks,'' in \emph{International Conference on Machine Learning},
  2019, pp. 6105--6114.

\bibitem{redmon2018yolov3}
J.~Redmon and A.~Farhadi, ``Yolov3: An incremental improvement,'' \emph{arXiv
  preprint arXiv:1804.02767}, 2018.

\bibitem{chen2017rethinking}
L.-C. Chen, G.~Papandreou, F.~Schroff, and H.~Adam, ``Rethinking atrous
  convolution for semantic image segmentation,'' \emph{arXiv preprint
  arXiv:1706.05587}, 2017.

\bibitem{tran2015c3d}
D.~Tran, L.~Bourdev, R.~Fergus, L.~Torresani, and M.~Paluri, ``Learning
  spatiotemporal features with 3d convolutional networks,'' in
  \emph{Proceedings of the IEEE international conference on computer vision},
  2015.

\bibitem{vaswani2017attention}
A.~Vaswani, N.~Shazeer, N.~Parmar, J.~Uszkoreit, L.~Jones, A.~N. Gomez,
  {\L}.~Kaiser, and I.~Polosukhin, ``Attention is all you need,'' in
  \emph{Advances in neural information processing systems}, 2017.

\bibitem{devlin2019bert}
J.~Devlin, M.-W. Chang, K.~Lee, and K.~Toutanova, ``Bert: Pre-training of deep
  bidirectional transformers for language understanding,'' in \emph{Proceedings
  of the 2019 Conference of the North American Chapter of the Association for
  Computational Linguistics: Human Language Technologies, Volume 1 (Long and
  Short Papers)}, 2019.

\bibitem{brown2020language}
T.~B. Brown, B.~Mann, N.~Ryder, M.~Subbiah, J.~Kaplan, P.~Dhariwal
  \emph{et~al.}, ``Language models are few-shot learners,'' \emph{arXiv
  preprint arXiv:2005.14165}, 2020.

\bibitem{goodfellow2014generative}
I.~Goodfellow, J.~Pouget-Abadie, M.~Mirza, B.~Xu, D.~Warde-Farley, S.~Ozair,
  A.~Courville, and Y.~Bengio, ``Generative adversarial nets,'' in
  \emph{Advances in neural information processing systems}, 2014.

\bibitem{park2018deep}
J.~Park, M.~Naumov, P.~Basu, S.~Deng, A.~Kalaiah, D.~Khudia, J.~Law, P.~Malani,
  A.~Malevich, S.~Nadathur \emph{et~al.}, ``Deep learning inference in facebook
  data centers: Characterization, performance optimizations and hardware
  implications,'' \emph{arXiv preprint arXiv:1811.09886}, 2018.

\bibitem{dlrm}
M.~Naumov, D.~Mudigere, H.-J.~M. Shi, J.~Huang, N.~Sundaraman, J.~Park,
  X.~Wang, U.~Gupta, C.-J. Wu, A.~G. Azzolini \emph{et~al.}, ``Deep learning
  recommendation model for personalization and recommendation systems,''
  \emph{arXiv preprint arXiv:1906.00091}, 2019.

\bibitem{litjens2017survey}
G.~Litjens, T.~Kooi, B.~E. Bejnordi, A.~A.~A. Setio, F.~Ciompi, M.~Ghafoorian,
  J.~A. Van Der~Laak, B.~Van~Ginneken, and C.~I. S{\'a}nchez, ``A survey on
  deep learning in medical image analysis,'' \emph{Medical image analysis},
  vol.~42, pp. 60--88, 2017.

\bibitem{kurth2018exascale}
T.~Kurth, S.~Treichler, J.~Romero, M.~Mudigonda, N.~Luehr, E.~Phillips
  \emph{et~al.}, ``Exascale deep learning for climate analytics,'' in
  \emph{SC18: International Conference for High Performance Computing,
  Networking, Storage and Analysis}.\hskip 1em plus 0.5em minus 0.4em\relax
  IEEE, 2018, pp. 649--660.

\bibitem{rezk2020recurrent}
N.~M. Rezk, M.~Purnaprajna, T.~Nordstr{\"o}m, and Z.~Ul-Abdin, ``Recurrent
  neural networks: an embedded computing perspective,'' \emph{IEEE Access},
  vol.~8, pp. 57\,967--57\,996, 2020.

\bibitem{zhang2019deep}
C.~Zhang, P.~Patras, and H.~Haddadi, ``Deep learning in mobile and wireless
  networking: A survey,'' \emph{IEEE Communications Surveys \& Tutorials},
  2019.

\bibitem{banbury2020benchmarking}
C.~R. Banbury, V.~J. Reddi, M.~Lam, W.~Fu, A.~Fazel, J.~Holleman \emph{et~al.},
  ``Benchmarking tinyml systems: Challenges and direction,'' \emph{arXiv
  preprint arXiv:2003.04821}, 2020.

\bibitem{dean20201}
J.~Dean, ``1.1 the deep learning revolution and its implications for computer
  architecture and chip design,'' in \emph{2020 IEEE International Solid-State
  Circuits Conference-(ISSCC)}.\hskip 1em plus 0.5em minus 0.4em\relax IEEE,
  2020, pp. 8--14.

\bibitem{olukotun2018designing}
K.~Olukotun, ``Designing computer systems for software 2.0,'' in
  \emph{Presentation at 2018 Conference on Neural Information Processing
  Systems}, 2018.

\bibitem{chen2016eyeriss}
Y.-H. Chen, T.~Krishna, J.~S. Emer, and V.~Sze, ``Eyeriss: An energy-efficient
  reconfigurable accelerator for deep convolutional neural networks,''
  \emph{IEEE Journal of Solid-State Circuits}, vol.~52, no.~1, 2016.

\bibitem{fleischer2018scalable}
B.~Fleischer, S.~Shukla, M.~Ziegler, J.~Silberman, J.~Oh, V.~Srinivasan,
  J.~Choi, S.~Mueller, A.~Agrawal, T.~Babinsky \emph{et~al.}, ``A scalable
  multi-teraops deep learning processor core for ai trainina and inference,''
  in \emph{2018 IEEE Symposium on VLSI Circuits}.\hskip 1em plus 0.5em minus
  0.4em\relax IEEE, 2018, pp. 35--36.

\bibitem{jouppi2017datacenter}
N.~P. Jouppi, C.~Young, N.~Patil, D.~Patterson, G.~Agrawal, R.~Bajwa
  \emph{et~al.}, ``In-datacenter performance analysis of a tensor processing
  unit,'' in \emph{2017 ACM/IEEE 44th Annual International Symposium on
  Computer Architecture (ISCA)}.\hskip 1em plus 0.5em minus 0.4em\relax IEEE,
  2017, pp. 1--12.

\bibitem{yang2018dnn}
X.~Yang, M.~Gao, J.~Pu, A.~Nayak, Q.~Liu, S.~E. Bell, J.~O. Setter, K.~Cao,
  H.~Ha, C.~Kozyrakis \emph{et~al.}, ``Dnn dataflow choice is overrated,''
  \emph{arXiv preprint arXiv:1809.04070}, 2018.

\bibitem{openaiblog}
D.~Amodei, D.~Hernandez, G.~Sastry, J.~Clark, G.~Brockman, and I.~Sutskever,
  ``Ai and compute,'' \url{https://openai.com/blog/ai-and-compute/}, May 2018.

\bibitem{han2015learning}
S.~Han, J.~Pool, J.~Tran, and W.~Dally, ``Learning both weights and connections
  for efficient neural network,'' in \emph{Advances in neural information
  processing systems}, 2015, pp. 1135--1143.

\bibitem{srivastava2014dropout}
N.~Srivastava, G.~Hinton, A.~Krizhevsky, I.~Sutskever, and R.~Salakhutdinov,
  ``Dropout: a simple way to prevent neural networks from overfitting,''
  \emph{The journal of machine learning research}, 2014.

\bibitem{han2015deep}
S.~Han, H.~Mao, and W.~J. Dally, ``Deep compression: Compressing deep neural
  networks with pruning, trained quantization and huffman coding,'' \emph{arXiv
  preprint arXiv:1510.00149}, 2015.

\bibitem{wen2016learning}
W.~Wen, C.~Wu, Y.~Wang, Y.~Chen, and H.~Li, ``Learning structured sparsity in
  deep neural networks,'' in \emph{Advances in neural information processing
  systems}, 2016, pp. 2074--2082.

\bibitem{frankle2018lottery}
J.~Frankle and M.~Carbin, ``The lottery ticket hypothesis: Finding sparse,
  trainable neural networks,'' in \emph{International Conference on Learning
  Representations}, 2018.

\bibitem{mishra2017fine}
A.~K. Mishra, E.~Nurvitadhi, G.~Venkatesh, J.~Pearce, and D.~Marr,
  ``Fine-grained accelerators for sparse machine learning workloads,'' in
  \emph{2017 22nd Asia and South Pacific Design Automation Conference
  (ASP-DAC)}.\hskip 1em plus 0.5em minus 0.4em\relax IEEE, 2017, pp. 635--640.

\bibitem{han2020survey}
B.~L. {Deng}, G.~{Li}, S.~{Han}, L.~{Shi}, and Y.~{Xie}, ``Model compression
  and hardware acceleration for neural networks: A comprehensive survey,''
  \emph{Proceedings of the IEEE}, vol. 108, no.~4, pp. 485--532, 2020.

\bibitem{sandler2018mobilenetv2}
M.~Sandler, A.~Howard, M.~Zhu, A.~Zhmoginov, and L.-C. Chen, ``Mobilenetv2:
  Inverted residuals and linear bottlenecks,'' in \emph{Proceedings of the IEEE
  Conference on Computer Vision and Pattern Recognition}, 2018, pp. 4510--4520.

\bibitem{iandola2016squeezenet}
F.~N. Iandola, S.~Han, M.~W. Moskewicz, K.~Ashraf, W.~J. Dally, and K.~Keutzer,
  ``Squeezenet: Alexnet-level accuracy with 50x fewer parameters and< 0.5 mb
  model size,'' \emph{arXiv:1602.07360}, 2016.

\bibitem{cichocki2015tensor}
A.~Cichocki, D.~Mandic, L.~De~Lathauwer, G.~Zhou, Q.~Zhao, C.~Caiafa, and H.~A.
  Phan, ``Tensor decompositions for signal processing applications: From
  two-way to multiway component analysis,'' \emph{IEEE signal processing
  magazine}, vol.~32, no.~2, pp. 145--163, 2015.

\bibitem{ragged_tensors}
\BIBentryALTinterwordspacing
L.~Moroney. (2018) Introducing ragged tensors. [Online]. Available:
  \url{https://blog.tensorflow.org/2018/12/introducing-ragged-tensors.html}
\BIBentrySTDinterwordspacing

\bibitem{krishnamoorthi2018quantizing}
R.~Krishnamoorthi, ``Quantizing deep convolutional networks for efficient
  inference: A whitepaper,'' \emph{arXiv preprint arXiv:1806.08342}, 2018.

\bibitem{zhou2018cambricon}
X.~Zhou, Z.~Du, Q.~Guo, S.~Liu, C.~Liu, C.~Wang, X.~Zhou, L.~Li, T.~Chen, and
  Y.~Chen, ``Cambricon-s: Addressing irregularity in sparse neural networks
  through a cooperative software/hardware approach,'' in \emph{2018 51st Annual
  IEEE/ACM International Symposium on Microarchitecture (MICRO)}.\hskip 1em
  plus 0.5em minus 0.4em\relax IEEE, 2018, pp. 15--28.

\bibitem{ren2019admm}
A.~Ren, T.~Zhang, S.~Ye, J.~Li, W.~Xu, X.~Qian, X.~Lin, and Y.~Wang, ``Admm-nn:
  An algorithm-hardware co-design framework of dnns using alternating direction
  methods of multipliers,'' in \emph{Proceedings of the Twenty-Fourth
  International Conference on Architectural Support for Programming Languages
  and Operating Systems}, 2019, pp. 925--938.

\bibitem{narang2017exploring}
S.~Narang, E.~Elsen, G.~Diamos, and S.~Sengupta, ``Exploring sparsity in
  recurrent neural networks,'' \emph{arXiv preprint arXiv:1704.05119}, 2017.

\bibitem{yang2017designing}
T.-J. Yang, Y.-H. Chen, and V.~Sze, ``Designing energy-efficient convolutional
  neural networks using energy-aware pruning,'' in \emph{Proceedings of the
  IEEE Conference on Computer Vision and Pattern Recognition}, 2017, pp.
  5687--5695.

\bibitem{zhang2016cambricon}
S.~Zhang, Z.~Du, L.~Zhang, H.~Lan, S.~Liu, L.~Li, Q.~Guo, T.~Chen, and Y.~Chen,
  ``Cambricon-x: An accelerator for sparse neural networks,'' in \emph{The 49th
  Annual IEEE/ACM International Symposium on Microarchitecture}.\hskip 1em plus
  0.5em minus 0.4em\relax IEEE Press, 2016, p.~20.

\bibitem{han2016eie}
S.~Han, X.~Liu, H.~Mao, J.~Pu, A.~Pedram, M.~A. Horowitz, and W.~J. Dally,
  ``Eie: efficient inference engine on compressed deep neural network,'' in
  \emph{2016 ACM/IEEE 43rd Annual International Symposium on Computer
  Architecture (ISCA)}.\hskip 1em plus 0.5em minus 0.4em\relax IEEE, 2016, pp.
  243--254.

\bibitem{chen2019eyeriss}
Y.-H. Chen, T.-J. Yang, J.~Emer, and V.~Sze, ``Eyeriss v2: A flexible
  accelerator for emerging deep neural networks on mobile devices,'' \emph{IEEE
  Journal on Emerging and Selected Topics in Circuits and Systems}, vol.~9,
  no.~2, pp. 292--308, 2019.

\bibitem{sze2017efficient}
V.~Sze, Y.-H. Chen, T.-J. Yang, and J.~S. Emer, ``Efficient processing of deep
  neural networks: A tutorial and survey,'' \emph{Proceedings of the IEEE},
  vol. 105, no.~12, pp. 2295--2329, 2017.

\bibitem{pouyanfar2018survey}
S.~Pouyanfar, S.~Sadiq, Y.~Yan, H.~Tian, Y.~Tao, M.~P. Reyes \emph{et~al.}, ``A
  survey on deep learning: Algorithms, techniques, and applications,''
  \emph{ACM Computing Surveys (CSUR)}, vol.~51, no.~5, pp. 1--36, 2018.

\bibitem{alom2019state}
M.~Z. Alom, T.~M. Taha, C.~Yakopcic, S.~Westberg, P.~Sidike, M.~S. Nasrin
  \emph{et~al.}, ``A state-of-the-art survey on deep learning theory and
  architectures,'' \emph{Electronics}, vol.~8, no.~3, p. 292, 2019.

\bibitem{hamilton2017representation}
W.~L. Hamilton, R.~Ying, and J.~Leskovec, ``Representation learning on graphs:
  Methods and applications,'' \emph{arXiv:1709.05584}, 2017.

\bibitem{paszke2019pytorch}
A.~Paszke, S.~Gross, F.~Massa, A.~Lerer, J.~Bradbury, G.~Chanan, T.~Killeen,
  Z.~Lin, N.~Gimelshein, L.~Antiga \emph{et~al.}, ``Pytorch: An imperative
  style, high-performance deep learning library,'' in \emph{Advances in Neural
  Information Processing Systems}, 2019, pp. 8024--8035.

\bibitem{abadi2016tensorflow}
M.~Abadi, P.~Barham, J.~Chen, Z.~Chen, A.~Davis, J.~Dean, M.~Devin,
  S.~Ghemawat, G.~Irving, M.~Isard \emph{et~al.}, ``Tensorflow: A system for
  large-scale machine learning,'' in \emph{12th $\{$USENIX$\}$ Symposium on
  Operating Systems Design and Implementation ($\{$OSDI$\}$ 16)}, 2016.

\bibitem{wang2014intel}
E.~Wang, Q.~Zhang, B.~Shen, G.~Zhang, X.~Lu, Q.~Wu, and Y.~Wang, ``Intel math
  kernel library,'' in \emph{High-Performance Computing on the
  Intel{\textregistered} Xeon Phi™}.\hskip 1em plus 0.5em minus 0.4em\relax
  Springer, 2014, pp. 167--188.

\bibitem{chen2014dadiannao}
Y.~Chen, T.~Luo, S.~Liu, S.~Zhang, L.~He, J.~Wang, L.~Li, T.~Chen, Z.~Xu,
  N.~Sun \emph{et~al.}, ``Dadiannao: A machine-learning supercomputer,'' in
  \emph{Proceedings of the 47th Annual IEEE/ACM International Symposium on
  Microarchitecture}.\hskip 1em plus 0.5em minus 0.4em\relax IEEE Computer
  Society, 2014, pp. 609--622.

\bibitem{xDNN}
K.~Khan, ``Xilinx dnn processor (xdnn), accelerating ai in datacenters,'' 2018.

\bibitem{fowers2018configurable}
J.~Fowers, K.~Ovtcharov, M.~Papamichael, T.~Massengill, M.~Liu, D.~Lo
  \emph{et~al.}, ``A configurable cloud-scale dnn processor for real-time ai,''
  in \emph{2018 ACM/IEEE 45th Annual International Symposium on Computer
  Architecture (ISCA)}.\hskip 1em plus 0.5em minus 0.4em\relax IEEE, 2018, pp.
  1--14.

\bibitem{suda2016throughput}
N.~Suda, V.~Chandra, G.~Dasika, A.~Mohanty, Y.~Ma, S.~Vrudhula, J.-s. Seo, and
  Y.~Cao, ``Throughput-optimized opencl-based fpga accelerator for large-scale
  convolutional neural networks,'' in \emph{Proceedings of the 2016 ACM/SIGDA
  International Symposium on Field-Programmable Gate Arrays}.\hskip 1em plus
  0.5em minus 0.4em\relax ACM, 2016, pp. 16--25.

\bibitem{fleming2019apparatus}
K.~E. Fleming, K.~D. Glossop, and S.~C. Steely, ``Apparatus, methods, and
  systems with a configurable spatial accelerator,'' Oct.~15 2019, uS Patent
  10,445,250.

\bibitem{dave2019dmazerunner}
S.~Dave, Y.~Kim, S.~Avancha, K.~Lee, and A.~Shrivastava, ``Dmazerunner:
  Executing perfectly nested loops on dataflow accelerators,'' \emph{ACM
  Transactions on Embedded Computing Systems (TECS)}, vol.~18, no.~5s, pp.
  1--27, 2019.

\bibitem{kwon2019understanding}
H.~Kwon, P.~Chatarasi, M.~Pellauer, A.~Parashar, V.~Sarkar, and T.~Krishna,
  ``Understanding reuse, performance, and hardware cost of dnn dataflow: A
  data-centric approach,'' in \emph{Proceedings of the 52nd Annual IEEE/ACM
  International Symposium on Microarchitecture}, 2019, pp. 754--768.

\bibitem{srivastava2019joint}
G.~Srivastava, D.~Kadetotad, S.~Yin, V.~Berisha, C.~Chakrabarti, and J.-s. Seo,
  ``Joint optimization of quantization and structured sparsity for compressed
  deep neural networks,'' in \emph{ICASSP 2019-2019 IEEE International
  Conference on Acoustics, Speech and Signal Processing (ICASSP)}.\hskip 1em
  plus 0.5em minus 0.4em\relax IEEE, 2019, pp. 1393--1397.

\bibitem{yu2017scalpel}
J.~Yu, A.~Lukefahr, D.~Palframan, G.~Dasika, R.~Das, and S.~Mahlke, ``Scalpel:
  Customizing dnn pruning to the underlying hardware parallelism,'' \emph{ACM
  SIGARCH Computer Architecture News}, 2017.

\bibitem{kang2019accelerator}
H.-J. Kang, ``Accelerator-aware pruning for convolutional neural networks,''
  \emph{IEEE Transactions on Circuits and Systems for Video Technology}, 2019.

\bibitem{ampere}
R.~Krashinsky, O.~Giroux, S.~Jones, N.~Stam, and S.~Ramaswamy, ``Nvidia ampere
  architecture in-depth,''
  \url{https://devblogs.nvidia.com/nvidia-ampere-architecture-in-depth/}, 2020.

\bibitem{liu2020systolic}
Z.-G. Liu, P.~N. Whatmough, and M.~Mattina, ``Systolic tensor array: An
  efficient structured-sparse gemm accelerator for mobile cnn inference,''
  \emph{IEEE Computer Architecture Letters}, vol.~19, no.~1, 2020.

\bibitem{vooturi2018hierarchical}
D.~T. Vooturi, D.~Mudigree, and S.~Avancha, ``Hierarchical block sparse neural
  networks,'' \emph{arXiv preprint arXiv:1808.03420}, 2018.

\bibitem{cao2019seernet}
S.~Cao, L.~Ma, W.~Xiao, C.~Zhang, Y.~Liu, L.~Zhang, L.~Nie, and Z.~Yang,
  ``Seernet: Predicting convolutional neural network feature-map sparsity
  through low-bit quantization,'' in \emph{Proceedings of the IEEE Conference
  on Computer Vision and Pattern Recognition}, 2019.

\bibitem{li2019squeezeflow}
J.~Li, S.~Jiang, S.~Gong, J.~Wu, J.~Yan, G.~Yan, and X.~Li, ``Squeezeflow: A
  sparse cnn accelerator exploiting concise convolution rules,'' \emph{IEEE
  Transactions on Computers}, vol.~68, no.~11, pp. 1663--1677, 2019.

\bibitem{lee20197}
J.~Lee, J.~Lee, D.~Han, J.~Lee, G.~Park, and H.-J. Yoo, ``7.7 lnpu: A 25.3
  tflops/w sparse deep-neural-network learning processor with fine-grained
  mixed precision of fp8-fp16,'' in \emph{2019 IEEE International Solid-State
  Circuits Conference-(ISSCC)}.\hskip 1em plus 0.5em minus 0.4em\relax IEEE,
  2019, pp. 142--144.

\bibitem{elsen2020fast}
E.~Elsen, M.~Dukhan, T.~Gale, and K.~Simonyan, ``Fast sparse convnets,'' in
  \emph{Proceedings of the IEEE/CVF conference on computer vision and pattern
  recognition}, 2020, pp. 14\,629--14\,638.

\bibitem{han2017ese}
S.~Han, J.~Kang, H.~Mao, Y.~Hu, X.~Li, Y.~Li, D.~Xie, H.~Luo, S.~Yao, Y.~Wang
  \emph{et~al.}, ``Ese: Efficient speech recognition engine with sparse lstm on
  fpga,'' in \emph{Proceedings of the 2017 ACM/SIGDA International Symposium on
  Field-Programmable Gate Arrays}, 2017, pp. 75--84.

\bibitem{zhu2017prune}
M.~Zhu and S.~Gupta, ``To prune, or not to prune: exploring the efficacy of
  pruning for model compression,'' \emph{arXiv:1710.01878}, 2017.

\bibitem{gale2019state}
T.~Gale, E.~Elsen, and S.~Hooker, ``The state of sparsity in deep neural
  networks,'' \emph{arXiv preprint arXiv:1902.09574}, 2019.

\bibitem{huggingface}
T.~Wolf, L.~Debut, V.~Sanh, J.~Chaumond, C.~Delangue, A.~Moi \emph{et~al.},
  ``Transformers: State-of-the-art natural language processing,'' in
  \emph{Proceedings of the 2020 Conference on Empirical Methods in Natural
  Language Processing: System Demonstrations}.\hskip 1em plus 0.5em minus
  0.4em\relax Association for Computational Linguistics, Oct. 2020, pp. 38--45.

\bibitem{albericio2016cnvlutin}
J.~Albericio, P.~Judd, T.~Hetherington, T.~Aamodt, N.~E. Jerger, and
  A.~Moshovos, ``Cnvlutin: Ineffectual-neuron-free deep neural network
  computing,'' \emph{ACM SIGARCH Computer Architecture News}, 2016.

\bibitem{yang2019dasnet}
Q.~Yang, J.~Mao, Z.~Wang, and H.~Li, ``Dasnet: Dynamic activation sparsity for
  neural network efficiency improvement,'' \emph{arXiv preprint
  arXiv:1909.06964}, 2019.

\bibitem{reagen2016minerva}
B.~Reagen, P.~Whatmough, R.~Adolf, S.~Rama, H.~Lee, S.~K. Lee, J.~M.
  Hern{\'a}ndez-Lobato, G.-Y. Wei, and D.~Brooks, ``Minerva: Enabling
  low-power, highly-accurate deep neural network accelerators,'' in \emph{2016
  ACM/IEEE 43rd Annual International Symposium on Computer Architecture
  (ISCA)}.\hskip 1em plus 0.5em minus 0.4em\relax IEEE, 2016, pp. 267--278.

\bibitem{georgiadis2019accelerating}
G.~Georgiadis, ``Accelerating convolutional neural networks via activation map
  compression,'' in \emph{Proceedings of the IEEE Conference on Computer Vision
  and Pattern Recognition}, 2019, pp. 7085--7095.

\bibitem{shi2017speeding}
S.~Shi and X.~Chu, ``Speeding up convolutional neural networks by exploiting
  the sparsity of rectifier units,'' \emph{arXiv preprint arXiv:1704.07724},
  2017.

\bibitem{gupta2019masr}
U.~Gupta, B.~Reagen, L.~Pentecost, M.~Donato, T.~Tambe, A.~M. Rush, G.-Y. Wei,
  and D.~Brooks, ``Masr: A modular accelerator for sparse rnns,'' in \emph{2019
  28th International Conference on Parallel Architectures and Compilation
  Techniques (PACT)}.\hskip 1em plus 0.5em minus 0.4em\relax IEEE, 2019, pp.
  1--14.

\bibitem{wang2020spatten}
H.~Wang, Z.~Zhang, and S.~Han, ``Spatten: Efficient sparse attention
  architecture with cascade token and head pruning,'' \emph{arXiv preprint
  arXiv:2012.09852}, 2020.

\bibitem{yazdanbakhsh2018ganax}
A.~Yazdanbakhsh, K.~Samadi, N.~S. Kim, and H.~Esmaeilzadeh, ``Ganax: A unified
  mimd-simd acceleration for generative adversarial networks,'' in
  \emph{Proceedings of the 45th Annual International Symposium on Computer
  Architecture}.\hskip 1em plus 0.5em minus 0.4em\relax IEEE Press, 2018, pp.
  650--661.

\bibitem{radford2015unsupervised}
A.~Radford, L.~Metz, and S.~Chintala, ``Unsupervised representation learning
  with deep convolutional generative adversarial networks,'' \emph{arXiv
  preprint arXiv:1511.06434}, 2015.

\bibitem{acorns}
X.~F. Xiao~Dong, Lei~Liu, ``Acorns: A framework for accelerating deep neural
  networks with input sparsity,'' in \emph{Proceedings of the 2019
  International Conference on Parallel Architecture and Compilation (PACT)},
  ser. PACT '19, 2019.

\bibitem{ren2018sbnet}
M.~Ren, A.~Pokrovsky, B.~Yang, and R.~Urtasun, ``Sbnet: Sparse blocks network
  for fast inference,'' in \emph{Proceedings of the IEEE Conference on Computer
  Vision and Pattern Recognition}, 2018, pp. 8711--8720.

\bibitem{engelcke2017vote3deep}
M.~Engelcke, D.~Rao, D.~Z. Wang, C.~H. Tong, and I.~Posner, ``Vote3deep: Fast
  object detection in 3d point clouds using efficient convolutional neural
  networks,'' in \emph{2017 IEEE International Conference on Robotics and
  Automation (ICRA)}.\hskip 1em plus 0.5em minus 0.4em\relax IEEE, 2017, pp.
  1355--1361.

\bibitem{lin2018deep}
Y.~Lin, S.~Han, H.~Mao, Y.~Wang, and B.~Dally, ``Deep gradient compression:
  Reducing the communication bandwidth for distributed training,'' in
  \emph{International Conference on Learning Representations}, 2018.

\bibitem{gupta2020fast}
V.~Gupta, D.~Choudhary, P.~T.~P. Tang, X.~Wei, X.~Wang, Y.~Huang, A.~Kejariwal,
  K.~Ramchandran, and M.~W. Mahoney, ``Fast distributed training of deep neural
  networks: Dynamic communication thresholding for model and data
  parallelism,'' \emph{arXiv:2010.08899}, 2020.

\bibitem{he2017neural}
X.~He, L.~Liao, H.~Zhang, L.~Nie, X.~Hu, and T.-S. Chua, ``Neural collaborative
  filtering,'' in \emph{Proceedings of the 26th international conference on
  world wide web}, 2017, pp. 173--182.

\bibitem{acun2020understanding}
B.~Acun, M.~Murphy, X.~Wang, J.~Nie, C.-J. Wu, and K.~Hazelwood,
  ``Understanding training efficiency of deep learning recommendation models at
  scale,'' \emph{arXiv preprint arXiv:2011.05497}, 2020.

\bibitem{hygcn}
M.~{Yan}, L.~{Deng}, X.~{Hu}, L.~{Liang}, Y.~{Feng}, X.~{Ye}, Z.~{Zhang},
  D.~{Fan}, and Y.~{Xie}, ``Hygcn: A gcn accelerator with hybrid
  architecture,'' in \emph{2020 IEEE International Symposium on High
  Performance Computer Architecture (HPCA)}, 2020, pp. 15--29.

\bibitem{geng2020awb}
T.~Geng, A.~Li, R.~Shi, C.~Wu, T.~Wang, Y.~Li, P.~Haghi, A.~Tumeo, S.~Che,
  S.~Reinhardt \emph{et~al.}, ``Awb-gcn: A graph convolutional network
  accelerator with runtime workload rebalancing,'' in \emph{2020 53rd Annual
  IEEE/ACM International Symposium on Microarchitecture (MICRO)}.\hskip 1em
  plus 0.5em minus 0.4em\relax IEEE, 2020, pp. 922--936.

\bibitem{zeng2020graphact}
H.~Zeng and V.~Prasanna, ``Graphact: Accelerating gcn training on cpu-fpga
  heterogeneous platforms,'' in \emph{Proceedings of the 2020 ACM/SIGDA
  International Symposium on Field-Programmable Gate Arrays}, 2020, pp.
  255--265.

\bibitem{liang2020engn}
S.~Liang, Y.~Wang, C.~Liu, L.~He, L.~Huawei, D.~Xu, and X.~Li, ``Engn: A
  high-throughput and energy-efficient accelerator for large graph neural
  networks,'' \emph{IEEE Transactions on Computers}, 2020.

\bibitem{duff1977survey}
I.~S. Duff, ``A survey of sparse matrix research,'' \emph{Proceedings of the
  IEEE}, vol.~65, no.~4, pp. 500--535, 1977.

\bibitem{hegde2019extensor}
K.~Hegde, H.~Asghari-Moghaddam, M.~Pellauer, N.~Crago, A.~Jaleel, E.~Solomonik,
  J.~Emer, and C.~W. Fletcher, ``Extensor: An accelerator for sparse tensor
  algebra,'' in \emph{Proceedings of the 52nd Annual IEEE/ACM International
  Symposium on Microarchitecture}, 2019.

\bibitem{horowitz20141}
M.~Horowitz, ``1.1 computing's energy problem (and what we can do about it),''
  in \emph{2014 IEEE International Solid-State Circuits Conference Digest of
  Technical Papers (ISSCC)}.\hskip 1em plus 0.5em minus 0.4em\relax IEEE, 2014,
  pp. 10--14.

\bibitem{xie2017aggregated}
S.~Xie, R.~Girshick, P.~Doll{\'a}r, Z.~Tu, and K.~He, ``Aggregated residual
  transformations for deep neural networks,'' in \emph{Proceedings of the IEEE
  conference on computer vision and pattern recognition}, 2017.

\bibitem{szegedy2015going}
C.~Szegedy, W.~Liu, Y.~Jia, P.~Sermanet, S.~Reed, D.~Anguelov, D.~Erhan,
  V.~Vanhoucke, and A.~Rabinovich, ``Going deeper with convolutions,'' in
  \emph{Proceedings of the IEEE conference on computer vision and pattern
  recognition}, 2015, pp. 1--9.

\bibitem{szegedy2016rethinking}
C.~Szegedy, V.~Vanhoucke, S.~Ioffe, J.~Shlens, and Z.~Wojna, ``Rethinking the
  inception architecture for computer vision,'' in \emph{Proceedings of the
  IEEE conference on computer vision and pattern recognition}, 2016.

\bibitem{hegde2018ucnn}
K.~Hegde, J.~Yu, R.~Agrawal, M.~Yan, M.~Pellauer, and C.~Fletcher, ``Ucnn:
  Exploiting computational reuse in deep neural networks via weight
  repetition,'' in \emph{2018 ACM/IEEE 45th Annual International Symposium on
  Computer Architecture (ISCA)}, 2018, pp. 674--687.

\bibitem{riera2018computation}
M.~Riera, J.-M. Arnau, and A.~Gonz{\'a}lez, ``Computation reuse in dnns by
  exploiting input similarity,'' in \emph{2018 ACM/IEEE 45th Annual
  International Symposium on Computer Architecture (ISCA)}, 2018.

\bibitem{quantization8bitPeter}
P.~Warden, ``Why are eight bits enough for deep neural networks?''
  \url{https://petewarden.com/2015/05/23/why-are-eight-bits-enough-for-deep-neural-networks/},
  2015.

\bibitem{kalamkar2019study}
D.~Kalamkar, D.~Mudigere, N.~Mellempudi, D.~Das, K.~Banerjee, S.~Avancha, D.~T.
  Vooturi, N.~Jammalamadaka, J.~Huang, H.~Yuen \emph{et~al.}, ``A study of
  bfloat16 for deep learning training,'' \emph{arXiv preprint
  arXiv:1905.12322}, 2019.

\bibitem{hubara2017quantized}
I.~Hubara, M.~Courbariaux, D.~Soudry, R.~El-Yaniv, and Y.~Bengio, ``Quantized
  neural networks: Training neural networks with low precision weights and
  activations,'' \emph{The Journal of Machine Learning Research}, vol.~18,
  no.~1, pp. 6869--6898, 2017.

\bibitem{lee2017lognet}
E.~H. Lee, D.~Miyashita, E.~Chai, B.~Murmann, and S.~S. Wong, ``Lognet:
  Energy-efficient neural networks using logarithmic computation,'' in
  \emph{2017 IEEE International Conference on Acoustics, Speech and Signal
  Processing (ICASSP)}.\hskip 1em plus 0.5em minus 0.4em\relax IEEE, 2017, pp.
  5900--5904.

\bibitem{chen2014diannao}
T.~Chen, Z.~Du, N.~Sun, J.~Wang, C.~Wu, Y.~Chen, and O.~Temam, ``Diannao: A
  small-footprint high-throughput accelerator for ubiquitous
  machine-learning,'' in \emph{ACM Sigplan Notices}, vol.~49, no.~4, 2014.

\bibitem{qin2020sigma}
E.~Qin, A.~Samajdar, H.~Kwon, V.~Nadella, S.~Srinivasan, D.~Das, B.~Kaul, and
  T.~Krishna, ``Sigma: A sparse and irregular gemm accelerator with flexible
  interconnects for dnn training,'' in \emph{2020 IEEE International Symposium
  on High Performance Computer Architecture (HPCA)}.\hskip 1em plus 0.5em minus
  0.4em\relax IEEE, 2020, pp. 58--70.

\bibitem{adelman2018faster}
M.~Adelman and M.~Silberstein, ``Faster neural network training with
  approximate tensor operations,'' \emph{arXiv:1805.08079}, 2018.

\bibitem{zhang2019snap}
J.-F. Zhang, C.-E. Lee, C.~Liu, Y.~S. Shao, S.~W. Keckler, and Z.~Zhang,
  ``Snap: A 1.67—21.55 tops/w sparse neural acceleration processor for
  unstructured sparse deep neural network inference in 16nm cmos,'' in
  \emph{2019 Symposium on VLSI Circuits}.\hskip 1em plus 0.5em minus
  0.4em\relax IEEE, 2019, pp. C306--C307.

\bibitem{albericio2017bit}
J.~Albericio, A.~Delm{\'a}s, P.~Judd, S.~Sharify, G.~O'Leary, R.~Genov, and
  A.~Moshovos, ``Bit-pragmatic deep neural network computing,'' in
  \emph{Proceedings of the 50th Annual IEEE/ACM International Symposium on
  Microarchitecture}, 2017, pp. 382--394.

\bibitem{moons201714}
B.~Moons, R.~Uytterhoeven, W.~Dehaene, and M.~Verhelst, ``14.5 envision: A
  0.26-to-10tops/w subword-parallel dynamic-voltage-accuracy-frequency-scalable
  convolutional neural network processor in 28nm fdsoi,'' in \emph{2017 IEEE
  International Solid-State Circuits Conference (ISSCC)}.\hskip 1em plus 0.5em
  minus 0.4em\relax IEEE, 2017, pp. 246--247.

\bibitem{dadu2019towards}
V.~Dadu, J.~Weng, S.~Liu, and T.~Nowatzki, ``Towards general purpose
  acceleration by exploiting common data-dependence forms,'' in
  \emph{Proceedings of the 52nd Annual IEEE/ACM International Symposium on
  Microarchitecture}, 2019, pp. 924--939.

\bibitem{zhang2019compact}
J.~J. Zhang, P.~Raj, S.~Zarar, A.~Ambardekar, and S.~Garg, ``Compact: On-chip
  compression of activations for low power systolic array based cnn
  acceleration,'' \emph{ACM Transactions on Embedded Computing Systems (TECS)},
  vol.~18, no.~5s, p.~47, 2019.

\bibitem{judd2017cnvlutin2}
P.~Judd, A.~Delmas, S.~Sharify, and A.~Moshovos, ``Cnvlutin2:
  Ineffectual-activation-and-weight-free deep neural network computing,''
  \emph{arXiv preprint arXiv:1705.00125}, 2017.

\bibitem{zheng2018kernelxform}
S.~Zheng, Y.~Liu, S.~Yin, L.~Liu, and S.~Wei, ``An efficient kernel
  transformation architecture for binary-and ternary-weight neural network
  inference,'' in \emph{Proceedings of the 55th Annual Design Automation
  Conference}.\hskip 1em plus 0.5em minus 0.4em\relax ACM, 2018, p. 137.

\bibitem{struharik2018conna}
R.~Struharik, B.~Vukobratovi{\'c}, A.~Erdeljan, and D.~Rakanovi{\'c},
  ``Conna--compressed cnn hardware accelerator,'' in \emph{2018 21st Euromicro
  Conference on Digital System Design (DSD)}.\hskip 1em plus 0.5em minus
  0.4em\relax IEEE, 2018, pp. 365--372.

\bibitem{parashar2017scnn}
A.~Parashar, M.~Rhu, A.~Mukkara, A.~Puglielli, R.~Venkatesan, B.~Khailany,
  J.~Emer, S.~W. Keckler, and W.~J. Dally, ``Scnn: An accelerator for
  compressed-sparse convolutional neural networks,'' in \emph{2017 ACM/IEEE
  44th Annual International Symposium on Computer Architecture (ISCA)}.\hskip
  1em plus 0.5em minus 0.4em\relax IEEE, 2017, pp. 27--40.

\bibitem{gondimalla2019sparten}
A.~Gondimalla, N.~Chesnut, M.~Thottethodi, and T.~Vijaykumar, ``Sparten: A
  sparse tensor accelerator for convolutional neural networks,'' in
  \emph{Proceedings of the 52nd Annual IEEE/ACM International Symposium on
  Microarchitecture}.\hskip 1em plus 0.5em minus 0.4em\relax ACM, 2019, pp.
  151--165.

\bibitem{yuan2018sticker}
Z.~Yuan, J.~Yue, H.~Yang, Z.~Wang, J.~Li, Y.~Yang, Q.~Guo, X.~Li, M.-F. Chang,
  H.~Yang \emph{et~al.}, ``Sticker: A 0.41-62.1 tops/w 8bit neural network
  processor with multi-sparsity compatible convolution arrays and online tuning
  acceleration for fully connected layers,'' in \emph{2018 IEEE Symposium on
  VLSI Circuits}.\hskip 1em plus 0.5em minus 0.4em\relax IEEE, 2018, pp.
  33--34.

\bibitem{choles2018parsecore}
S.~Chole, R.~Tadishetti, and S.~Reddy, ``Sparsecore: An accelerator for
  structurally sparse cnns,'' in \emph{SysML Conference}, 2018.

\bibitem{yavits2017accelerator}
L.~Yavits and R.~Ginosar, ``Accelerator for sparse machine learning,''
  \emph{IEEE Computer Architecture Letters}, vol.~17, no.~1, pp. 21--24, 2017.

\bibitem{nvdla}
{NVIDIA Corporation}, ``Nvidia deep learning accelerator (nvdla),''
  \url{http://nvdla.org}, accessed: 2018-11-05.

\bibitem{aimar2018nullhop}
A.~Aimar, H.~Mostafa, E.~Calabrese, A.~Rios-Navarro, R.~Tapiador-Morales, I.-A.
  Lungu, M.~B. Milde, F.~Corradi, A.~Linares-Barranco, S.-C. Liu \emph{et~al.},
  ``Nullhop: A flexible convolutional neural network accelerator based on
  sparse representations of feature maps,'' \emph{IEEE transactions on neural
  networks and learning systems}, 2018.

\bibitem{page2017sparcnet}
A.~Page, A.~Jafari, C.~Shea, and T.~Mohsenin, ``Sparcnet: A hardware
  accelerator for efficient deployment of sparse convolutional networks,''
  \emph{ACM Journal on Emerging Technologies in Computing Systems (JETC)},
  vol.~13, no.~3, pp. 1--32, 2017.

\bibitem{lu2018spwa}
L.~Lu and Y.~Liang, ``Spwa: an efficient sparse winograd convolutional neural
  networks accelerator on fpgas,'' in \emph{2018 55th ACM/ESDA/IEEE Design
  Automation Conference (DAC)}.\hskip 1em plus 0.5em minus 0.4em\relax IEEE,
  2018, pp. 1--6.

\bibitem{lu2019efficient}
L.~Lu, J.~Xie, R.~Huang, J.~Zhang, W.~Lin, and Y.~Liang, ``An efficient
  hardware accelerator for sparse convolutional neural networks on fpgas,'' in
  \emph{2019 IEEE 27th Annual International Symposium on Field-Programmable
  Custom Computing Machines (FCCM)}, 2019.

\bibitem{gao2018deltarnn}
C.~Gao, D.~Neil, E.~Ceolini, S.-C. Liu, and T.~Delbruck, ``Deltarnn: A
  power-efficient recurrent neural network accelerator,'' in \emph{Proceedings
  of the 2018 ACM/SIGDA International Symposium on Field-Programmable Gate
  Arrays}, 2018, pp. 21--30.

\bibitem{asgari2019eridanus}
B.~Asgari, R.~Hadidi, H.~Kim, and S.~Yalamanchili, ``Eridanus: Efficiently
  running inference of dnns using systolic arrays,'' \emph{IEEE Micro},
  vol.~39, no.~5, pp. 46--54, 2019.

\bibitem{kung2018adaptive}
H.~Kung, B.~McDanel, and S.~Q. Zhang, ``Adaptive tiling: Applying fixed-size
  systolic arrays to sparse convolutional neural networks,'' in \emph{2018 24th
  International Conference on Pattern Recognition (ICPR)}.\hskip 1em plus 0.5em
  minus 0.4em\relax IEEE, 2018, pp. 1006--1011.

\bibitem{yin2017thinker}
S.~Yin, P.~Ouyang, S.~Tang, F.~Tu, X.~Li, S.~Zheng, T.~Lu, J.~Gu, L.~Liu, and
  S.~Wei, ``A high energy efficient reconfigurable hybrid neural network
  processor for deep learning applications,'' \emph{IEEE Journal of Solid-State
  Circuits}, vol.~53, no.~4, pp. 968--982, 2017.

\bibitem{lee2018stitch}
C.-E. Lee, Y.~S. Shao, J.-F. Zhang, A.~Parashar, J.~Emer, S.~W. Keckler
  \emph{et~al.}, ``Stitch-x: An accelerator architecture for exploiting
  unstructured sparsity in deep neural networks,'' in \emph{SysML Conference},
  2018.

\bibitem{kim2017zena}
D.~Kim, J.~Ahn, and S.~Yoo, ``Zena: Zero-aware neural network accelerator,''
  \emph{IEEE Design \& Test}, vol.~35, no.~1, pp. 39--46, 2017.

\bibitem{jang2019mnnfast}
H.~Jang, J.~Kim, J.-E. Jo, J.~Lee, and J.~Kim, ``Mnnfast: a fast and scalable
  system architecture for memory-augmented neural networks,'' in
  \emph{Proceedings of the 46th International Symposium on Computer
  Architecture}, 2019, pp. 250--263.

\bibitem{mcdanel2019full}
B.~McDanel, S.~Q. Zhang, H.~Kung, and X.~Dong, ``Full-stack optimization for
  accelerating cnns using powers-of-two weights with fpga validation,'' in
  \emph{Proceedings of the ACM International Conference on Supercomputing},
  2019, pp. 449--460.

\bibitem{whatmough2018dnn}
P.~N. Whatmough, S.~K. Lee, D.~Brooks, and G.-Y. Wei, ``Dnn engine: A 28-nm
  timing-error tolerant sparse deep neural network processor for iot
  applications,'' \emph{IEEE Journal of Solid-State Circuits}, 2018.

\bibitem{venkatesh2017accelerating}
G.~Venkatesh, E.~Nurvitadhi, and D.~Marr, ``Accelerating deep convolutional
  networks using low-precision and sparsity,'' in \emph{2017 IEEE International
  Conference on Acoustics, Speech and Signal Processing (ICASSP)}.\hskip 1em
  plus 0.5em minus 0.4em\relax IEEE, 2017, pp. 2861--2865.

\bibitem{ham20203}
T.~J. Ham, S.~J. Jung, S.~Kim, Y.~H. Oh, Y.~Park, Y.~Song, J.-H. Park, S.~Lee,
  K.~Park, J.~W. Lee \emph{et~al.}, ``A$^{3}$: Accelerating attention
  mechanisms in neural networks with approximation,'' \emph{arXiv preprint
  arXiv:2002.10941}, 2020.

\bibitem{he2020sparse}
X.~He, S.~Pal, A.~Amarnath, S.~Feng, D.-H. Park, A.~Rovinski, H.~Ye, Y.~Chen,
  R.~Dreslinski, and T.~Mudge, ``Sparse-tpu: Adapting systolic arrays for
  sparse matrices,'' in \emph{Proceedings of the 34th ACM International
  Conference on Supercomputing}, 2020, pp. 1--12.

\bibitem{shi2020csb}
R.~Shi, P.~Dong, T.~Geng, Y.~Ding, X.~Ma, H.~K.-H. So, M.~Herbordt, A.~Li, and
  Y.~Wang, ``Csb-rnn: a faster-than-realtime rnn acceleration framework with
  compressed structured blocks,'' in \emph{Proceedings of the 34th ACM
  International Conference on Supercomputing}, 2020.

\bibitem{rajpurkar2016squad}
P.~Rajpurkar, J.~Zhang, K.~Lopyrev, and P.~Liang, ``Squad: 100,000+ questions
  for machine comprehension of text,'' in \emph{Proceedings of the 2016
  Conference on Empirical Methods in Natural Language Processing}, 2016, pp.
  2383--2392.

\bibitem{chou2018format}
S.~Chou, F.~Kjolstad, and S.~Amarasinghe, ``Format abstraction for sparse
  tensor algebra compilers,'' \emph{Proceedings of the ACM on Programming
  Languages}, vol.~2, no. OOPSLA, pp. 1--30, 2018.

\bibitem{frosttTensorFormat}
\BIBentryALTinterwordspacing
S.~Smith, J.~W. Choi, J.~Li, R.~Vuduc, J.~Park, X.~Liu, and G.~Karypis. (2017)
  Frostt file format. [Online]. Available:
  \url{http://frostt.io/tensors/file-formats.html}
\BIBentrySTDinterwordspacing

\bibitem{MatrixMarket}
N.~I. of~Standards and Technology, ``Matrix market exchange formats,''
  \url{https://math.nist.gov/MatrixMarket/formats.html}, 2013.

\bibitem{jones2001scipy}
E.~Jones, T.~Oliphant, and P.~Peterson, ``Scipy: Open source scientific tools
  for python,'' 2001.

\bibitem{saad1990sparskit}
Y.~Saad, ``Sparskit: A basic tool kit for sparse matrix computations,'' 1990.

\bibitem{buluc2008representation}
A.~Buluc and J.~R. Gilbert, ``On the representation and multiplication of
  hypersparse matrices,'' in \emph{2008 IEEE International Symposium on
  Parallel and Distributed Processing}.\hskip 1em plus 0.5em minus 0.4em\relax
  IEEE, 2008, pp. 1--11.

\bibitem{im1998model}
E.-J. Im and K.~Yelick, ``Model-based memory hierarchy optimizations for sparse
  matrices,'' in \emph{Workshop on Profile and Feedback-Directed Compilation},
  vol. 139, 1998.

\bibitem{smith2015tensor}
S.~Smith and G.~Karypis, ``Tensor-matrix products with a compressed sparse
  tensor,'' in \emph{Proceedings of the 5th Workshop on Irregular Applications:
  Architectures and Algorithms}, 2015, pp. 1--7.

\bibitem{tew2016investigation}
P.~A. Tew, ``An investigation of sparse tensor formats for tensor libraries,''
  Ph.D. dissertation, Massachusetts Institute of Technology, 2016.

\bibitem{simonyan2014very}
K.~Simonyan and A.~Zisserman, ``Very deep convolutional networks for
  large-scale image recognition,'' \emph{arXiv preprint arXiv:1409.1556}, 2014.

\bibitem{buckler2018eva2}
M.~Buckler, P.~Bedoukian, S.~Jayasuriya, and A.~Sampson, ``Eva$^2$: Exploiting
  temporal redundancy in live computer vision,'' in \emph{2018 ACM/IEEE 45th
  Annual International Symposium on Computer Architecture (ISCA)}.\hskip 1em
  plus 0.5em minus 0.4em\relax IEEE, 2018, pp. 533--546.

\bibitem{guo2017software}
K.~Guo, S.~Han, S.~Yao, Y.~Wang, Y.~Xie, and H.~Yang, ``Software-hardware
  codesign for efficient neural network acceleration,'' \emph{IEEE Micro},
  vol.~37, no.~2, pp. 18--25, 2017.

\bibitem{ma2019non}
X.~Ma, S.~Lin, S.~Ye, Z.~He, L.~Zhang, G.~Yuan, S.~H. Tan, Z.~Li, D.~Fan,
  X.~Qian \emph{et~al.}, ``Non-structured dnn weight pruning--is it beneficial
  in any platform?'' \emph{arXiv preprint arXiv:1907.02124}, 2019.

\bibitem{bulucc2009parallel}
A.~Bulu{\c{c}}, J.~T. Fineman, M.~Frigo, J.~R. Gilbert, and C.~E. Leiserson,
  ``Parallel sparse matrix-vector and matrix-transpose-vector multiplication
  using compressed sparse blocks,'' in \emph{Proceedings of the twenty-first
  annual symposium on Parallelism in algorithms and architectures}, 2009, pp.
  233--244.

\bibitem{chang2011libsvm}
C.-C. Chang and C.-J. Lin, ``Libsvm: A library for support vector machines,''
  \emph{ACM transactions on intelligent systems and technology (TIST)}, vol.~2,
  no.~3, pp. 1--27, 2011.

\bibitem{kincaid1984itpack}
D.~R. Kincaid and D.~M. Young, ``The itpack project: Past, present, and
  future,'' in \emph{Elliptic Problem Solvers}.\hskip 1em plus 0.5em minus
  0.4em\relax Elsevier, 1984, pp. 53--63.

\bibitem{saad2003iterative}
Y.~Saad, \emph{Iterative methods for sparse linear systems}.\hskip 1em plus
  0.5em minus 0.4em\relax siam, 2003.

\bibitem{king2016dynamic}
J.~King, T.~Gilray, R.~M. Kirby, and M.~Might, ``Dynamic sparse-matrix
  allocation on gpus,'' in \emph{International Conference on High Performance
  Computing}.\hskip 1em plus 0.5em minus 0.4em\relax Springer, 2016, pp.
  61--80.

\bibitem{willcock2006accelerating}
J.~Willcock and A.~Lumsdaine, ``Accelerating sparse matrix computations via
  data compression,'' in \emph{Proceedings of the 20th annual international
  conference on Supercomputing}, 2006, pp. 307--316.

\bibitem{baskaran2012efficient}
M.~Baskaran, B.~Meister, N.~Vasilache, and R.~Lethin, ``Efficient and scalable
  computations with sparse tensors,'' in \emph{2012 IEEE Conference on High
  Performance Extreme Computing}.\hskip 1em plus 0.5em minus 0.4em\relax IEEE,
  2012, pp. 1--6.

\bibitem{bader2008efficient}
B.~W. Bader and T.~G. Kolda, ``Efficient matlab computations with sparse and
  factored tensors,'' \emph{SIAM Journal on Scientific Computing}, vol.~30,
  no.~1, pp. 205--231, 2008.

\bibitem{vuduc2003automatic}
R.~W. Vuduc and J.~W. Demmel, \emph{Automatic performance tuning of sparse
  matrix kernels}.\hskip 1em plus 0.5em minus 0.4em\relax University of
  California, Berkeley, 2003, vol.~1.

\bibitem{hong2019adaptive}
C.~Hong, A.~Sukumaran-Rajam, I.~Nisa, K.~Singh, and P.~Sadayappan, ``Adaptive
  sparse tiling for sparse matrix multiplication,'' in \emph{Proceedings of the
  24th Symposium on Principles and Practice of Parallel Programming}.\hskip 1em
  plus 0.5em minus 0.4em\relax ACM, 2019, pp. 300--314.

\bibitem{TTB_Software}
\BIBentryALTinterwordspacing
B.~W. Bader, T.~G. Kolda \emph{et~al.}, ``Matlab tensor toolbox version 2.6,''
  Available online, February 2015. [Online]. Available:
  \url{http://www.sandia.gov/~tgkolda/TensorToolbox/}
\BIBentrySTDinterwordspacing

\bibitem{nvidiasparse}
\BIBentryALTinterwordspacing
{NVIDIA}. cusparse, the cuda sparse matrix library. [Online]. Available:
  \url{http://docs. nvidia. com/cuda/cusparse}
\BIBentrySTDinterwordspacing

\bibitem{yuan2019sticker}
Z.~Yuan, Y.~Liu, J.~Yue, Y.~Yang, J.~Wang, X.~Feng, J.~Zhao, X.~Li, and
  H.~Yang, ``Sticker: An energy-efficient multi-sparsity compatible accelerator
  for convolutional neural networks in 65-nm cmos,'' \emph{IEEE Journal of
  Solid-State Circuits}, 2019.

\bibitem{ding2017circnn}
C.~Ding, S.~Liao, Y.~Wang, Z.~Li, N.~Liu, Y.~Zhuo, C.~Wang, X.~Qian, Y.~Bai,
  G.~Yuan \emph{et~al.}, ``C ir cnn: accelerating and compressing deep neural
  networks using block-circulant weight matrices,'' in \emph{Proceedings of the
  50th Annual IEEE/ACM International Symposium on Microarchitecture}.\hskip 1em
  plus 0.5em minus 0.4em\relax ACM, 2017, pp. 395--408.

\bibitem{wang2018c}
S.~Wang, Z.~Li, C.~Ding, B.~Yuan, Q.~Qiu, Y.~Wang, and Y.~Liang, ``C-lstm:
  Enabling efficient lstm using structured compression techniques on fpgas,''
  in \emph{Proceedings of the 2018 ACM/SIGDA International Symposium on
  Field-Programmable Gate Arrays}, 2018, pp. 11--20.

\bibitem{yan2019alleviating}
M.~Yan, X.~Hu, S.~Li, A.~Basak, H.~Li, X.~Ma, I.~Akgun, Y.~Feng, P.~Gu, L.~Deng
  \emph{et~al.}, ``Alleviating irregularity in graph analytics acceleration: a
  hardware/software co-design approach,'' in \emph{Proceedings of the 52nd
  Annual IEEE/ACM International Symposium on Microarchitecture}, 2019, pp.
  615--628.

\bibitem{hegde2018morph}
K.~Hegde, R.~Agrawal, Y.~Yao, and C.~W. Fletcher, ``Morph: Flexible
  acceleration for 3d cnn-based video understanding,'' in \emph{2018 51st
  Annual IEEE/ACM International Symposium on Microarchitecture (MICRO)}.\hskip
  1em plus 0.5em minus 0.4em\relax IEEE, 2018, pp. 933--946.

\bibitem{kim2011high}
Y.~Kim, J.~Lee, A.~Shrivastava, J.~W. Yoon, D.~Cho, and Y.~Paek, ``High
  throughput data mapping for coarse-grained reconfigurable architectures,''
  \emph{IEEE Transactions on Computer-Aided Design of Integrated Circuits and
  Systems}, vol.~30, no.~11, pp. 1599--1609, 2011.

\bibitem{weste2015cmos}
N.~H. Weste and D.~Harris, \emph{CMOS VLSI design: a circuits and systems
  perspective}.\hskip 1em plus 0.5em minus 0.4em\relax Pearson Education India,
  2015.

\bibitem{shen2017maximizing}
Y.~Shen, M.~Ferdman, and P.~Milder, ``Maximizing cnn accelerator efficiency
  through resource partitioning,'' in \emph{2017 ACM/IEEE 44th Annual
  International Symposium on Computer Architecture (ISCA)}.\hskip 1em plus
  0.5em minus 0.4em\relax IEEE, 2017, pp. 535--547.

\bibitem{azizimazreah2019shortcut}
A.~Azizimazreah and L.~Chen, ``Shortcut mining: exploiting cross-layer shortcut
  reuse in dcnn accelerators,'' in \emph{2019 IEEE International Symposium on
  High Performance Computer Architecture (HPCA)}.\hskip 1em plus 0.5em minus
  0.4em\relax IEEE, 2019, pp. 94--105.

\bibitem{alwani2016fused}
M.~Alwani, H.~Chen, M.~Ferdman, and P.~Milder, ``Fused-layer cnn
  accelerators,'' in \emph{2016 49th Annual IEEE/ACM International Symposium on
  Microarchitecture (MICRO)}.\hskip 1em plus 0.5em minus 0.4em\relax IEEE,
  2016, pp. 1--12.

\bibitem{kwon2017rethinking}
H.~Kwon, A.~Samajdar, and T.~Krishna, ``Rethinking nocs for spatial neural
  network accelerators,'' in \emph{2017 Eleventh IEEE/ACM International
  Symposium on Networks-on-Chip (NOCS)}, 2017, pp. 1--8.

\bibitem{vainbrand2010network}
D.~Vainbrand and R.~Ginosar, ``Network-on-chip architectures for neural
  networks,'' in \emph{2010 Fourth ACM/IEEE International Symposium on
  Networks-on-Chip}.\hskip 1em plus 0.5em minus 0.4em\relax IEEE, 2010, pp.
  135--144.

\bibitem{xu2020autodnnchip}
P.~Xu, X.~Zhang, C.~Hao, Y.~Zhao, Y.~Zhang, Y.~Wang, C.~Li, Z.~Guan, D.~Chen,
  and Y.~Lin, ``Autodnnchip: An automated dnn chip predictor and builder for
  both fpgas and asics,'' in \emph{The 2020 ACM/SIGDA International Symposium
  on Field-Programmable Gate Arrays}, 2020.

\bibitem{kwon2018maeri}
H.~Kwon, A.~Samajdar, and T.~Krishna, ``Maeri: Enabling flexible dataflow
  mapping over dnn accelerators via reconfigurable interconnects,'' \emph{ACM
  SIGPLAN Notices}, vol.~53, no.~2, pp. 461--475, 2018.

\bibitem{lu2017flexflow}
W.~Lu, G.~Yan, J.~Li, S.~Gong, Y.~Han, and X.~Li, ``Flexflow: A flexible
  dataflow accelerator architecture for convolutional neural networks,'' in
  \emph{2017 IEEE International Symposium on High Performance Computer
  Architecture (HPCA)}.\hskip 1em plus 0.5em minus 0.4em\relax IEEE, 2017, pp.
  553--564.

\bibitem{dave2020dmazerunner}
S.~Dave, A.~Shrivastava, Y.~Kim, S.~Avancha, and K.~Lee, ``dmazerunner:
  Optimizing convolutions on dataflow accelerators,'' in \emph{ICASSP 2020-2020
  IEEE International Conference on Acoustics, Speech and Signal Processing
  (ICASSP)}.\hskip 1em plus 0.5em minus 0.4em\relax IEEE, 2020, pp. 1544--1548.

\bibitem{andri2017yodann}
R.~Andri, L.~Cavigelli, D.~Rossi, and L.~Benini, ``Yodann: An architecture for
  ultralow power binary-weight cnn acceleration,'' \emph{IEEE Transactions on
  Computer-Aided Design of Integrated Circuits and Systems}, vol.~37, no.~1,
  pp. 48--60, 2017.

\bibitem{tann2017hardware}
H.~Tann, S.~Hashemi, R.~I. Bahar, and S.~Reda, ``Hardware-software codesign of
  accurate, multiplier-free deep neural networks,'' in \emph{2017 54th
  ACM/EDAC/IEEE Design Automation Conference (DAC)}, 2017.

\bibitem{sharma2018bit}
H.~Sharma, J.~Park, N.~Suda, L.~Lai, B.~Chau, V.~Chandra, and H.~Esmaeilzadeh,
  ``Bit fusion: Bit-level dynamically composable architecture for accelerating
  deep neural networks,'' in \emph{Proceedings of the 45th Annual International
  Symposium on Computer Architecture}, 2018.

\bibitem{sharify2019laconic}
S.~Sharify, A.~D. Lascorz, M.~Mahmoud, M.~Nikolic, K.~Siu, D.~M. Stuart,
  Z.~Poulos, and A.~Moshovos, ``Laconic deep learning inference acceleration,''
  in \emph{Proceedings of the 46th International Symposium on Computer
  Architecture}.\hskip 1em plus 0.5em minus 0.4em\relax ACM, 2019, pp.
  304--317.

\bibitem{sharma2016high}
H.~Sharma, J.~Park, D.~Mahajan, E.~Amaro, J.~K. Kim, C.~Shao, A.~Mishra, and
  H.~Esmaeilzadeh, ``From high-level deep neural models to fpgas,'' in
  \emph{The 49th Annual IEEE/ACM International Symposium on
  Microarchitecture}.\hskip 1em plus 0.5em minus 0.4em\relax IEEE Press, 2016,
  p.~17.

\bibitem{delmas2019bit}
A.~Delmas~Lascorz, P.~Judd, D.~M. Stuart, Z.~Poulos, M.~Mahmoud, S.~Sharify,
  M.~Nikolic, K.~Siu, and A.~Moshovos, ``Bit-tactical: A software/hardware
  approach to exploiting value and bit sparsity in neural networks,'' in
  \emph{Proceedings of the Twenty-Fourth International Conference on
  Architectural Support for Programming Languages and Operating Systems}.\hskip
  1em plus 0.5em minus 0.4em\relax ACM, 2019, pp. 749--763.

\bibitem{parashar2019timeloop}
A.~Parashar, P.~Raina, Y.~S. Shao, Y.-H. Chen, V.~A. Ying, A.~Mukkara,
  R.~Venkatesan, B.~Khailany, S.~W. Keckler, and J.~Emer, ``Timeloop: A
  systematic approach to dnn accelerator evaluation,'' in \emph{2019 IEEE
  International Symposium on Performance Analysis of Systems and Software
  (ISPASS)}.\hskip 1em plus 0.5em minus 0.4em\relax IEEE, 2019, pp. 304--315.

\bibitem{song2019hypar}
L.~Song, J.~Mao, Y.~Zhuo, X.~Qian, H.~Li, and Y.~Chen, ``Hypar: Towards hybrid
  parallelism for deep learning accelerator array,'' in \emph{2019 IEEE
  International Symposium on High Performance Computer Architecture
  (HPCA)}.\hskip 1em plus 0.5em minus 0.4em\relax IEEE, 2019, pp. 56--68.

\bibitem{gonccalves2019aggressive}
L.~R. Gon{\c{c}}alves, R.~F.~D. Moura, and L.~Carro, ``Aggressive energy
  reduction for video inference with software-only strategies,'' \emph{ACM
  Transactions on Embedded Computing Systems (TECS)}, 2019.

\bibitem{mahdiani2019deltann}
H.~Mahdiani, A.~Khadem, A.~Ghanbari, M.~Modarressi, F.~Fattahi, and
  M.~Daneshtalab, ``$\delta$nn: Power-efficient neural network acceleration
  using differential weights,'' \emph{IEEE Micro}, 2019.

\bibitem{mahmoud2018diffy}
M.~Mahmoud, K.~Siu, and A.~Moshovos, ``Diffy: A d{\'e}j{\`a} vu-free
  differential deep neural network accelerator,'' in \emph{2018 51st Annual
  IEEE/ACM International Symposium on Microarchitecture (MICRO)}.\hskip 1em
  plus 0.5em minus 0.4em\relax IEEE, 2018, pp. 134--147.

\bibitem{silfa2019neuron}
F.~Silfa, G.~Dot, J.-M. Arnau, and A.~Gonz{\`a}lez, ``Neuron-level fuzzy
  memoization in rnns,'' in \emph{Proceedings of the 52nd Annual IEEE/ACM
  International Symposium on Microarchitecture}, 2019, pp. 782--793.

\bibitem{wang2019none}
Y.~Wang, S.~Liang, H.~Li, and X.~Li, ``A none-sparse inference accelerator that
  distills and reuses the computation redundancy in cnns,'' in
  \emph{Proceedings of the 56th Annual Design Automation Conference
  2019}.\hskip 1em plus 0.5em minus 0.4em\relax ACM, 2019, p. 202.

\bibitem{zhu2018euphrates}
Y.~Zhu, A.~Samajdar, M.~Mattina, and P.~Whatmough, ``Euphrates: Algorithm-soc
  co-design for low-power mobile continuous vision,'' in \emph{2018 ACM/IEEE
  45th Annual International Symposium on Computer Architecture (ISCA)}.\hskip
  1em plus 0.5em minus 0.4em\relax IEEE, 2018, pp. 547--560.

\bibitem{akhlaghi2018snapea}
V.~Akhlaghi, A.~Yazdanbakhsh, K.~Samadi, R.~K. Gupta, and H.~Esmaeilzadeh,
  ``Snapea: Predictive early activation for reducing computation in deep
  convolutional neural networks,'' in \emph{2018 ACM/IEEE 45th Annual
  International Symposium on Computer Architecture (ISCA)}.\hskip 1em plus
  0.5em minus 0.4em\relax IEEE, 2018, pp. 662--673.

\bibitem{zhu2018sparsenn}
J.~Zhu, J.~Jiang, X.~Chen, and C.-Y. Tsui, ``Sparsenn: An energy-efficient
  neural network accelerator exploiting input and output sparsity,'' in
  \emph{2018 Design, Automation \& Test in Europe Conference \& Exhibition
  (DATE)}.\hskip 1em plus 0.5em minus 0.4em\relax IEEE, 2018, pp. 241--244.

\bibitem{lee2018compend}
D.~Lee, S.~Kang, and K.~Choi, ``Compend: Computation pruning through early
  negative detection for relu in a deep neural network accelerator,'' in
  \emph{Proceedings of the 2018 International Conference on Supercomputing},
  2018, pp. 139--148.

\bibitem{miao2015eesen}
Y.~Miao, M.~Gowayyed, and F.~Metze, ``Eesen: End-to-end speech recognition
  using deep rnn models and wfst-based decoding,'' in \emph{2015 IEEE Workshop
  on Automatic Speech Recognition and Understanding (ASRU)}.\hskip 1em plus
  0.5em minus 0.4em\relax IEEE, 2015, pp. 167--174.

\bibitem{bojarski2016end}
M.~Bojarski, D.~Del~Testa, D.~Dworakowski, B.~Firner, B.~Flepp, P.~Goyal
  \emph{et~al.}, ``End to end learning for self-driving cars,'' \emph{arXiv
  preprint arXiv:1604.07316}, 2016.

\bibitem{amodei2016deep}
D.~Amodei, S.~Ananthanarayanan, R.~Anubhai, J.~Bai, E.~Battenberg, C.~Case,
  J.~Casper, B.~Catanzaro, Q.~Cheng, G.~Chen \emph{et~al.}, ``Deep speech 2:
  End-to-end speech recognition in english and mandarin,'' in
  \emph{International conference on machine learning}, 2016, pp. 173--182.

\bibitem{real2017youtube}
E.~Real, J.~Shlens, S.~Mazzocchi, X.~Pan, and V.~Vanhoucke,
  ``Youtube-boundingboxes: A large high-precision human-annotated data set for
  object detection in video,'' in \emph{Proceedings of the IEEE Conference on
  Computer Vision and Pattern Recognition}, 2017, pp. 5296--5305.

\bibitem{krizhevsky2014one}
A.~Krizhevsky, ``One weird trick for parallelizing convolutional neural
  networks,'' \emph{arXiv preprint arXiv:1404.5997}, 2014.

\bibitem{zmora2019neural}
N.~Zmora, G.~Jacob, L.~Zlotnik, B.~Elharar, and G.~Novik, ``Neural network
  distiller: A python package for dnn compression research,'' \emph{arXiv
  preprint arXiv:1910.12232}, 2019.

\bibitem{MKLMemoryFormats}
\BIBentryALTinterwordspacing
{Intel}. Understanding memory formats, intel mkl-dnn. Accessed: 2020-03-03.
  [Online]. Available:
  \url{https://intel.github.io/mkl-dnn/understanding\_memory\_formats.html}
\BIBentrySTDinterwordspacing

\bibitem{jia2014caffe}
Y.~Jia, E.~Shelhamer, J.~Donahue, S.~Karayev, J.~Long, R.~Girshick,
  S.~Guadarrama, and T.~Darrell, ``Caffe: Convolutional architecture for fast
  feature embedding,'' in \emph{Proceedings of the 22nd ACM international
  conference on Multimedia}.\hskip 1em plus 0.5em minus 0.4em\relax ACM, 2014,
  pp. 675--678.

\bibitem{baghdadi2019tiramisu}
R.~Baghdadi, J.~Ray, M.~B. Romdhane, E.~Del~Sozzo, A.~Akkas, Y.~Zhang,
  P.~Suriana, S.~Kamil, and S.~Amarasinghe, ``Tiramisu: A polyhedral compiler
  for expressing fast and portable code,'' in \emph{Proceedings of the 2019
  IEEE/ACM International Symposium on Code Generation and Optimization}.\hskip
  1em plus 0.5em minus 0.4em\relax IEEE Press, 2019, pp. 193--205.

\bibitem{chen2018tvm}
T.~Chen, T.~Moreau, Z.~Jiang, L.~Zheng, E.~Yan, H.~Shen, M.~Cowan, L.~Wang,
  Y.~Hu, L.~Ceze \emph{et~al.}, ``$\{$TVM$\}$: An automated end-to-end
  optimizing compiler for deep learning,'' in \emph{13th $\{$USENIX$\}$
  Symposium on Operating Systems Design and Implementation ($\{$OSDI$\}$ 18)},
  2018.

\bibitem{halide_2012}
J.~Ragan-Kelley, A.~Adams, S.~Paris, M.~Levoy, S.~Amarasinghe, and F.~Durand,
  ``Decoupling algorithms from schedules for easy optimization of image
  processing pipelines,'' \emph{ACM Transactions on Graphics (TOG)}, vol.~31,
  no.~4, pp. 1--12, 2012.

\bibitem{lattner2020mlir}
C.~Lattner, J.~Pienaar, M.~Amini, U.~Bondhugula, R.~Riddle, A.~Cohen,
  T.~Shpeisman, A.~Davis, N.~Vasilache, and O.~Zinenko, ``Mlir: A compiler
  infrastructure for the end of moore's law,'' 2020.

\bibitem{polyhedral}
F.~Paul and L.~Christian, ``The polyhedron model,'' in \emph{Encyclopedia of
  Parallel Computing}, D.~Padua, Ed.\hskip 1em plus 0.5em minus 0.4em\relax
  Springer, 2011, pp. 1581, 1592.

\bibitem{verdoolaege2010isl}
S.~Verdoolaege, ``isl: An integer set library for the polyhedral model,'' in
  \emph{International Congress on Mathematical Software}.\hskip 1em plus 0.5em
  minus 0.4em\relax Springer, 2010.

\bibitem{pencil__paper}
R.~Baghdadi, U.~Beaugnon, A.~Cohen, T.~Grosser, M.~Kruse, C.~Reddy,
  S.~Verdoolaege, A.~Betts, A.~F. Donaldson, J.~Ketema \emph{et~al.}, ``Pencil:
  A platform-neutral compute intermediate language for accelerator
  programming,'' in \emph{2015 International Conference on Parallel
  Architecture and Compilation (PACT)}.\hskip 1em plus 0.5em minus 0.4em\relax
  IEEE, 2015, pp. 138--149.

\bibitem{vasilache2018tensor}
N.~Vasilache, O.~Zinenko, T.~Theodoridis, P.~Goyal, Z.~DeVito, W.~S. Moses,
  S.~Verdoolaege, A.~Adams, and A.~Cohen, ``Tensor comprehensions:
  Framework-agnostic high-performance machine learning abstractions,''
  \emph{arXiv preprint arXiv:1802.04730}, 2018.

\bibitem{elango2018diesel}
V.~Elango, N.~Rubin, M.~Ravishankar, H.~Sandanagobalane, and V.~Grover,
  ``Diesel: Dsl for linear algebra and neural net computations on gpus,'' in
  \emph{Proceedings of the 2nd ACM SIGPLAN International Workshop on Machine
  Learning and Programming Languages}, 2018.

\bibitem{leary2017xla}
C.~Leary and T.~Wang, ``Xla: Tensorflow, compiled,'' \emph{TensorFlow Dev
  Summit}, 2017.

\bibitem{bondhugulapractical2008}
U.~Bondhugula, A.~Hartono, J.~Ramanujam, and P.~Sadayappan, ``A practical
  automatic polyhedral parallelizer and locality optimizer,'' in \emph{PLDI},
  2008, pp. 101--113.

\bibitem{polly}
\BIBentryALTinterwordspacing
T.~Grosser, A.~Groslinger, and C.~Lengauer, ``Polly - performing polyhedral
  optimizations on a low-level intermediate representation.'' \emph{Parallel
  Processing Letters}, vol.~22, no.~4, 2012. [Online]. Available:
  \url{http://dblp.uni-trier.de/db/journals/ppl/ppl22.html\#GrosserGL12}
\BIBentrySTDinterwordspacing

\bibitem{mullapudi2015polymage}
R.~T. Mullapudi, V.~Vasista, and U.~Bondhugula, ``Polymage: Automatic
  optimization for image processing pipelines,'' in \emph{Proceedings of the
  Twentieth International Conference on Architectural Support for Programming
  Languages and Operating Systems}, 2015, pp. 429--443.

\bibitem{yuki2012alphaz}
T.~Yuki, G.~Gupta, D.~Kim, T.~Pathan, and S.~Rajopadhye, ``Alphaz: A system for
  design space exploration in the polyhedral model,'' in \emph{International
  Workshop on Languages and Compilers for Parallel Computing}.\hskip 1em plus
  0.5em minus 0.4em\relax Springer, 2012, pp. 17--31.

\bibitem{chill}
C.~Chen, J.~Chame, and M.~Hall, ``Chill: A framework for composing high-level
  loop transformations,'' U. of Southern California, Tech. Rep. 08-897, 2008.

\bibitem{Benabderrahmane}
M.-W. Benabderrahmane, L.-N. Pouchet, A.~Cohen, and C.~Bastoul, ``The
  polyhedral model is more widely applicable than you think,'' in
  \emph{Proceedings of the 19th Joint European Conference on Theory and
  Practice of Software, International Conference on Compiler Construction},
  ser. CC'10/ETAPS'10.\hskip 1em plus 0.5em minus 0.4em\relax Springer-Verlag,
  2010.

\bibitem{hartono2009parametric}
A.~Hartono, M.~M. Baskaran, C.~Bastoul, A.~Cohen, S.~Krishnamoorthy, B.~Norris,
  J.~Ramanujam, and P.~Sadayappan, ``Parametric multi-level tiling of
  imperfectly nested loops,'' in \emph{Proceedings of the 23rd international
  conference on Supercomputing}, 2009, pp. 147--157.

\bibitem{pencil}
\BIBentryALTinterwordspacing
R.~Baghdadi, A.~Cohen, T.~Grosser, S.~Verdoolaege, A.~Lokhmotov, J.~Absar,
  S.~van Haastregt, A.~Kravets, and A.~F. Donaldson, ``{PENCIL} language
  specification,'' {INRIA}, Research Rep. RR-8706, 2015. [Online]. Available:
  \url{https://hal.inria.fr/hal-01154812}
\BIBentrySTDinterwordspacing

\bibitem{48842}
R.~Baghdadi and A.~Cohen, ``Scalable polyhedral compilation, syntax vs.
  semantics: 1–0 in the first round,'' in \emph{IMPACT 2020 workshop
  (associated with HIPEAC 2020)}, 2020, informal proceedings.

\bibitem{wei2017dlvm}
R.~Wei, L.~Schwartz, and V.~Adve, ``Dlvm: A modern compiler infrastructure for
  deep learning systems,'' \emph{arXiv:1711.03016}, 2017.

\bibitem{truong2016latte}
L.~Truong, R.~Barik, E.~Totoni, H.~Liu, C.~Markley, A.~Fox, and T.~Shpeisman,
  ``Latte: a language, compiler, and runtime for elegant and efficient deep
  neural networks,'' in \emph{Proceedings of the 37th ACM SIGPLAN Conference on
  Programming Language Design and Implementation}, 2016, pp. 209--223.

\bibitem{feautrier1991dataflow}
P.~Feautrier, ``Dataflow analysis of array and scalar references,''
  \emph{International Journal of Parallel Programming}, vol.~20, no.~1, 1991.

\bibitem{wolf1992improving}
M.~E. Wolf, ``Improving locality and parallelism in nested loops,'' Ph.D.
  dissertation, to the Department of Computer Science.Stanford University,
  1992.

\bibitem{hu2019taichi}
Y.~Hu, T.-M. Li, L.~Anderson, J.~Ragan-Kelley, and F.~Durand, ``Taichi: a
  language for high-performance computation on spatially sparse data
  structures,'' \emph{ACM Transactions on Graphics (TOG)}, pp. 1--16, 2019.

\bibitem{kjolstad2017tensor}
F.~Kjolstad, S.~Kamil, S.~Chou, D.~Lugato, and S.~Amarasinghe, ``The tensor
  algebra compiler,'' \emph{Proceedings of the ACM on Programming Languages},
  vol.~1, no. OOPSLA, pp. 1--29, 2017.

\bibitem{goto2008anatomy}
K.~Goto and R.~A. v.~d. Geijn, ``Anatomy of high-performance matrix
  multiplication,'' \emph{ACM Transactions on Mathematical Software (TOMS)},
  vol.~34, no.~3, pp. 1--25, 2008.

\bibitem{trifunovic2009polyhedral}
K.~Trifunovic, D.~Nuzman, A.~Cohen, A.~Zaks, and I.~Rosen, ``Polyhedral-model
  guided loop-nest auto-vectorization,'' in \emph{2009 18th International
  Conference on Parallel Architectures and Compilation Techniques}.\hskip 1em
  plus 0.5em minus 0.4em\relax IEEE, 2009, pp. 327--337.

\bibitem{agakov2006using}
F.~Agakov, E.~Bonilla, J.~Cavazos, B.~Franke, G.~Fursin, M.~F. O'Boyle,
  J.~Thomson, M.~Toussaint, and C.~K. Williams, ``Using machine learning to
  focus iterative optimization,'' in \emph{International Symposium on Code
  Generation and Optimization (CGO'06)}.\hskip 1em plus 0.5em minus 0.4em\relax
  IEEE, 2006.

\bibitem{adams2019learning}
A.~Adams, K.~Ma, L.~Anderson, R.~Baghdadi, T.-M. Li, M.~Gharbi, B.~Steiner,
  S.~Johnson, K.~Fatahalian, F.~Durand \emph{et~al.}, ``Learning to optimize
  halide with tree search and random programs,'' \emph{ACM Transactions on
  Graphics (TOG)}, vol.~38, no.~4, pp. 1--12, 2019.

\bibitem{mendis2019ithemal}
C.~Mendis, A.~Renda, S.~Amarasinghe, and M.~Carbin, ``Ithemal: Accurate,
  portable and fast basic block throughput estimation using deep neural
  networks,'' in \emph{International Conference on Machine Learning}, 2019, pp.
  4505--4515.

\bibitem{lattner2004llvm}
C.~Lattner and V.~Adve, ``Llvm: A compilation framework for lifelong program
  analysis \& transformation,'' in \emph{International Symposium on Code
  Generation and Optimization, 2004. CGO 2004.}, 2004.

\bibitem{venkataramani2017scaledeep}
S.~Venkataramani, A.~Ranjan, S.~Banerjee, D.~Das, S.~Avancha, A.~Jagannathan,
  A.~Durg, D.~Nagaraj, B.~Kaul, P.~Dubey \emph{et~al.}, ``Scaledeep: A scalable
  compute architecture for learning and evaluating deep networks,'' \emph{ACM
  SIGARCH Computer Architecture News}, 2017.

\bibitem{chen2019instruction}
Y.~Chen, H.~Lan, Z.~Du, S.~Liu, J.~Tao, D.~Han, T.~Luo, Q.~Guo, L.~Li, Y.~Xie
  \emph{et~al.}, ``An instruction set architecture for machine learning,''
  \emph{ACM Transactions on Computer Systems (TOCS)}, 2019.

\bibitem{gopinath2019compiling}
S.~Gopinath, N.~Ghanathe, V.~Seshadri, and R.~Sharma, ``Compiling kb-sized
  machine learning models to tiny iot devices,'' in \emph{Proceedings of the
  40th ACM SIGPLAN Conference on Programming Language Design and
  Implementation}, 2019, pp. 79--95.

\bibitem{moreau2018hardware}
T.~Moreau, T.~Chen, L.~Vega, J.~Roesch, E.~Yan, L.~Zheng, J.~Fromm, Z.~Jiang,
  L.~Ceze, C.~Guestrin \emph{et~al.}, ``A hardware-software blueprint for
  flexible deep learning specialization,'' \emph{arXiv:1807.04188}, 2018.

\bibitem{chen2015mxnet}
T.~Chen, M.~Li, Y.~Li, M.~Lin, N.~Wang, M.~Wang, T.~Xiao, B.~Xu, C.~Zhang, and
  Z.~Zhang, ``Mxnet: A flexible and efficient machine learning library for
  heterogeneous distributed systems,'' \emph{arXiv preprint arXiv:1512.01274},
  2015.

\bibitem{tung2018clip}
F.~Tung and G.~Mori, ``Clip-q: Deep network compression learning by in-parallel
  pruning-quantization,'' in \emph{Proceedings of the IEEE Conference on
  Computer Vision and Pattern Recognition}, 2018.

\bibitem{yang2019automatic}
H.~Yang, S.~Gui, Y.~Zhu, and J.~Liu, ``Automatic neural network compression by
  sparsity-quantization joint learning: A constrained optimization-based
  approach,'' 2019.

\bibitem{kwon2018co}
K.~Kwon, A.~Amid, A.~Gholami, B.~Wu, K.~Asanovic, and K.~Keutzer, ``Co-design
  of deep neural nets and neural net accelerators for embedded vision
  applications,'' in \emph{2018 55th ACM/ESDA/IEEE Design Automation Conference
  (DAC)}.\hskip 1em plus 0.5em minus 0.4em\relax IEEE, 2018, pp. 1--6.

\bibitem{marculescu2018hardware}
D.~Marculescu, D.~Stamoulis, and E.~Cai, ``Hardware-aware machine learning:
  Modeling and optimization,'' in \emph{Proceedings of the International
  Conference on Computer-Aided Design}, 2018, pp. 1--8.

\bibitem{zhang2019neural}
X.~Zhang, W.~Jiang, Y.~Shi, and J.~Hu, ``When neural architecture search meets
  hardware implementation: from hardware awareness to co-design,'' in
  \emph{2019 IEEE Computer Society Annual Symposium on VLSI (ISVLSI)}.\hskip
  1em plus 0.5em minus 0.4em\relax IEEE, 2019, pp. 25--30.

\bibitem{abdelfattah2020best}
M.~S. Abdelfattah, {\L}.~Dudziak, T.~Chau, R.~Lee, H.~Kim, and N.~D. Lane,
  ``Best of both worlds: Automl codesign of a cnn and its hardware
  accelerator,'' \emph{arXiv preprint arXiv:2002.05022}, 2020.

\bibitem{he2018amc}
Y.~He, J.~Lin, Z.~Liu, H.~Wang, L.-J. Li, and S.~Han, ``Amc: Automl for model
  compression and acceleration on mobile devices,'' in \emph{Proceedings of the
  European Conference on Computer Vision (ECCV)}, 2018.

\bibitem{ji2018recom}
H.~Ji, L.~Song, L.~Jiang, H.~H. Li, and Y.~Chen, ``Recom: An efficient
  resistive accelerator for compressed deep neural networks,'' in \emph{2018
  Design, Automation \& Test in Europe Conference \& Exhibition (DATE)}.\hskip
  1em plus 0.5em minus 0.4em\relax IEEE, 2018, pp. 237--240.

\bibitem{wang2018snrram}
P.~Wang, Y.~Ji, C.~Hong, Y.~Lyu, D.~Wang, and Y.~Xie, ``Snrram: an efficient
  sparse neural network computation architecture based on resistive
  random-access memory,'' in \emph{Proceedings of the 55th Annual Design
  Automation Conference}.\hskip 1em plus 0.5em minus 0.4em\relax ACM, 2018, p.
  106.

\bibitem{zhang2018dnnbuilder}
X.~Zhang, J.~Wang, C.~Zhu, Y.~Lin, J.~Xiong, W.-m. Hwu, and D.~Chen,
  ``Dnnbuilder: an automated tool for building high-performance dnn hardware
  accelerators for fpgas,'' in \emph{Proceedings of the International
  Conference on Computer-Aided Design}.\hskip 1em plus 0.5em minus 0.4em\relax
  ACM, 2018, p.~56.

\bibitem{srivastava2019t2s}
N.~Srivastava, H.~Rong, P.~Barua, G.~Feng, H.~Cao, Z.~Zhang \emph{et~al.},
  ``T2s-tensor: Productively generating high-performance spatial hardware for
  dense tensor computations,'' in \emph{2019 IEEE 27th Annual International
  Symposium on Field-Programmable Custom Computing Machines (FCCM)}.\hskip 1em
  plus 0.5em minus 0.4em\relax IEEE, 2019, pp. 181--189.

\bibitem{lai2019heterocl}
Y.-H. Lai, Y.~Chi, Y.~Hu, J.~Wang, C.~H. Yu, Y.~Zhou, J.~Cong, and Z.~Zhang,
  ``Heterocl: A multi-paradigm programming infrastructure for software-defined
  reconfigurable computing,'' in \emph{Proceedings of the 2019 ACM/SIGDA
  International Symposium on Field-Programmable Gate Arrays}.\hskip 1em plus
  0.5em minus 0.4em\relax ACM, 2019, pp. 242--251.

\bibitem{venkatesan2019magnet}
R.~Venkatesan, Y.~S. Shao, M.~Wang, J.~Clemons, S.~Dai, M.~Fojtik, B.~Keller,
  A.~Klinefelter, N.~Pinckney, P.~Raina \emph{et~al.}, ``Magnet: A modular
  accelerator generator for neural networks,'' in \emph{2019 IEEE/ACM
  International Conference on Computer-Aided Design (ICCAD)}, 2019.

\bibitem{bachrach2012chisel}
J.~Bachrach, H.~Vo, B.~Richards, Y.~Lee, A.~Waterman, R.~Avi{\v{z}}ienis
  \emph{et~al.}, ``Chisel: constructing hardware in a scala embedded
  language,'' in \emph{DAC Design Automation Conference 2012}, 2012, pp.
  1212--1221.

\bibitem{sharifian2019muir}
A.~Sharifian, R.~Hojabr, N.~Rahimi, S.~Liu, A.~Guha, T.~Nowatzki, and
  A.~Shriraman, ``$\mu$ir-an intermediate representation for transforming and
  optimizing the microarchitecture of application accelerators,'' in
  \emph{Proceedings of the 52nd Annual IEEE/ACM International Symposium on
  Microarchitecture}, 2019, pp. 940--953.

\bibitem{graphicionado}
T.~J. Ham, L.~Wu, N.~Sundaram, N.~Satish, and M.~Martonosi, ``Graphicionado: A
  high-performance and energy-efficient accelerator for graph analytics,'' in
  \emph{2016 49th Annual IEEE/ACM International Symposium on Microarchitecture
  (MICRO)}.\hskip 1em plus 0.5em minus 0.4em\relax IEEE, 2016, pp. 1--13.

\bibitem{tesseract}
J.~Ahn, S.~Hong, S.~Yoo, O.~Mutlu, and K.~Choi, ``A scalable
  processing-in-memory accelerator for parallel graph processing,'' in
  \emph{Proceedings of the 42nd Annual International Symposium on Computer
  Architecture}, 2015, pp. 105--117.

\bibitem{q100}
L.~Wu, A.~Lottarini, T.~K. Paine, M.~A. Kim, and K.~A. Ross, ``Q100: the
  architecture and design of a database processing unit,'' in \emph{Proceedings
  of the 19th international conference on Architectural support for programming
  languages and operating systems}, 2014.

\bibitem{darwin}
Y.~Turakhia, G.~Bejerano, and W.~J. Dally, ``Darwin: A genomics co-processor
  provides up to 15,000 x acceleration on long read assembly,'' in
  \emph{Proceedings of the Twenty-Third International Conference on
  Architectural Support for Programming Languages and Operating Systems}, 2018,
  pp. 199--213.

\bibitem{genax}
D.~Fujiki, A.~Subramaniyan, T.~Zhang, Y.~Zeng, R.~Das, D.~Blaauw, and
  S.~Narayanasamy, ``Genax: A genome sequencing accelerator,'' in \emph{2018
  ACM/IEEE 45th Annual International Symposium on Computer Architecture
  (ISCA)}.\hskip 1em plus 0.5em minus 0.4em\relax IEEE, 2018, pp. 69--82.

\bibitem{fowers2015comrpession}
J.~Fowers, J.-Y. Kim, D.~Burger, and S.~Hauck, ``A scalable high-bandwidth
  architecture for lossless compression on fpgas,'' in \emph{2015 IEEE 23rd
  Annual International Symposium on Field-Programmable Custom Computing
  Machines}.\hskip 1em plus 0.5em minus 0.4em\relax IEEE, 2015, pp. 52--59.

\bibitem{gu2018recent}
J.~Gu, Z.~Wang, J.~Kuen, L.~Ma, A.~Shahroudy, B.~Shuai, T.~Liu, X.~Wang,
  G.~Wang, J.~Cai \emph{et~al.}, ``Recent advances in convolutional neural
  networks,'' \emph{Pattern Recognition}, vol.~77, pp. 354--377, 2018.

\bibitem{wei2020review}
Y.~Wei, J.~Zhou, Y.~Wang, Y.~Liu, Q.~Liu, J.~Luo, C.~Wang, F.~Ren, and
  L.~Huang, ``A review of algorithm \& hardware design for ai-based biomedical
  applications.'' \emph{IEEE Transactions on Biomedical Circuits and Systems},
  vol.~14, no.~2, pp. 145--163, 2020.

\bibitem{elsken2019neural}
T.~Elsken, J.~H. Metzen, and F.~Hutter, ``Neural architecture search: A
  survey,'' \emph{Journal of Machine Learning Research}, 2019.

\bibitem{cheng2018model}
Y.~Cheng, D.~Wang, P.~Zhou, and T.~Zhang, ``Model compression and acceleration
  for deep neural networks: The principles, progress, and challenges,''
  \emph{IEEE Signal Processing Magazine}, vol.~35, no.~1, 2018.

\bibitem{wang2019deep}
E.~Wang, J.~J. Davis, R.~Zhao, H.-C. Ng, X.~Niu, W.~Luk, P.~Y. Cheung, and
  G.~A. Constantinides, ``Deep neural network approximation for custom
  hardware: Where we've been, where we're going,'' \emph{ACM Computing Surveys
  (CSUR)}, vol.~52, no.~2, p.~40, 2019.

\bibitem{shawahna2018fpga}
A.~Shawahna, S.~M. Sait, and A.~El-Maleh, ``Fpga-based accelerators of deep
  learning networks for learning and classification: A review,'' \emph{IEEE
  Access}, vol.~7, pp. 7823--7859, 2018.

\bibitem{venieris2018toolflows}
S.~I. Venieris, A.~Kouris, and C.-S. Bouganis, ``Toolflows for mapping
  convolutional neural networks on fpgas: A survey and future directions,''
  \emph{ACM Computing Surveys (CSUR)}, vol.~51, no.~3, 2018.

\bibitem{reuther2019survey}
A.~Reuther, P.~Michaleas, M.~Jones, V.~Gadepally, S.~Samsi, and J.~Kepner,
  ``Survey and benchmarking of machine learning accelerators,'' in \emph{2019
  IEEE High Performance Extreme Computing Conference (HPEC)}.\hskip 1em plus
  0.5em minus 0.4em\relax IEEE, 2019, pp. 1--9.

\bibitem{li2020deep}
M.~Li, Y.~Liu, X.~Liu, Q.~Sun, X.~You, H.~Yang \emph{et~al.}, ``The deep
  learning compiler: A comprehensive survey,'' \emph{arXiv:2002.03794}, 2020.

\bibitem{mittal2018survey}
S.~Mittal, ``A survey of fpga-based accelerators for convolutional neural
  networks,'' \emph{Neural computing and applications}, pp. 1--31, 2018.

\bibitem{du2020self}
B.~Du, Q.~Guo, Y.~Zhao, T.~Zhi, Y.~Chen, and Z.~Xu, ``Self-aware neural network
  systems: A survey and new perspective,'' \emph{Proceedings of the IEEE},
  2020.

\bibitem{ignatov2019ai}
A.~Ignatov, R.~Timofte, A.~Kulik, S.~Yang, K.~Wang, F.~Baum, M.~Wu, L.~Xu, and
  L.~Van~Gool, ``Ai benchmark: All about deep learning on smartphones in
  2019,'' \emph{arXiv preprint arXiv:1910.06663}, 2019.

\bibitem{gale2020sparse}
T.~Gale, M.~Zaharia, C.~Young, and E.~Elsen, ``Sparse gpu kernels for deep
  learning,'' in \emph{Proceedings of the International Conference for High
  Performance Computing, Networking, Storage and Analysis}, 2020.

\end{thebibliography}

\end{document}